%% file: main.tex
\documentclass{article}

\usepackage[english]{babel}

\usepackage[letterpaper,top=2cm,bottom=2cm,left=3cm,right=3cm,marginparwidth=1.75cm]{geometry}

\usepackage{amsmath}
\usepackage{graphicx}
\usepackage{subcaption}  
\usepackage[colorlinks=true, linkcolor=black, citecolor=blue, urlcolor=blue]{hyperref}

\usepackage[sort&compress]{natbib}
\usepackage{multirow}
\usepackage{rotating}
\usepackage{makecell}

\usepackage{xcolor}

\usepackage{listings}  

\definecolor{codegreen}{rgb}{0,0.6,0}
\definecolor{codegray}{rgb}{0.5,0.5,0.5}
\definecolor{codepurple}{rgb}{0.58,0,0.82}
\definecolor{backcolor}{rgb}{0.95,0.95,0.92}

\lstdefinestyle{mystyle}{
    backgroundcolor=\color{backcolor},   
    commentstyle=\color{codegreen},
    keywordstyle=\color{magenta},
    numberstyle=\tiny\color{codegray},
    stringstyle=\color{codepurple},
    basicstyle=\ttfamily\footnotesize,
    breakatwhitespace=false,         
    breaklines=true,                 
    captionpos=b,                    
    keepspaces=true,                 
    numbers=left,                    
    numbersep=5pt,                  
    showspaces=false,                
    showstringspaces=false,
    showtabs=false,                  
    tabsize=2
}

\usepackage[acronym, nonumberlist, toc, section=section,automake]{glossaries-extra}
\setabbreviationstyle[acronym]{long-short-user}
\input{abbreviations.tex}

\newcommand\vectorize[1]{\vec{\mathbf{#1}}}

\title{Gravitational Microlensing}
\author{Aikaterini Vandorou \& Stela {Ishitani Silva}}

\begin{document}
\maketitle

\begin{abstract}
Gravitational microlensing is a unique planet-detection method that does not rely on detecting light from the host star or planet. Instead, planets are discovered when their gravitational field perturbs the light coming from a distant source. 
The mechanics of this method offer several advantages. It has the ability to detect wide-orbit planets of various masses around host stars of different types, at and beyond the system's snow line. Thus, microlensing complements the other detecting techniques that are more sensitive to close-orbit or high-mass planets.
In this Chapter, we review the microlensing method and discuss how the physical properties, such as mass and distance, can be extracted from the microlensing light curve or from additional high-angular resolution observations. Microlensing is essential for the broader context of exoplanet demographics, as it can place constraints on planets, in wide orbits, beyond the {\it snow line}. Studying this parameter space can help us unravel the details of planet formation and how our own Solar System formed. Lastly, we provide an example of how to model a light curve and obtain accurate physical parameters for the host and planet.

\end{abstract}

\section{Introduction}

Gravitational lensing is a phenomenon that occurs when light is deflected due to the gravitational field of a massive body. Its origins stem from the theory of general relativity formulated by Albert Einstein in the early twentieth century. However, the concept that gravity can bend light far precedes this, with its roots tracing back to Sir Isaac Newton. In the early 1800s, the bending of light around the Sun was calculated by Johann Georg von Soldner using Newtonian mechanics, but it was Einstein's theory of general relativity that provided a more complete description: that massive objects cause a curvature in spacetime, and light follows this curvature. 

Gravitational microlensing is a type of lensing where multiple images are created due to the deflection of light by gravity but have separations of less than a few milliarcseconds and are therefore unresolved. The equations of microlensing by a foreground star aligning with a background star were first published by Einstein in 1936 \citep{Einstein1936}. Additional published references to the phenomenon of microlensing also exist \citep{Lodge1919, Eddington1920, Chwolson1924}. However, when Einstein presented his derivations, he dismissed the practical significance of the microlensing effect, remarking that ``there is no great chance of observing this phenomenon" \citep{Einstein1936}. 

Einstein's pessimism, however, was based in truth: there is ``no great chance" of observing a gravitational microlensing event. The chance alignment between an observer, foreground star and background star has a $10^{-6}$ probability of occurring when observing towards the Galactic Bulge \citep{alcock2000macho}, and can be even less for other lines of sight. This low probability partly contributed to gravitational microlensing laying mostly dormant for decades (however, see \citealt{Liebes1964} and \citealt{Refsdal1964}). 

In 1986, however, a paper by Bohdan Paczyński resurrected the idea and the hunt for microlensing events towards the Large and Small Magellanic Clouds and Galactic Bulge began in the early 1990s \citep{Alcock1993, Udalski1993, Aubourg1993}. His idea was based on the new developments being made in CCD detectors, which had larger field-of-views, enabling billions of stars to be observed simultaneously every night. In 1991, Mao and Paczy\'nski proposed that the microlensing method could also be used to detect exoplanets. They predicted that 10\% of the microlensing events  to-be observed could be binary-lens events, in which one of the lens components could be a planet \citep{Mao1991}.

The first microlensing surveys began in the early 1990's with a primary goal of detecting dark matter in our galaxy in the form of \glspl{MACHO}. This gave birth to the \gls{MACHO} project (1992-1999,\citealt{alcock2000macho}), along with the \gls{EROS} (1989-2003) and \gls{OGLE}. All these collaborations detected microlensing events and placed constraints on the number of Massive Compact Halo Objects in the Galaxy (see \citealt{alcock2000macho}, \citealt{Tisserand2007} and \citealt{Udalski1994}). 

 In the early 2000s, \textbf{the} \gls{MOA} collaboration joined forces with \gls{OGLE}, searching for planetary systems. One of the earliest planetary discoveries through microlensing was event OGLE-2003-BLG-235/MOA-2003-BLG-53 \citep{Bond2004}, detected in July 2003 by \gls{OGLE}'s 1.3m Warsaw telescope at Las Campanas Observatory in Chile, and \gls{MOA}'s 0.6 metre telescope at the Mt John observatory in New Zealand. In the mid-2010s, a network of three wide-field telescopes distributed across the globe also joined the effort of continuously monitoring the sky in search of microlensing events. The network, known as \gls{KMTNet}, has three 1.6 metre telescopes at \gls{CTIO}, Chile; \gls{SAAO}, South Africa; and at \gls{SSO} in Australia. In addition to these surveys, the \gls{PRIME} survey began operations in 2023. Its 1.8m telescope at \gls{SAAO} is the first dedicated wide-field near-infrared microlensing survey telescope.

Continuously monitoring the Galactic Bulge, where stellar density is the highest, is crucial for a microlensing event to be detected. However, the microlensing light curve also needs to be well-sampled for useful information to be extracted. Once survey collaborations have detected an event, the data is reduced in real time, and the event is alerted. This means that follow-up collaborations can monitor these probable microlensing events. These teams include \gls{PLANET}, Robonet \citep{Tsapras2009}, \gls{MiNDSTEp}, \gls{muFUN}, and OMEGA \citep{Bachelet_2024}.

In the following sections, we explain the microlensing method, with descriptions of a single lens and binary lens system, in addition to how the physical parameters can be measured. We then discuss the demographics of exoplanets discovered through microlensing, followed by an example of data analysis. The different microlensing parameters that we discuss in this chapter can be found in Table \ref{tab:lensing_parameters}.

\section{Method Description}

It is known that the gravitational field of an object can act similarly to an optical lens, magnifying light from a background star. This phenomenon can produce multiple images for a single source, whose number and characteristics depend on the configuration and properties of the lens system. Most microlensing events, however, do not produce resolvable images with current observational instruments. Thus, the phenomenon is typically studied by analyzing the total magnification of these images by monitoring the changes in brightness of the background star over time. Nevertheless, advanced interferometric observations, such as those performed with GRAVITY in 2019 \citep{dong2019first}, PIONIER in 2022 \citep{cassan2022microlensing}, and GRAVITY in 2024 \cite{wu2024first} (the first resolution of microlensed images of a binary-lens event), have succeeded in resolving microlensed images in exceptional cases. In these observations, the main challenge lies not in angular resolution, but rather in the brightness of the target and the availability of suitable nearby fringe-tracker stars.

Unlike other lensing effects, such as strong lensing, where the resulting images can persist for years, planetary microlensing events typically last only a few weeks. Due to the relative motion of stars in our Galaxy, different combinations of source stars and lens systems can become aligned over time, enabling the observation of many different microlensing events. Observing regions with high stellar density, such as the Galactic Bulge, significantly increases the likelihood of detecting such events and is therefore preferred for microlensing-dedicated surveys.

For didactic purposes, this section organizes the description of the physics behind microlensing according to the number of objects in the lens system. Table ~\ref{tab:lensing_parameters} presents the main parameters used in microlensing studies, along with their explanations.

\begin{figure}[b!]
    \centering
    \includegraphics[width=0.8\linewidth]{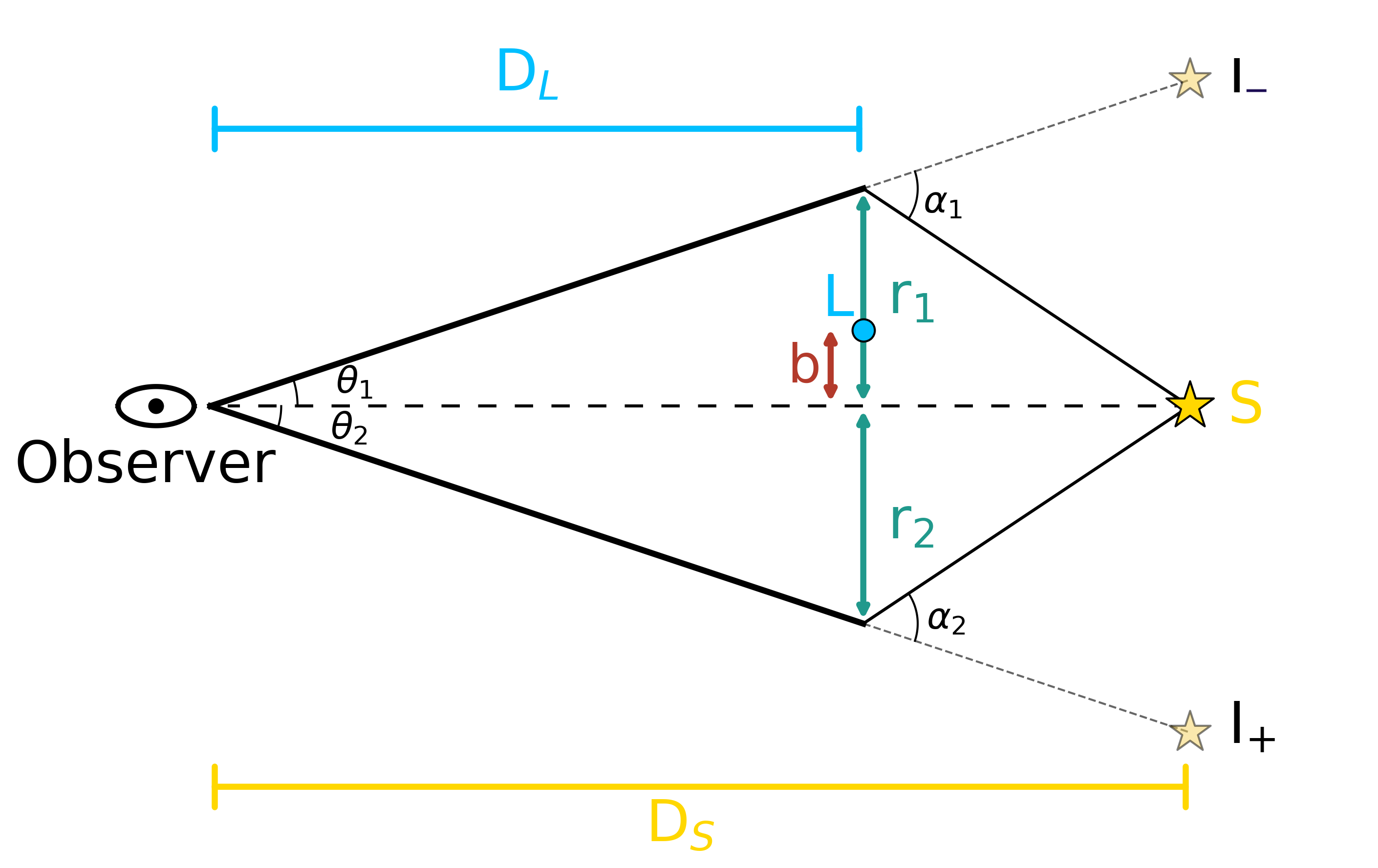}
    \caption{Geometry of a microlensing event with a single lens and a single source. The gravitational field of one object can act similarly to an optical lens and create two images of a single source star. During microlensing events, it is typical that the instrumental resolution is not enough to resolve both images that were created by the presence of the lens when aligned with the source star. What is commonly studied is the magnification of the source, as it is the sum of the brightness of the two images. Details of the parameters shown can be found in Section \ref{subsec:single_lens}.}
    \label{fig:geometry_lens}
\end{figure}

\subsection{Single Lens Event}\label{subsec:single_lens}

As usual, the simplest scenario is that in which the lens system is composed of one object, as shown in Figure~\ref{fig:geometry_lens}. In the microlensing community, the lens is often denoted by L and the source by S, a convention we follow throughout this Chapter. Accordingly, the distance from the observer to the lens is denoted $D_\mathrm{L}$, and the distance from the observer to the source is $D_\mathrm{S}$. In the case of one object acting as a lens, two images are created, denoted $I_+$ and $I_-$. When the source star and the lens object of a microlensing event are infinitesimal points, this event is commonly called a \gls{PSPL} event. A \gls{PSPL} is a type of \gls{1L1S} event.

Modelling a microlensing event essentially involves determining the set of parameters that best reproduce the observed magnification. Therefore, in this subsection, we walk through how we derive the magnification (see Figure~\ref{fig:paczynski_curve}) of a \gls{PSPL} event, considering the lensing geometry shown in Figure~\ref{fig:geometry_lens}. We base our calculations on the works referenced in \citet{bennett2008detection} and \citet{gaudi2012microlensing}.

Based on the geometry presented in Figure~\ref{fig:geometry_lens}, 
we see $\theta_1 \approx \frac{r_1}{D_\mathrm{L}} $ and $\alpha_1 ( D_\mathrm{S} - D_\mathrm{L}) \approx \theta_1 D_\mathrm{S}$. Therefore, the geometry gives an $\alpha_1$:
\begin{equation}
\alpha_1 = \left[\frac{r_1}{D_\mathrm{L}}\right] D_\mathrm{S} \frac{1}{(D_\mathrm{S} - D_\mathrm{L})}.
\label{eq:geometry_bend}
\end{equation}

According to General Relativity, a light ray passing a distance $r$ from a mass $M$, here $M_\mathrm{L}$ as the mass of the lens, would be bent by an angle $\alpha$, which is given by:

\begin{equation}
\label{eq:gravity_bend}
\alpha = \frac{4 G M_\mathrm{L}}{c^2 r},
\end{equation}
where $G$ is the gravitational constant, $G = 6.67430(15)\times 10^{-11} \; \mathrm{m}^3 \; \mathrm{kg}^{-1} \; \mathrm{s}^{-2}$, and $c$ is the speed of light, $c = 299792458 \; \mathrm{m} \; \mathrm{s}^{-1}$. By combining Equation~\ref{eq:geometry_bend} and Equation~\ref{eq:gravity_bend}, we obtain Equation~\ref{eq:theory_geometry}, which can then be rearranged into Equation~\ref{eq:theory_geometry_rearrange}:

\begin{equation}
 \alpha_1 = \frac{4 G M_\mathrm{L}}{c^2 (r_1 - b)} = \frac{r_1}{D_\mathrm{L}} \frac{D_\mathrm{S}}{D_\mathrm{S} - D_\mathrm{L}},
 \label{eq:theory_geometry}
\end{equation}
 
\begin{equation}
 r_1(r_1 - b) = \frac{D_\mathrm{L}(D_\mathrm{S} - D_\mathrm{L})}{D_\mathrm{S}}\frac{4 G M_\mathrm{L}}{c^2}.
\label{eq:theory_geometry_rearrange}
\end{equation}

Note that a special case occurs when the source is perfectly aligned right behind the lens (that is, when $b = 0$). In this scenario, the resulting images, $I_+$ and $I_-$, merge to form an Einstein ring with an Einstein radius $R_\mathrm{E}$. When $b = 0$, Equation~\ref{eq:theory_geometry_rearrange} can be rewritten as:

    \begin{equation}
    R_\mathrm{E} = \sqrt {\frac{D_\mathrm{L}(D_\mathrm{S} - D_\mathrm{L})}{D_\mathrm{S} }\frac{4 G M_\mathrm{L}}{c^2}}.
    \end{equation}
    
The angular Einstein radius $\theta_\mathrm{E}$ is geometrically defined as $    \theta_\mathrm{E} = \frac{R_\mathrm{E}}{D_\mathrm{L}}$. Consequently, $\theta_\mathrm{E}$, a principal parameter that sets the angular scale of a gravitational microlensing event, can be expressed as a function of the lensing phenomenon as follows:

\begin{equation}
\theta_\mathrm{E} = \sqrt{\frac{4GM_L}{c^2D_S}\left(\frac{D_S}{D_L}-1\right)}.
\label{eqn:einstein_radius}
\end{equation}

By substituting $R_\mathrm{E}$ into Equation~\ref{eq:theory_geometry_rearrange}, we obtain a second-order equation, the two solutions of which indicate the position of the images:

\begin{equation}
r_1^2 - r_1 b - R_\mathrm{E}^2 = 0
\label{eqn:second_order_einstein_radius}
\end{equation}

\begin{equation}
r_+ = \frac{b + \sqrt{b^2 + 4 R_\mathrm{E}^2}}{2}
\end{equation}

\begin{equation}
r_- = \frac{b - \sqrt{b^2 + 4 R_\mathrm{E}^2}}{2}
\end{equation}

{Repeating the steps of Equations~1--7 after rewriting Equation~1 in terms of $r_2$ yields the same solutions, $r_+$ and $r_-$, as for $r_1$. Be mindful that if the orientation of the vertical axis associated with $r_2$ is not taken into account (that is, $r_2$ is in the opposite direction of $r_1$ in this figure), these may appear as $-r_+$ and $-r_-$.  

Although our earlier steps used indexes 1 and 2 to follow the geometry, the image positions are consistently referred to as $r_+$ and $r_-$, as these represent the solutions to the quadratic equation. Referring to the images as either positive or negative follows a microlensing convention: the positive ($+$, major) image always lies outside the Einstein ring, whereas the negative ($-$, minor) image lies inside it. In Figure~\ref{fig:geometry_lens}, we represent the negative image as $I_-$, and the positive image as $I_+$.

Defining a dimensionless lens-source separation $u\equiv \frac{b}{R_\mathrm{E}}$ and a dimensionless position of the images $y\equiv \frac{r}{R_\mathrm{E}}$, the equation for the image positions can be rewritten as:

\begin{align}
\frac{r_+}{R_\mathrm{E}} \equiv y_+ &= \frac{u + \sqrt{u^2 + 4}}{2} & \textnormal{ and}\\
\frac{r_-}{R_\mathrm{E}} \equiv y_- &= \frac{u - \sqrt{u^2 + 4}}{2}.
\end{align}

In this form, the Einstein radius {$R_\mathrm{E}$ acts as a scaling factor, and the image positions depend only on the normalized separation $u$. As a consequence, microlensing observables provide access to quantities expressed in units of the Einstein radius, rather than to the physical scale itself. This normalization also explains why binary-lens models are commonly described using parameters normalized by $R_\mathrm{E}$, and why independent measurements of $R_\mathrm{E}$ or $\theta_\mathrm{E}$ are required to determine the physical properties of the lens system.

To calculate the magnification $A_\pm$ of each image ($+$ and $-$) of the source star, we use a Jacobian matrix $J$ that acts as a conversion factor between an area in the source plane and the corresponding area in the lens plane, which is distorted by the microlens.  The magnification of a point-source image is formally the Jacobian of the inverse transformation. Thus, we can calculate the magnification by taking the inverse of the determinant of the Jacobian $J$ of the lens equation evaluated at the location of each image.

\begin{equation}
\label{eq:amplification_jacobiano}
 A_\pm = \frac{1}{\mathrm{det}J}\biggr|_{r = r_\pm} = \biggl|\frac{y_\pm}{u} \frac{dy_\pm}{du}\biggr| = 
 \frac{1}{2}\biggl[\frac{u^2 + 2}{u \sqrt{u ^2 + 4}} \pm 1 \biggr].
\end{equation}

The total magnification is the sum of $A_+$ and $A_-$:

\begin{equation}
    A(u) = A_+ + A_- = \frac{u^2 + 2}{u \sqrt{u ^2 + 4}}.
    \label{eq:amplification}
\end{equation}

Because both the lens and the source move relative to the observer, the dimensionless lens-source separation $u$ changes over time, making the amplification time dependent.  For that, we define $u_0$ as the minimum separation that occurs at the time of the peak $t_0$. We can also define $\tau(t)$ as a normalized time $\tau(t) = \frac{t - t_0}{t_\mathrm{E}}$. Notice that when $t = t_0$, $u = u_0$.

\begin{equation}
u(t) \equiv   \sqrt{\Bigl(\frac{t - t_0}{t_\mathrm{E}}\Bigr)^2 + u_0^2} = \sqrt{\tau(t)^2 + u_0^2}, 
\end{equation}

Since the magnification of a \gls{PSPL} was first calculated by \cite{paczynski1986gravitational_curve}, this magnification curve is also known as the Paczy\'nski curve (see Figure~\ref{fig:paczynski_curve}). 

\begin{figure}[h]
    \centering
    \begin{subfigure}[t]{0.45\textwidth}
        \centering
        \includegraphics[width=\linewidth]{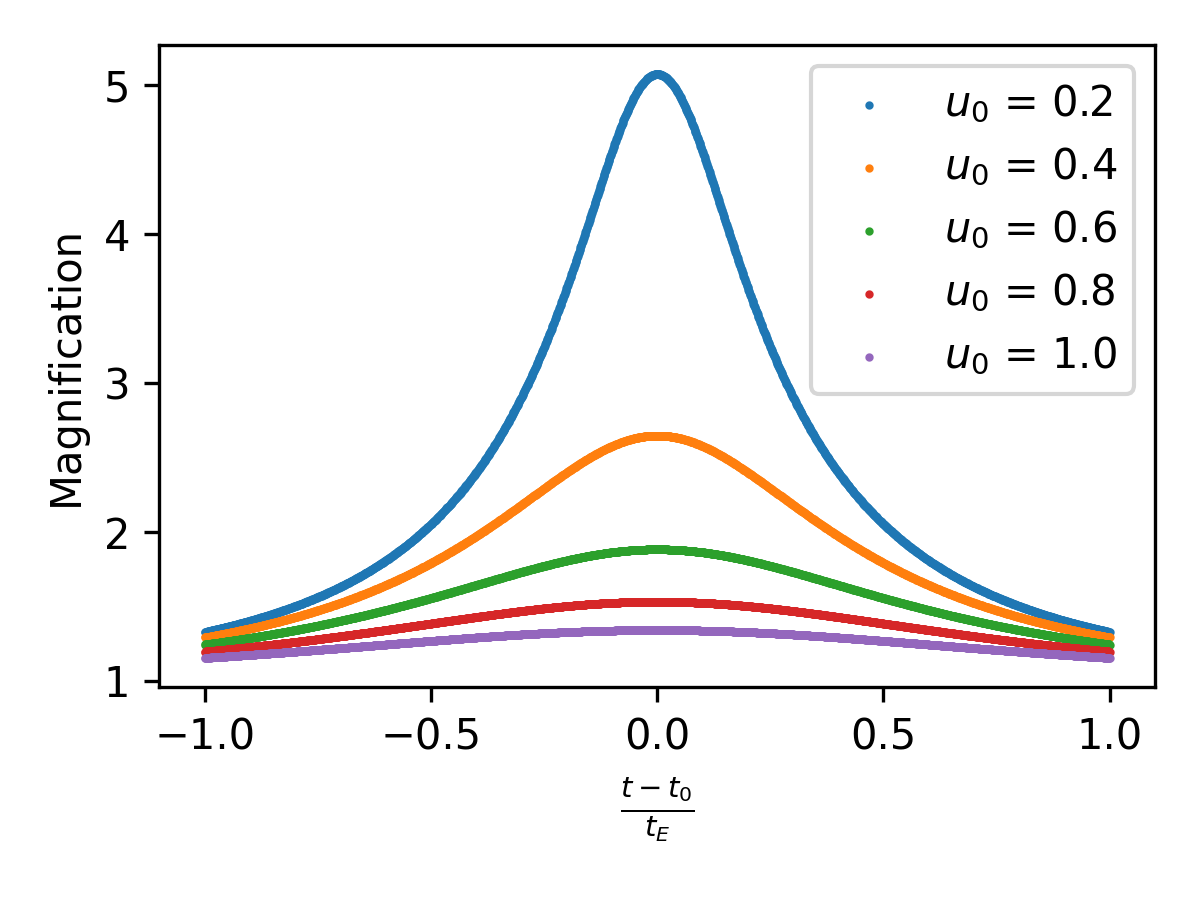}
    \end{subfigure}
    \begin{subfigure}[t]{0.45\textwidth}
        \centering
        \includegraphics[width=\linewidth]{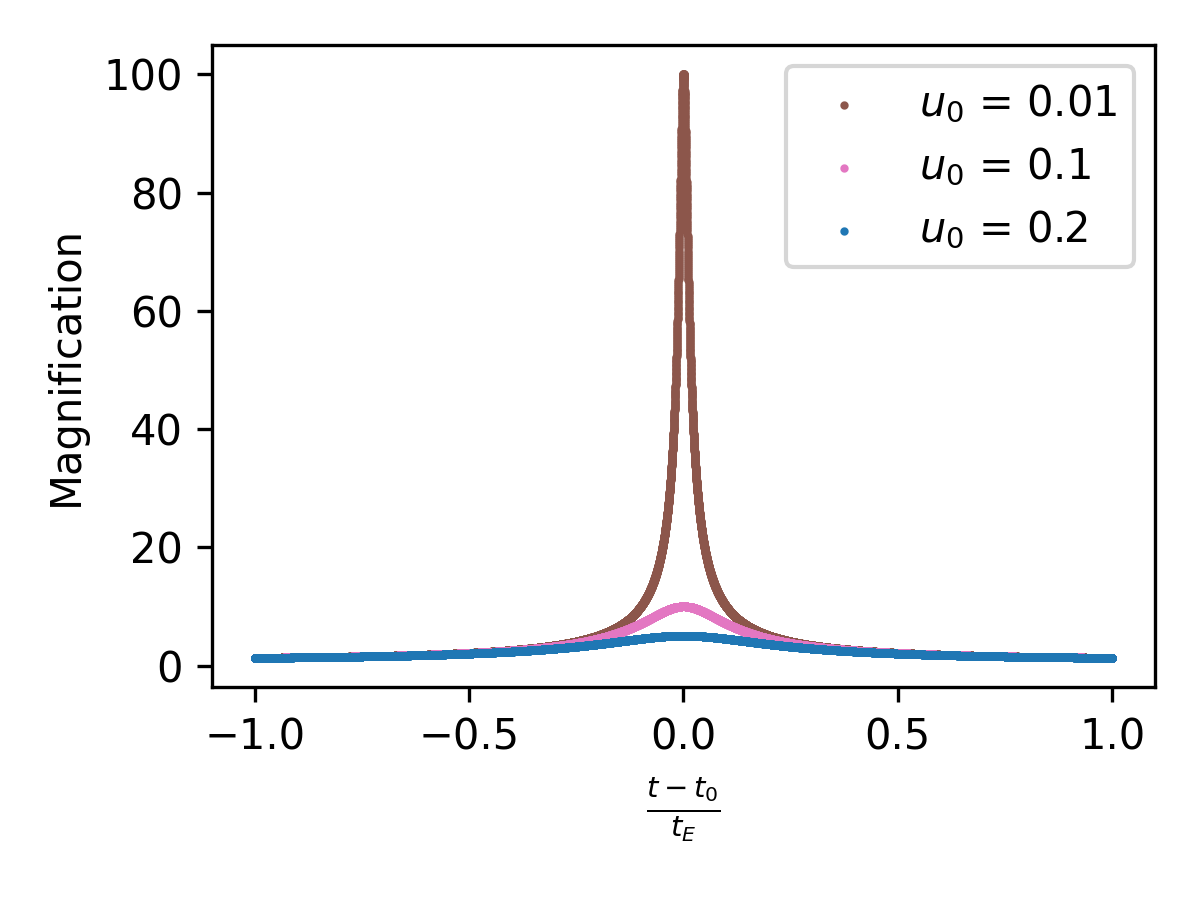} 
        \label{fig:subfig2}
    \end{subfigure}
    \caption{The Paczy\'nski curve, also known as \gls{PSPL} curve, predicts how much a source star's brightness can be amplified over time during a point-source point-lens microlensing event. The magnification is given by $A(u) = \frac{u^2 + 2}{u \sqrt{u ^2 + 4}}$, where $u(t) = \sqrt{(\frac{t - t_0}{t_\mathrm{E}})^2 + u_0^2}$, for different values of $u_0$.}
    \label{fig:paczynski_curve}
\end{figure}

The magnification function is used to model the observed flux during a microlensing event. The total flux for an instrument with a passband $i$ at any given time $t$ is expressed as:

\begin{equation}
F_i(t) = A \cdot f_{s,i} + f_{b,i}
\label{eq:flux}
\end{equation}
where $A$ is the derived magnification (equation~\ref{eq:amplification}) of the event at any given time and for any given set of non-linear parameters ($t_\mathrm{E}, t_0, u_0$), $f_{s,i}$ is the unlensed source flux in a specific passband $i$, and $f_{b,i}$ is its excess flux, accounting for blended light. The latter represents fluxes from other unrelated nearby stars or objects along the line of sight of the observer to the microlensing system and is constant as it is not affected by the lensing event.

Since microlensing is a highly sensitive effect, it is often necessary to consider that the source star is not a point source, which is referred to in microlensing as a finite source. This type of \gls{1L1S} event is named a \gls{FSPL} event. The microlensing effect causes the images of the source to appear more or less distorted depending on how close the source approaches the lens from the observer’s viewpoint. This effect is illustrated in Figure~\ref{fig:finite_source_lens_plane}, where the contour of the source is mapped and its image is calculated in the lens plane. See Section~\ref{subsec:2nd_order_effect} for more details.

\begin{figure}
    \centering
    \includegraphics[width=0.8\linewidth]{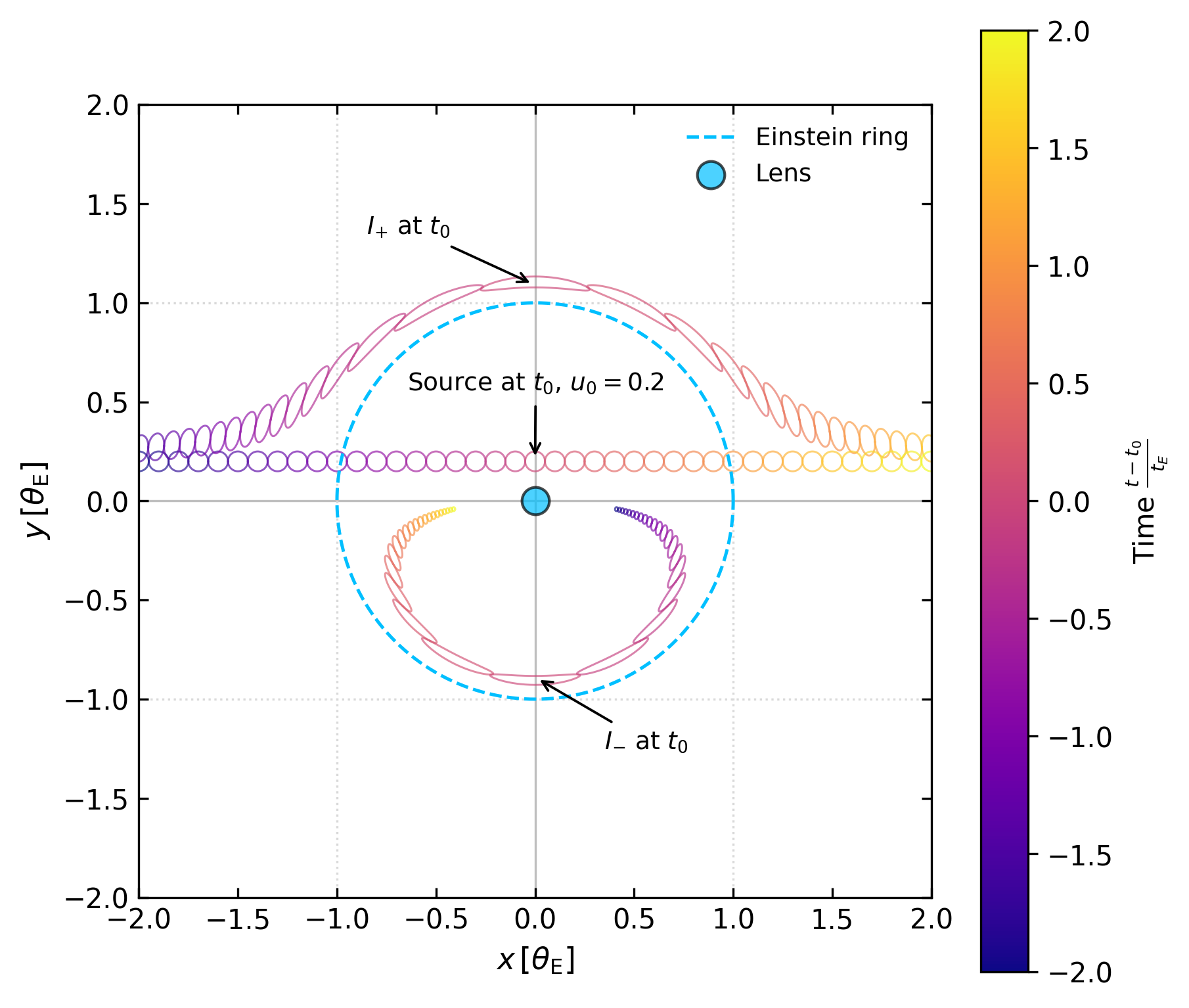}
    \caption{This figure illustrates the gravitational lensing phenomenon centred on a single-point lens (marked in pink), shown in angular coordinates normalized by the angular Einstein radius $\theta_E$. The lens remains stationary in this representation, while the path of the source S and the evolving positions of the two created images ($I_+$ and $I_-$) are displayed over time. The contours of the source and images are displayed at different stages of the event, with matching colours indicating the corresponding instant in time. In most microlensing events, the individual images cannot be resolved due to instrumental limitations; however, the brightness from the sum of the two images is observable. }
    \label{fig:finite_source_lens_plane}
\end{figure}

Exoplanet detection has traditionally focused on bound planets, which for microlensing detection involves a multi-lens system. Even so, the Paczy\'nski curve remains very relevant, as many planetary microlensing events are initially identified using \gls{PSPL} or single-lens search techniques before being further modelled as multi-lens systems. Moreover, single-lens models are particularly valuable for detecting \glspl{FFP} \citep{sumi2023free} as well as neutron stars and stellar-mass black holes \citep{Lam_2022}. It is also worth noting that a wide-orbit planet could orbit far enough from its host star to appear as an \gls{FFP}. See Section \ref{demographics} for more information on exoplanet demographics with microlensing.

\subsection{Binary Lens Event}

During a microlensing event, multiple objects can collectively act as a lens. Common lensing systems include combinations such as two stars, a star and a planet, a star with two planets, two stars with a planet, star and Brown dwarf, or Brown dwarf binaries - all of which can magnify the light from a background star. Among these, a binary lens system composed of a star and a planet is, nowadays, the most frequently observed type for exoplanet detection in microlensing studies. An event involving two lenses and one source is referred to as a \gls{2L1S} event, while an event with two lenses and two sources is referred to as a \gls{2L2S} event.

According to the International Astronomical Union, any object that orbits stars, brown dwarfs, or stellar remnants and has a mass ratio of $q < 2/(25 + \sqrt{621}) \approx 0.04$ qualifies as an exoplanet \citep{des2022iau}. Therefore, a binary lens event where the mass ratio $q$ is significantly small is usually considered a planetary microlensing event, which is of primary interest to the exoplanet research community. Importantly, microlensing naturally provides a direct measurement of the lens mass ratio $q$, and the recent adoption of a mass-ratio-based definition of planets further emphasizes the physical significance of this parameter in planet formation scenarios. This reflects a significant conceptual shift, as the mass ratio itself is used to identify the planetary nature of the secondary body in the lens system, whereas converting mass ratios into absolute masses was previously thought necessary to determine the nature of the secondary lens.

In a binary lens system, the deflection of light can be determined by the vector sum of the bending angles caused by each mass. Similarly to equation~\ref{eq:gravity_bend} derived in Subsection \ref{subsec:single_lens} for one lens, we can rewrite the Einstein bending angle as a sum of the deviation caused by two different masses:

\begin{equation}
\label{eq:binary_lens_alpha}
\vectorize\alpha = \frac{4 G M_1}{c^2}\frac{\vectorize r - \vectorize r_{M_1}}{|\vectorize r - \vectorize r_{M_1}|^2} + \frac{4 G M_2}{c^2}\frac{\vectorize r - \vectorize r_{M_2}}{|\vectorize r - \vectorize r_{M_2}|^2},
\end{equation}
where $\vectorize r$ indicates where the light ray hits the lens \textbf{plane} and $\vectorize r_{M_1}$ and $\vectorize r_{M_2}$ are the positions of the mass $M_1$ and $M_2$, respectively, onto the lens plane. 

This equation can be derived into the normalized lens equation given by \cite{witt1990investigation,bourassa1973spheroidal,rhie1997infimum}:

\begin{equation}
\label{eq:binary_lens}
w = z - \frac{\epsilon_1}{\overline{z} - r_{M_1}} - \frac{\epsilon_2}{\overline{z} - r_{M_2}},
\end{equation}
where $w$ is the source position in complex coordinates, $z$ is the image position in complex coordinates, and $\epsilon_{1}$ and $\epsilon_{2}$ are the fractional mass of the lenses ($\epsilon_{1} = M_{1}/M_{total}$, $\epsilon_{2} = M_{2}/M_{total}$ and $\epsilon_1 + \epsilon_2 = 1$). The quantities $r_{M_1}$ and $r_{M_2}$ denote the positions of lenses 1 and 2, respectively, and $\overline{z}$ is the complex conjugate of $z$.
Here, the coordinate system is chosen such that both lens components lie along the real axis, so $r_{M_1}$ and $r_{M_2}$ are real quantities.
The lens mass ratio is defined as
$q=M_2/M_1$, therefore, $q=\epsilon_2/\epsilon_1=\epsilon_2/(1-\epsilon_2)$.

Currently, there is no analytical solution for the magnification resulting from a binary lens system. However, the lens equation can be solved numerically. The conjugate of Equation~\ref{eq:binary_lens} can be rearranged and $\overline{z}$ substituted back into the same equation, allowing the binary lens equation to be rewritten as a fifth-order complex polynomial in $z$ \citep{witt1990investigation,witt1995minimum}. 
The five polynomial
coefficients, detailed in \citet{witt1995minimum}, enable the numerical determination of five roots. Note that these solutions do not always correspond to physical images of the source \citep{witt1990investigation,witt1995minimum,rhie1997infimum}. In particular, sources located outside a caustic loop produce three images, whereas sources inside the caustic loop produce five images \citep{rhie1997infimum,witt1990investigation}. To determine whether a root of the fifth-order polynomial corresponds to a physical image, the solution must be checked against the original lens equation. This is because the polynomial equation is implied by the lens equation, but is not equivalent to it, as the equivalence is lost during the mathematical derivation \citep{schneider1986two,witt1990investigation,rhie1997infimum}.

As in the case with single lenses, the magnification of the background star is calculated by $A = \frac{1}{\mathrm{det}J}$. The set of source positions where $\mathrm{det}J = 0$ defines what is known as \textbf{caustic curves} \citep{gaudi2012microlensing}. Note that in the point-source approximation, the magnification formally diverges when the source lies exactly on a caustic. However, real stars have a finite angular size, so different regions of the stellar disc experience different magnifications when the source is crossing or approaching a caustic. The observed magnification, therefore, corresponds to the surface-brightness-weighted average over the source profile, which smooths the formal divergence predicted by the point-source approximation and produces a very large, but finite, magnification, making finite-source effects particularly important to be taken into account.

For a binary lens system, caustic curves are of critical importance because they mark the boundaries within the source plane where multiple images (three or five) are created. The caustic curves in a binary lens system are defined by this new set of parameters: the mass ratio $q$, which represents the ratio between the secondary lens and the primary lens, and the projected separation $s$ between the two lenses, which is normalized by the Einstein radius $\theta_\mathrm{E}$.  We usually define three caustic topologies: close, intermediate/resonant, and wide, which are defined by the shape of the curves themselves. As shown in Figure~\ref{fig:caustics}, the close topology is composed of a central four-cusped caustic along with two planetary three-cusped caustics. The resonant topology consists of a single central six-cusped caustic. The wide topology includes a central four-cusped caustic and a planetary four-cusped caustic. Figure~\ref{fig:caustics}(b) shows how the caustic topology varies as a function of the projected lens separation $s$ and mass ratio $q$. The topology transition limits are given in \cite{dominik1999binary}, where $s_w$ is the wide--resonant separation limit and $s_c$ is the close--resonant separation limit:

\begin{equation}
 s_w = \left( \frac{(1 + q^{1/3})^3}{1+q} \right)^{1/2}
\end{equation}

and

\begin{equation}
\frac{s_c^8}{(1-s_c^4)^3} = \frac{(1+q)^2}{27q}.
\end{equation}

\begin{figure}[h]
    \centering

    \begin{subfigure}[t]{\textwidth}
        \centering
        \includegraphics[width=\textwidth]{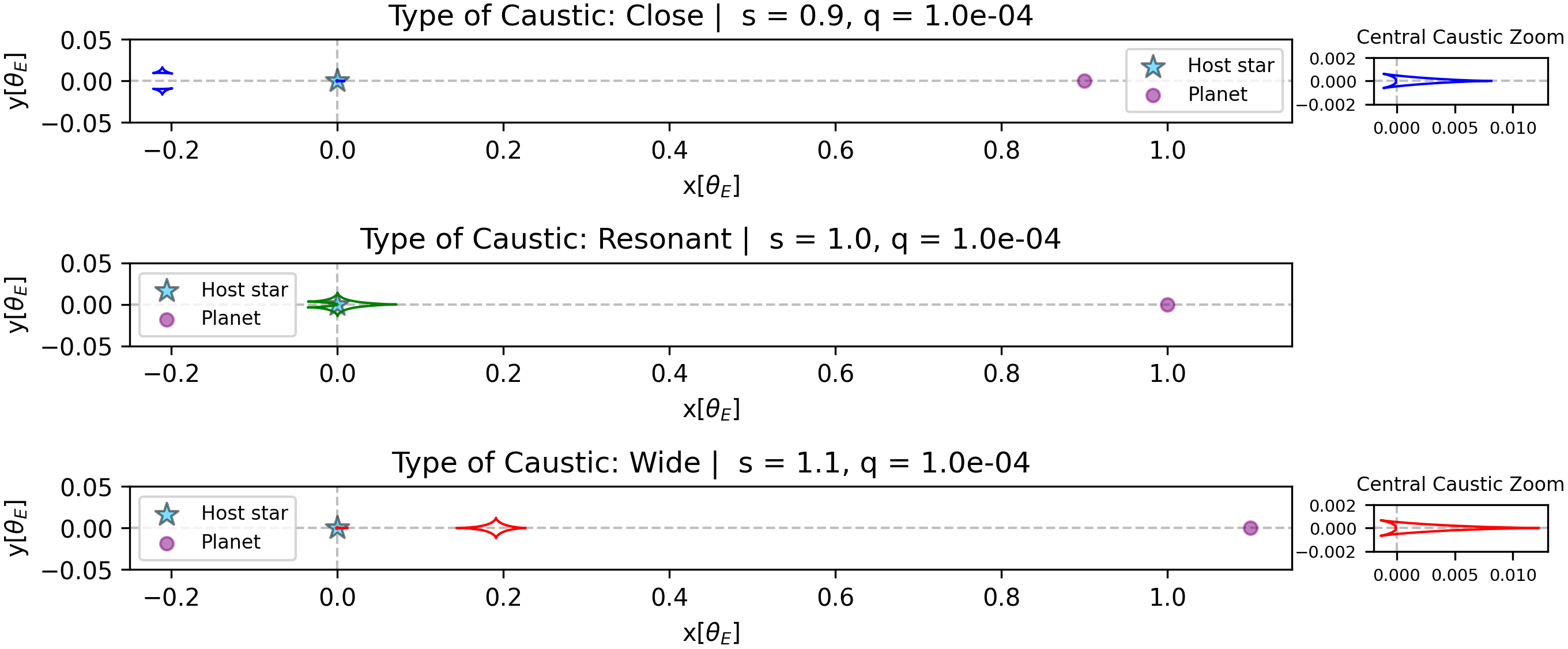}
        \caption{Caustic curves for different binary lens topologies, where each plot corresponds to a different projected separation $s$ and fixed mass ratio 
        $q$. The primary lens is the host star, while the secondary lens is the planet. The $(x,y)$ coordinates are centred on the centre of mass of the lens system, which, for this small mass ratio, lies very close to the host star. The colour of the curves represents the topology of the binary lens system: blue for close, green for resonant/intermediate, and red for wide. This plot was made using \texttt{VBMicrolensing}.}
    \end{subfigure}

    \vspace{0.5cm}

    \begin{subfigure}[t]{\textwidth}
        \centering
        \includegraphics[width=0.4\linewidth]{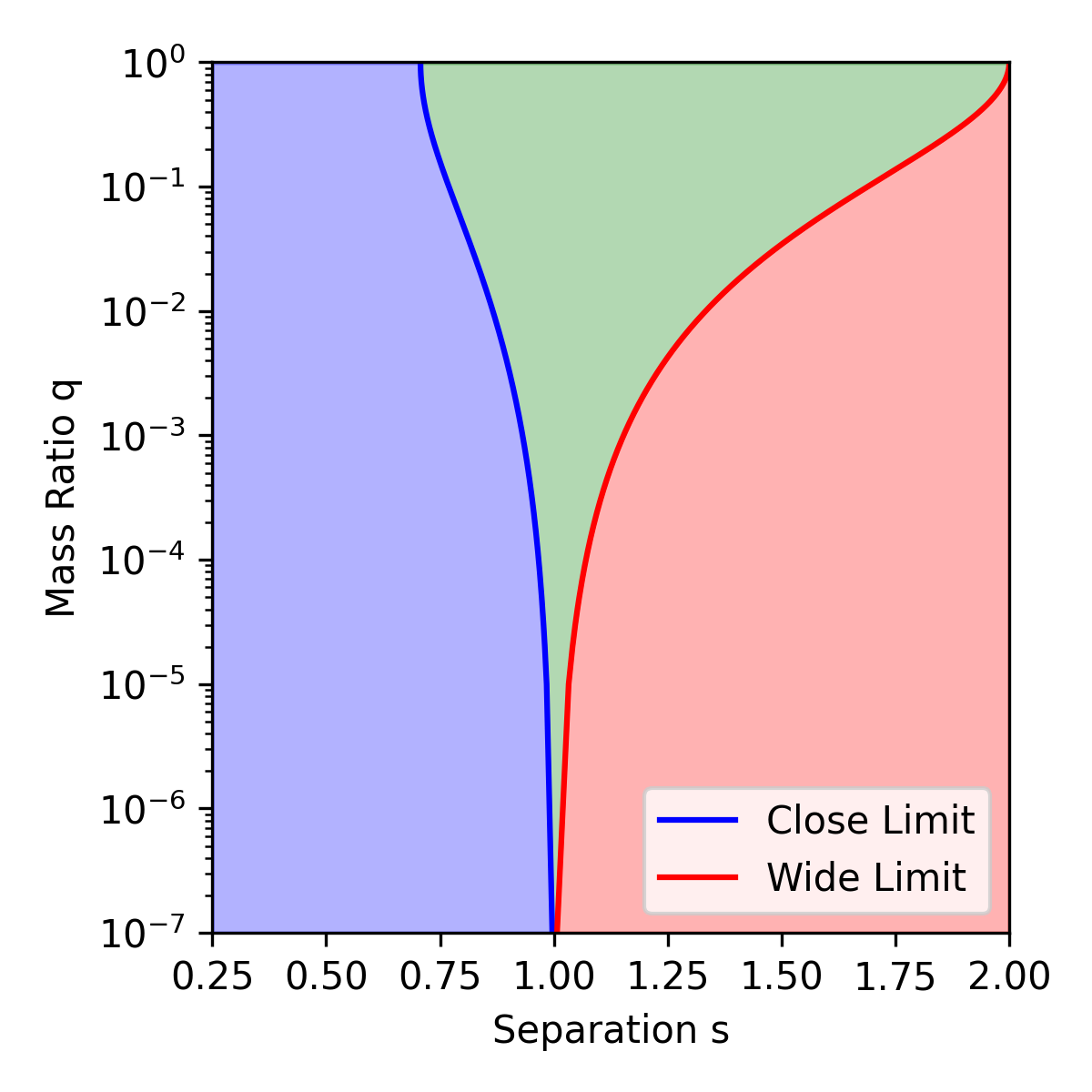}
        \caption{The plot of the binary lens system’s projected separation $s$ versus the mass ratio $q$, showing the regions defined by the close and wide limits. The blue region represents the close regime, the green region represents the resonant regime, and the red region represents the wide regime. Plot made with \texttt{MOAna}.}
    \end{subfigure}

    \caption{These plots illustrate how the projected separation $s$ and mass ratio $q$ influence the caustic structure derived from a binary lens system. The caustic curves represent positions in the source plane where light from a point source would experience infinite magnification due to gravitational lensing.}
    
    \label{fig:caustics}
\end{figure}

 While the caustic topology is determined by $s$ and $q$, modelling the event also requires describing the trajectory of the source relative to the lens system. A third parameter is introduced: the angle $\alpha$, which measures the angle between the relative motion of the source and the line connecting the two masses of the binary lens system, typically measured counter-clockwise. For planetary microlensing events, the primary lens is typically the host star, while the secondary lens is the planet. As mentioned previously, finite-source effects play an important role in binary-lensing models and must therefore be included. These effects can be explicitly included as the normalized size of the source $\rho_*$, or implicitly through the parameter $t_*$, which represents the time required for the source radius to cross the Einstein radius of the lens. This relationship is given by $t_* = \rho_* \cdot t_\mathrm{E}$.

It is possible to say that the source trajectory in relation to the caustic curves determines the shape of the light curve. To better illustrate this, we present two sets of examples showing light curves along with their corresponding caustic structures and source trajectories. An example of how to calculate and plot the magnification and caustic curves using \texttt{VBMicrolensing} is presented at the end of this subsection, in the Code Listing ~\ref{lst:microlensing_plot}.

In Figures~\ref{fig:caustic_different_trajectory_1}, \ref{fig:caustic_different_trajectory_2}, and~\ref{fig:caustic_different_trajectory_3}, all configurations share the same mass ratio $q=2.4 \times 10^{-3}$, lens separation $s = 1.0 $, Einstein timescale $t_\mathrm{E} = 20 \; \mathrm{days}$, and source size $\rho = 0.01$. We vary only the impact parameter $u_0$ and the trajectory angle $\alpha$ to demonstrate how changes in the source trajectory, while keeping the caustic structure fixed, affect the resulting light curve. A comparison of a \gls{FSPL} light curve and a \gls{2L1S} light curve is shown in Figure~\ref{fig:triple-lens vs binary-lens}.

\clearpage

\begin{figure}[h]
    \centering

    \begin{subfigure}[t]{\textwidth}
        \centering
        \includegraphics[width=0.9\linewidth]{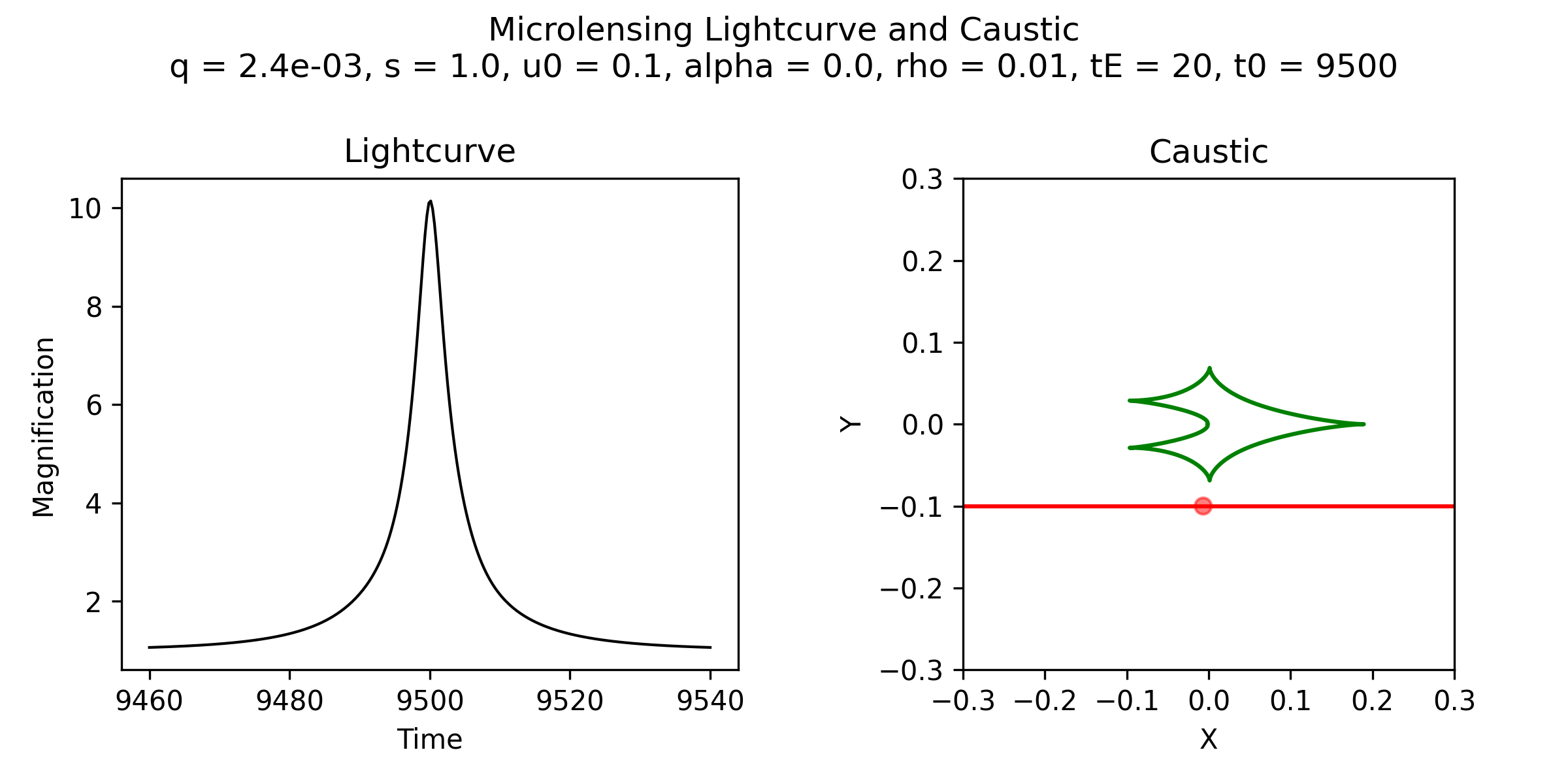}
    \end{subfigure}

    \vspace{0.5em} 
    
    \begin{subfigure}[t]{\textwidth}
        \centering
        \includegraphics[width=0.9\linewidth]{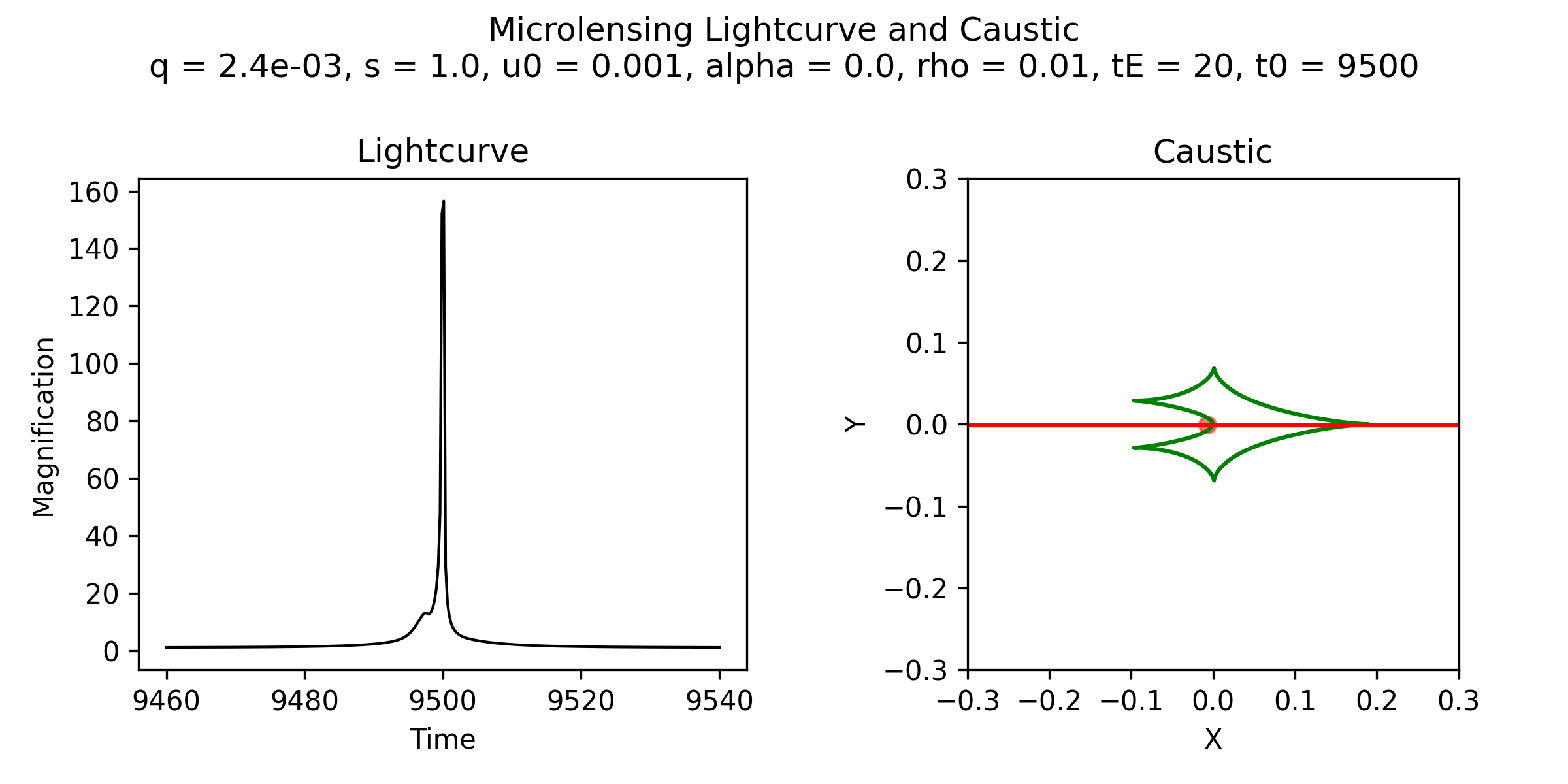}
    \end{subfigure}
    \caption{Effect of source trajectory on microlensing light curves for a fixed binary lens configuration. All panels share the same physical parameters: mass ratio $q = 2.4 \times 10^{-3}$, lens separation $s = 1.0$, Einstein timescale $t_\mathrm{E} = 20$ days, source size $\rho = 0.01$, and trajectory angle $\alpha = 0.0$. Only the impact parameter $u_0$ is varied between plots. These examples highlight how changes in the source trajectory, while keeping the caustic structure fixed, influence the shape of the resulting light curves. The right panels are centred on the centre of mass of the lens system, composed of the host star and planet. The planet is not shown in order to avoid making the plots excessively wide, but its location can be inferred from Figure~\ref{fig:caustics}(a). Since the planet is omitted, we opted not to plot the host star, which can also be inferred from Figure~\ref{fig:caustics}(a). Plot made with \texttt{VBMicrolensing}}
    \label{fig:caustic_different_trajectory_1}
\end{figure}

\begin{figure}[h]
    \centering
    \begin{subfigure}[t]{\textwidth}
        \centering
        \includegraphics[width=0.9\linewidth]{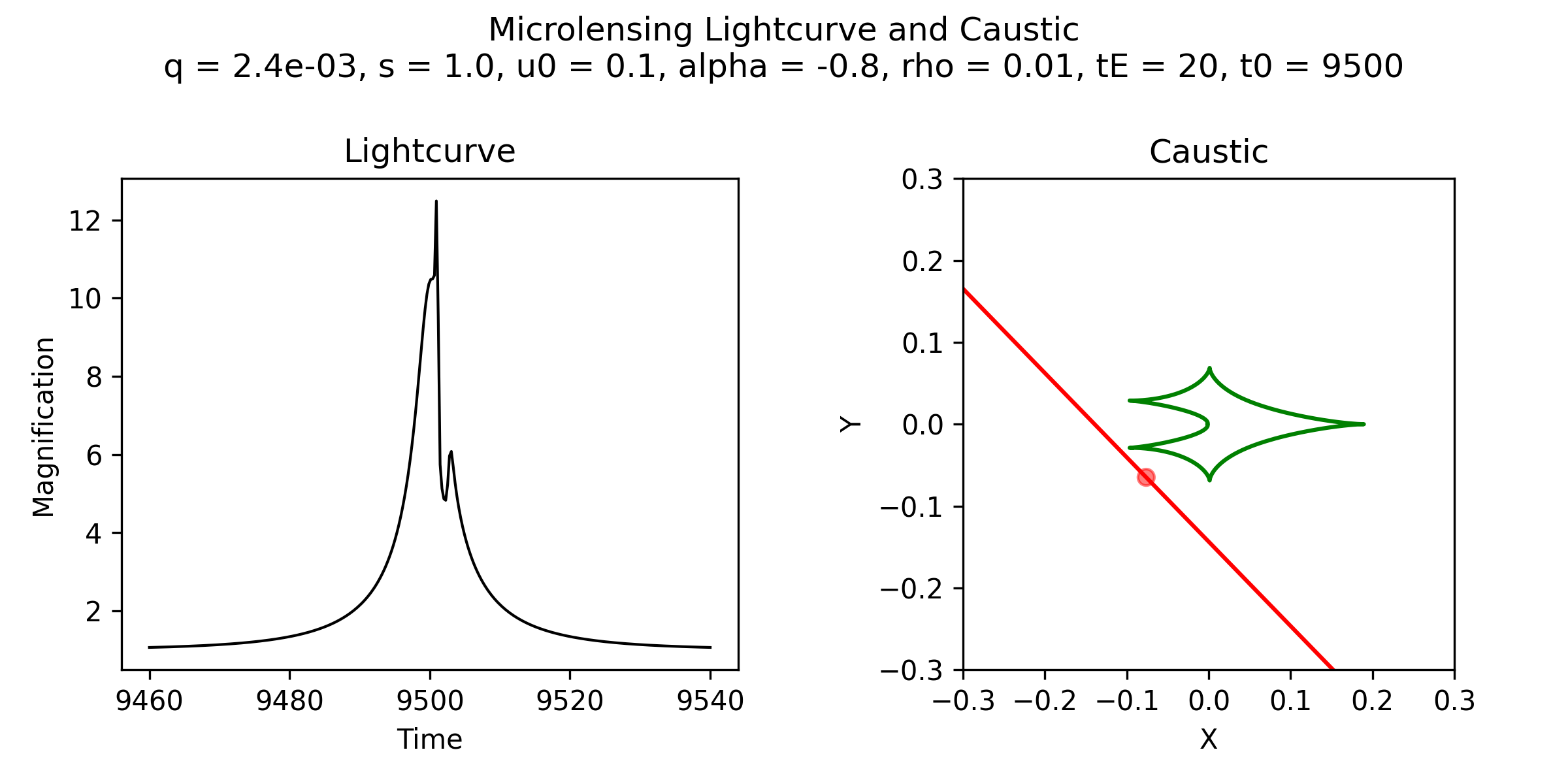} 
    \end{subfigure}

    \vspace{0.5em} 

    \begin{subfigure}[t]{\textwidth}
        \centering
        \includegraphics[width=0.9\linewidth]{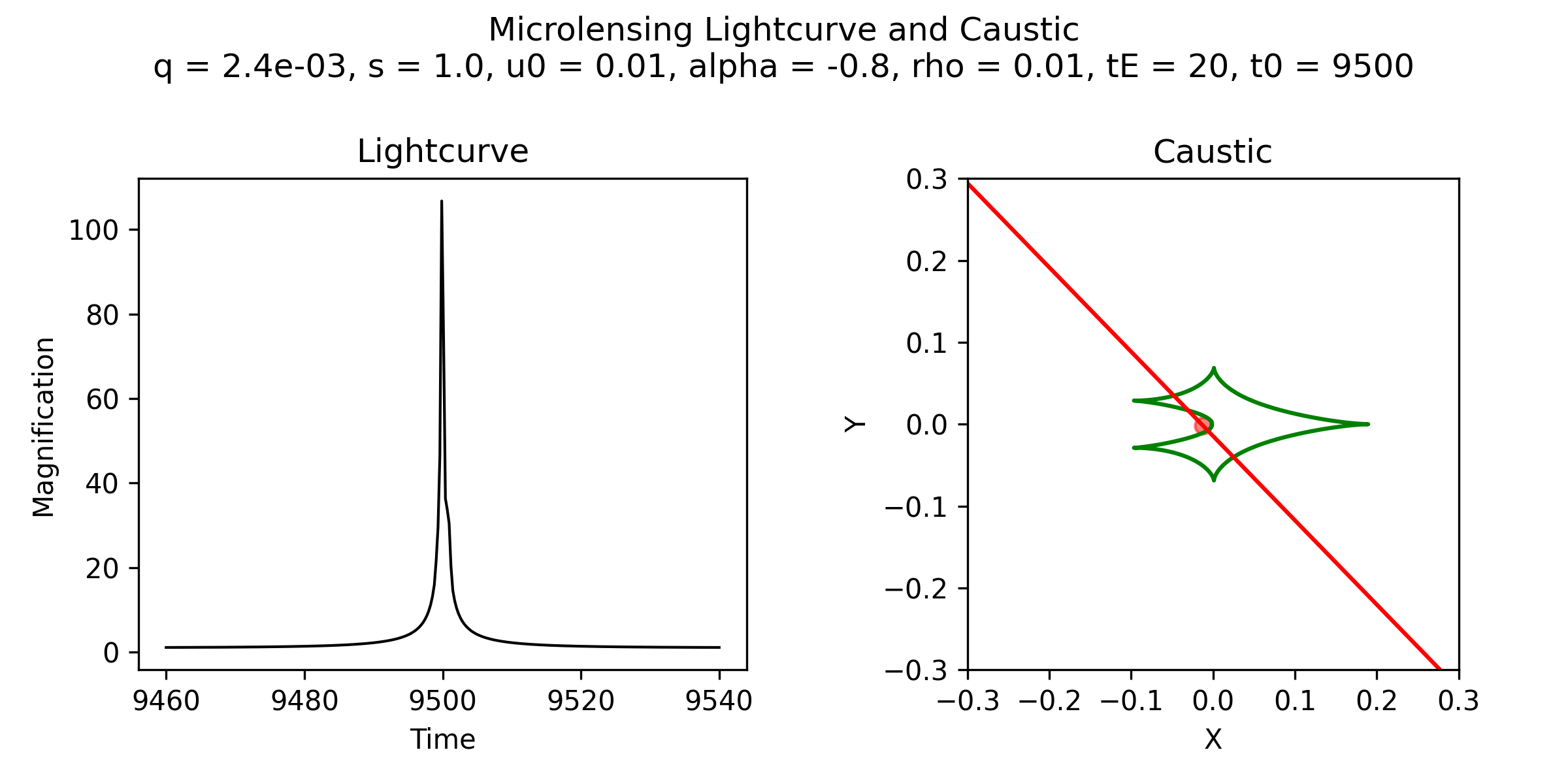}
    \end{subfigure}

    \caption{Similar to previous images, all panels share the same physical parameters: mass ratio $q = 2.4 \times 10^{-3}$, lens separation $s = 1.0$, Einstein timescale $t_\mathrm{E} = 20$ days, and source size $\rho = 0.01$. Now with trajectory angle $\alpha = -0.8$. The impact parameter $u_0$ varies between plots. These examples highlight how changes in the source trajectory, while keeping the caustic structure fixed, influence the shape of the resulting light curves. Plot made with \texttt{VBMicrolensing}}
    \label{fig:caustic_different_trajectory_2}
\end{figure}

\clearpage

In the second set of examples, Figures~\ref {fig:caustic_mass_ratio_variation_1} and ~\ref{fig:caustic_mass_ratio_variation_2}, we fix $u_0 = 0.1$, $\alpha = - 0.8$, $s = 1.0$, $t_\mathrm{E} = 20 \; \mathrm{days}$, and vary only the mass ratio $q$ across the following values: $q= \{3.3\times 10^{-6}, 9.5\times 10^{-4}, 1.52\times 10^{-2}, 8.25\times 10^{-6}, 2.375\times 10^{-3}, 3.8\times 10^{-2}\}$. The first three values correspond approximately to the mass ratios of an Earth–Sun, Jupiter–Sun, and a brown dwarf (16 Jupiter masses)–Sun system. The remaining values represent the same secondary lens, but hosted by a lower-mass star, an M dwarf with 0.4 solar masses.

\begin{figure}[h]
    \centering
    \begin{subfigure}[t]{\textwidth}
        \centering
        \includegraphics[width=0.9\linewidth]{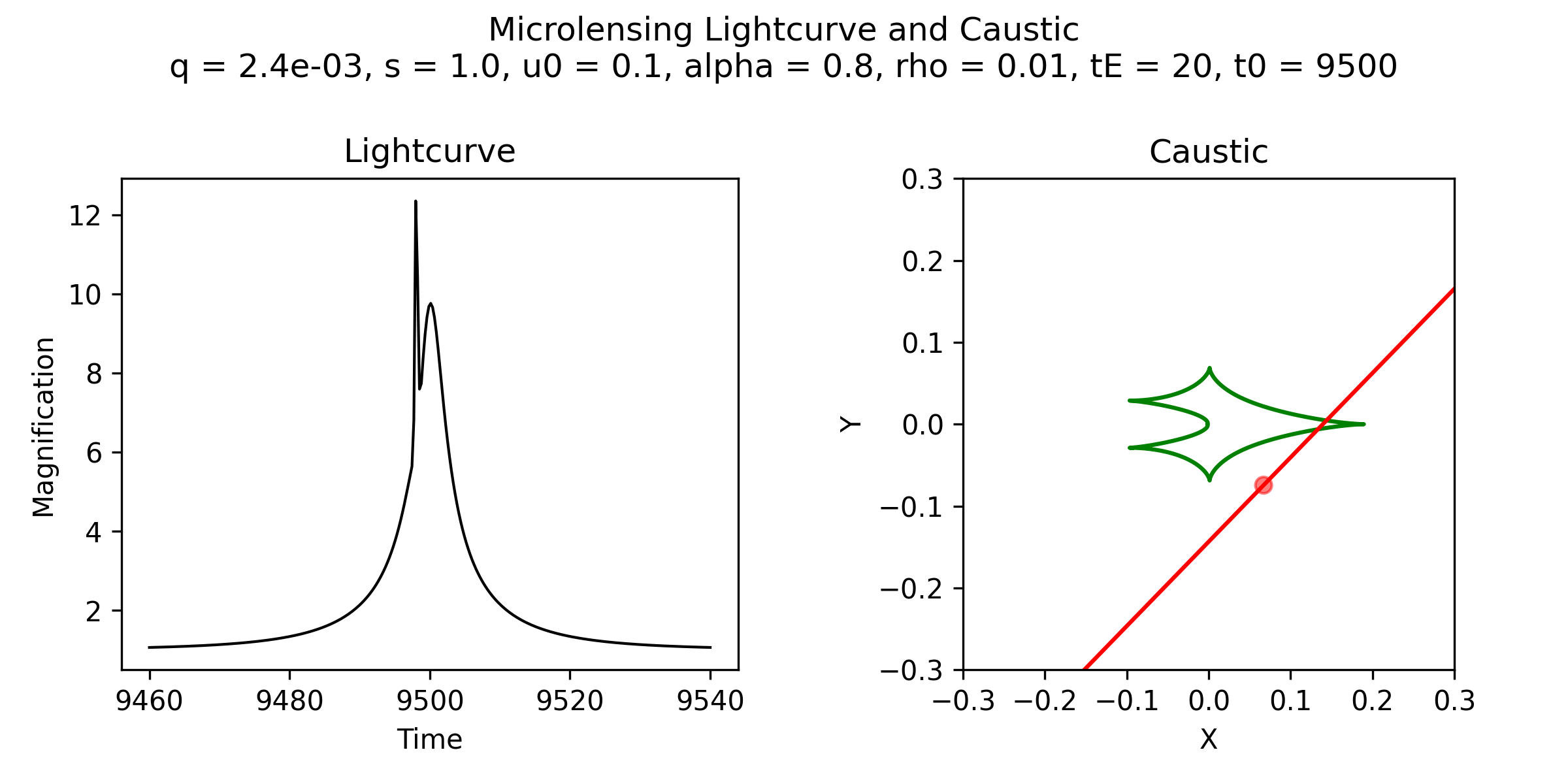} 
    \end{subfigure}

    \vspace{0.5em} 

    \begin{subfigure}[t]{\textwidth}
        \centering
        \includegraphics[width=0.9\linewidth]{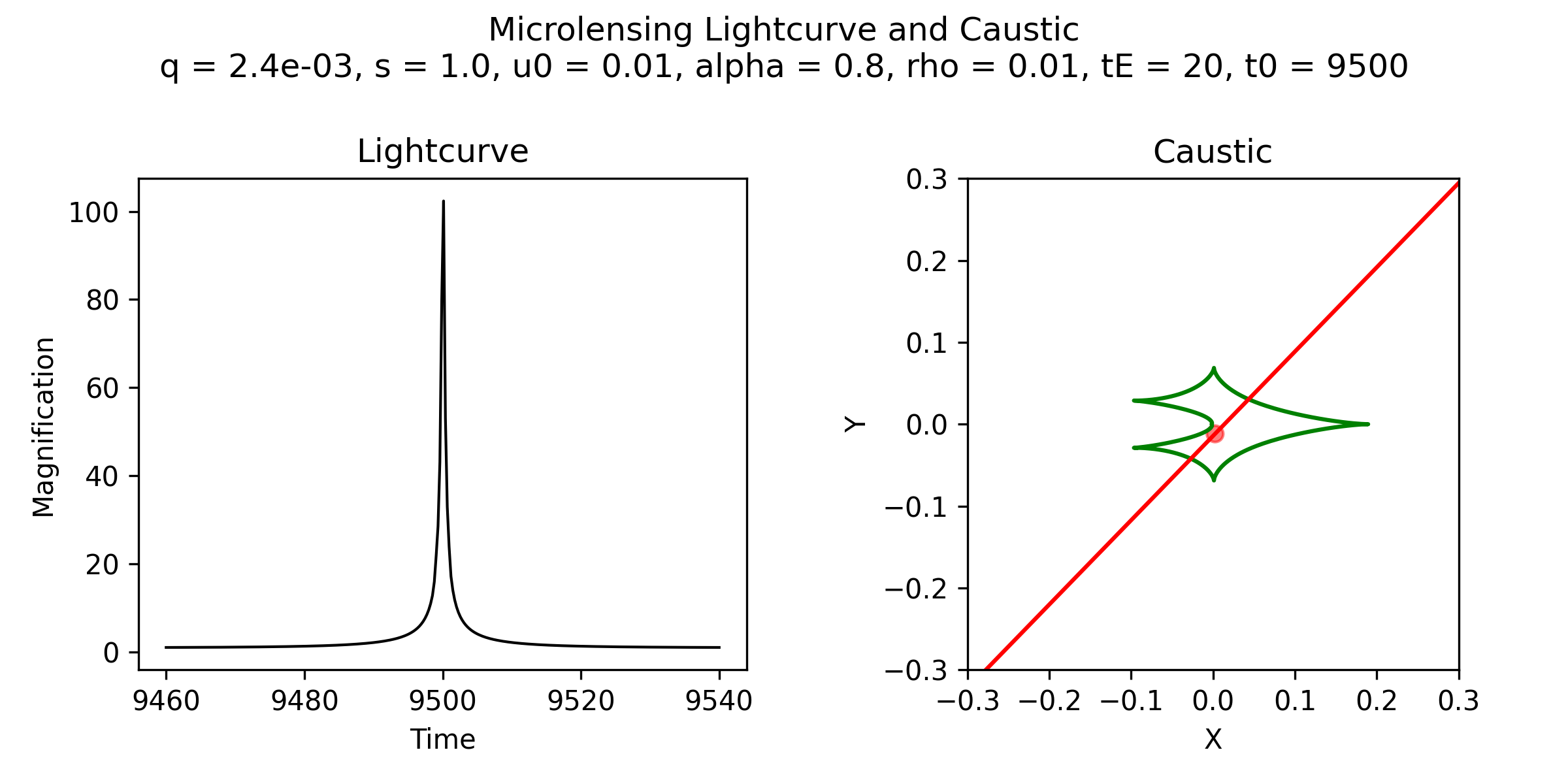}
    \end{subfigure}

    \caption{Similar to previous images, all panels share the same physical parameters: mass ratio $q = 2.4 \times 10^{-3}$, lens separation $s = 1.0$, Einstein timescale $t_\mathrm{E} = 20$ days, and source size $\rho = 0.01$. Now with trajectory angle $\alpha = 0.8$. The impact parameter $u_0$ varies between plots. These examples highlight how changes in the source trajectory, while keeping the caustic structure fixed, influence the shape of the resulting light curves. Plot made with \texttt{VBMicrolensing}}
    \label{fig:caustic_different_trajectory_3}
\end{figure}

\clearpage

\begin{figure}[h]
    \centering
    \begin{subfigure}[t]{\textwidth}
        \centering
        \includegraphics[width=0.8\linewidth]{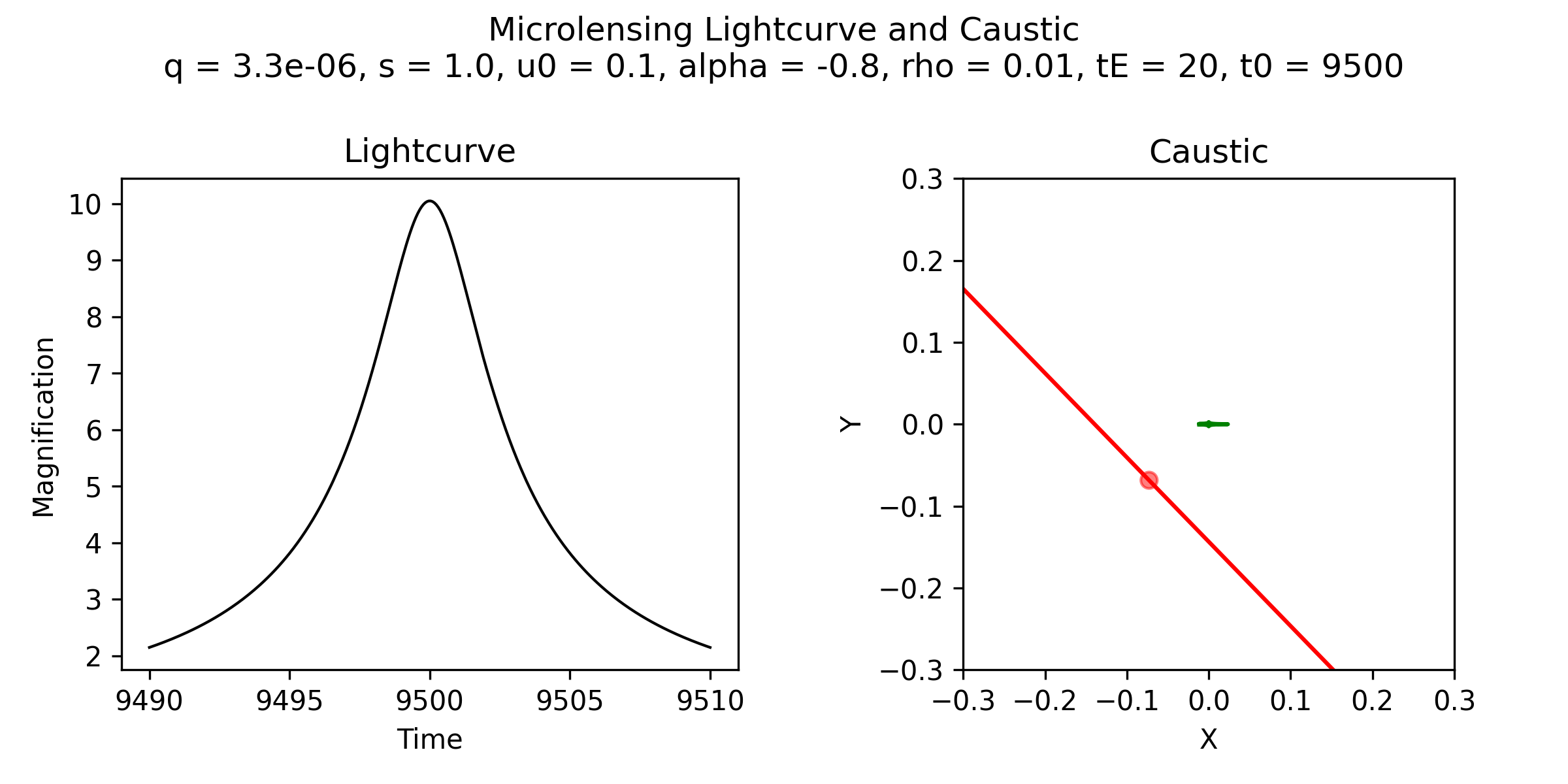}
        \caption{Earth–Sun mass ratio: $q = 3.3 \times 10^{-6}$}
    \end{subfigure}%
    \vspace{0.5em} 
    \begin{subfigure}[t]{\textwidth}
        \centering
        \includegraphics[width=0.8\linewidth]{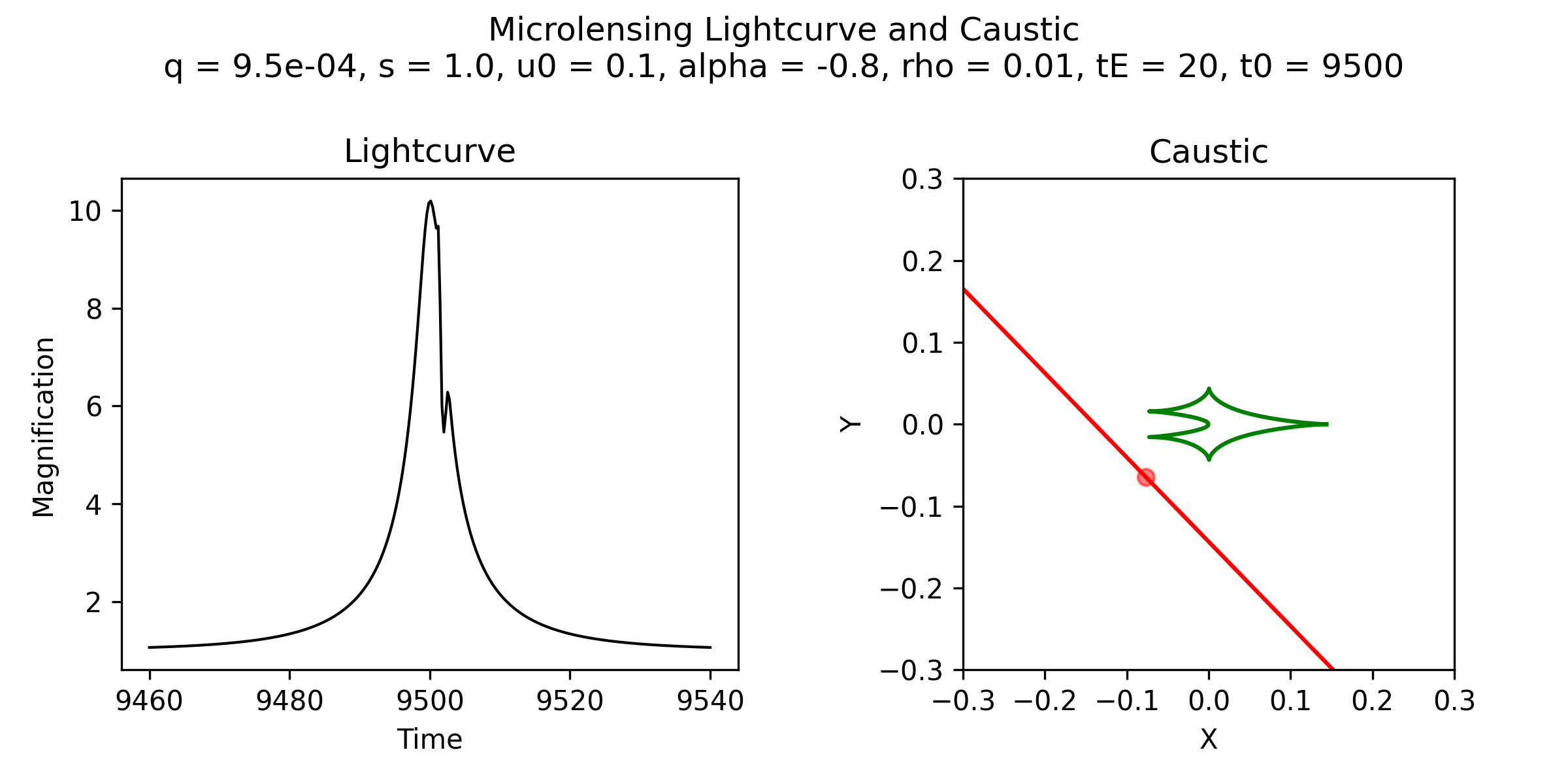}
        \caption{Jupiter–Sun mass ratio: $q = 9.5 \times 10^{-4}$}
    \end{subfigure}%
    \vspace{0.5em} 
    \begin{subfigure}[t]{\textwidth}
        \centering
        \includegraphics[width=0.8\linewidth]{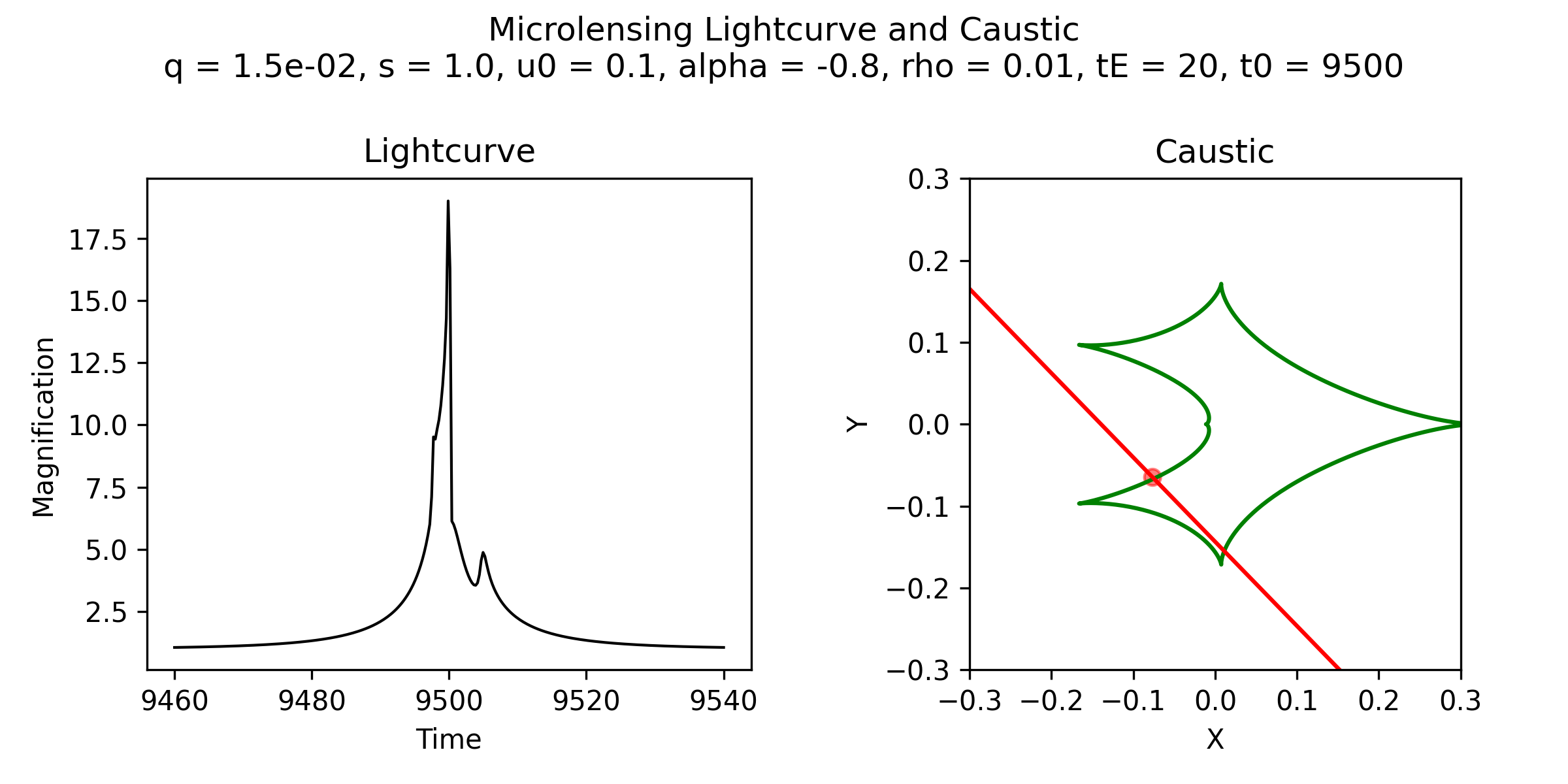}
        \caption{Brown Dwarf ($\sim$16 $M_\mathrm{Jup}$) – Sun mass ratio: $q = 1.52 \times 10^{-2}$}
    \end{subfigure}

    \caption{Microlensing light curves for different mass ratios $q$ with fixed trajectory parameters: $u_0 = 0.1$, $\alpha = -0.8$, lens separation $s = 1.0$, Einstein timescale $t_\mathrm{E} = 20$ days, and source size $\rho = 0.01$. The mass ratios span a range of planetary to substellar companions orbiting a solar-mass host. Plot made with \texttt{VBMicrolensing}}
    \label{fig:caustic_mass_ratio_variation_1}
\end{figure}

\begin{figure}[h]
    \centering

    \begin{subfigure}[t]{\textwidth}
        \centering
        \includegraphics[width=0.8\linewidth]{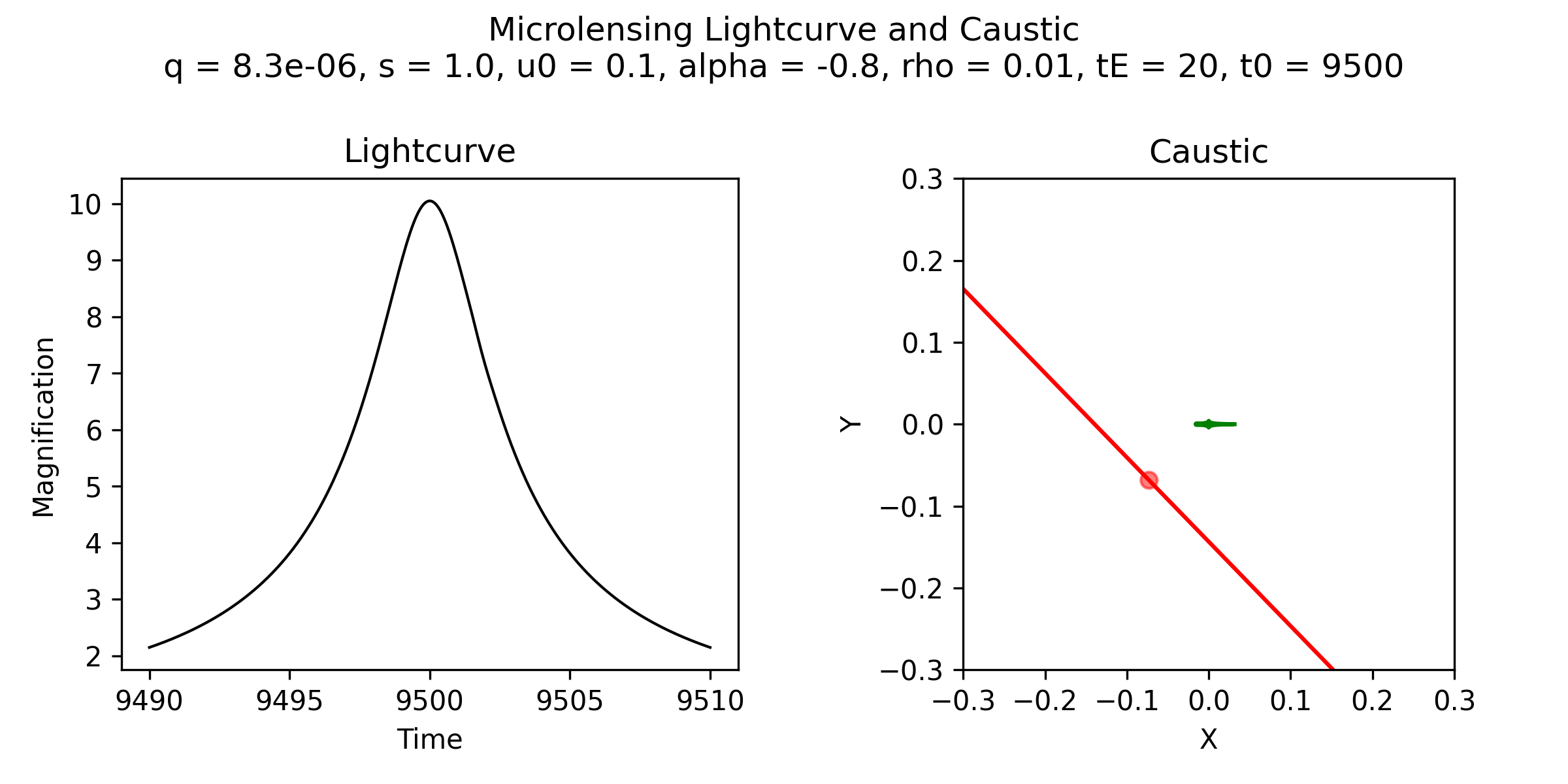}
        \caption{Earth–M dwarf mass (0.4 Solar Mass) ratio: $q = 8.25 \times 10^{-6}$}
    \end{subfigure}%
    \vspace{0.5em} 
    \begin{subfigure}[t]{\textwidth}
        \centering
        \includegraphics[width=0.8\linewidth]{Figures/microlensing_and_caustics/microlensing_curve_q0.002375_s1.0_u00.1_alpha-0.8_rho0.01_tE20.png}
        \caption{Jupiter–M dwarf mass ratio: $q = 2.375 \times 10^{-3}$}
    \end{subfigure}%
    \vspace{0.5em} 
    \begin{subfigure}[t]{\textwidth}
        \centering
        \includegraphics[width=0.8\linewidth]{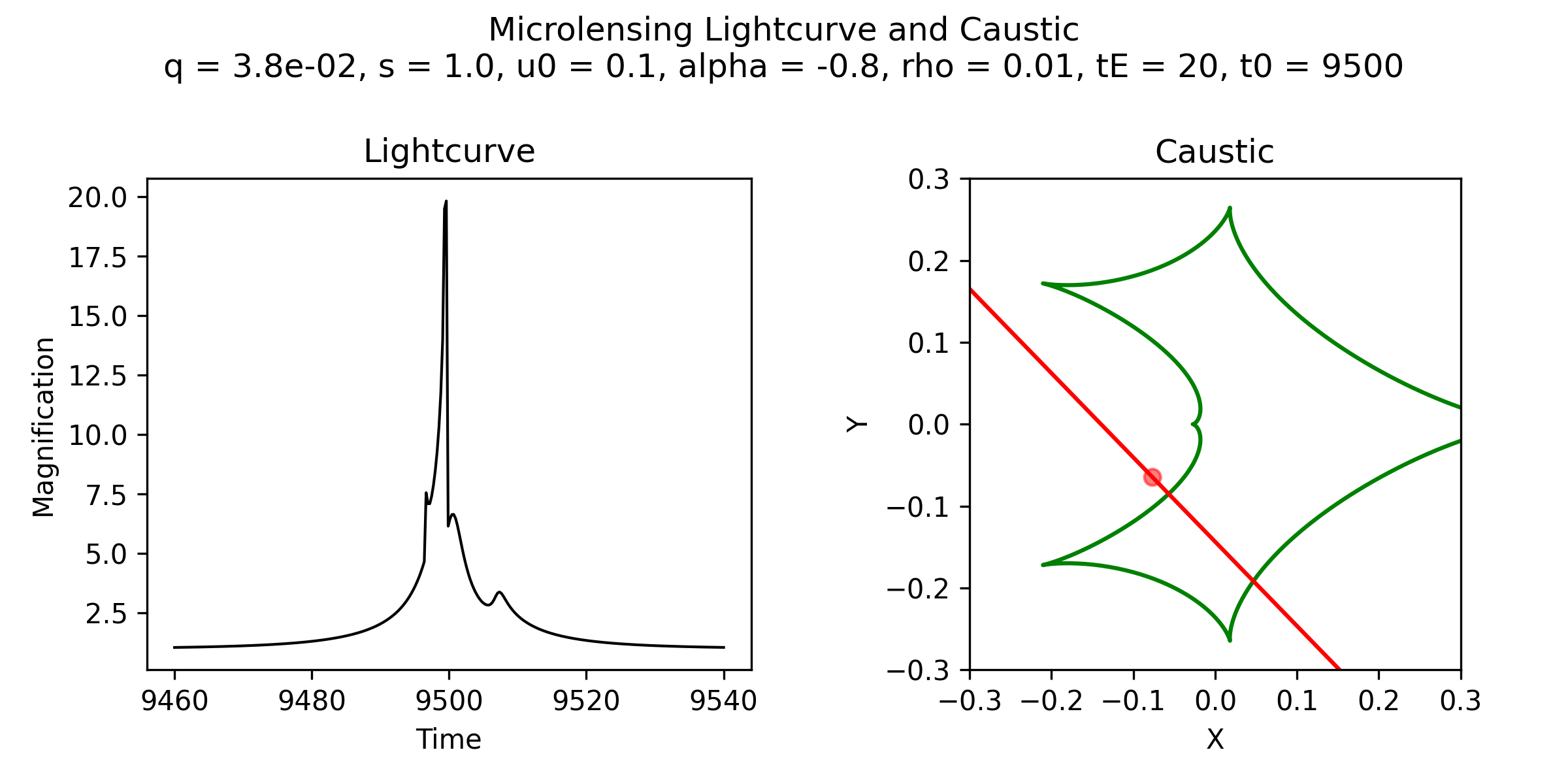}
        \caption{Brown Dwarf–M dwarf mass ratio: $q = 3.8 \times 10^{-2}$}
    \end{subfigure}

    \caption{Same as Figure~\ref{fig:caustic_mass_ratio_variation_1}, but with the companion hosted by a lower-mass star ($0.4 M_\odot$ M dwarf). Plot made with \texttt{VBMicrolensing}.}
    \label{fig:caustic_mass_ratio_variation_2}
\end{figure}

\clearpage

\begin{lstlisting}[language=Python, caption={Function to plot microlensing lightcurve and caustic curves using \texttt{VBMicrolensing}. This code plots the top panel on Figure 7.}, label={lst:microlensing_plot}]

import VBMicrolensing
import numpy as np
import matplotlib.pyplot as plt
from matplotlib.patches import Circle
import math

def plot_lightcurve_and_caustics(s, q, u0, alpha, rho, tE, t0):
    # Create a figure with 2 subplots (1 rows, 2 columns):
    fig, axes = plt.subplots(1, 2, figsize=(8, 4))

    # An array for the parameters:
    parameters = [math.log(s), math.log(q), u0, alpha, math.log(rho), math.log(tE), t0]
    # Generate time array:
    times = np.linspace(t0 - 2 * tE, t0 + 2 * tE, 300)

    # Create an instance of the VBMicrolensing class:
    vbm = VBMicrolensing.VBMicrolensing()
    # Calculate the binary light curve:
    results = vbm.BinaryLightCurve(parameters, times)

    # Plot the light curve
    fig.suptitle(f"Microlensing Lightcurve and Caustic\nq = {q:.1e}, s = {s}, u0 = {u0}, alpha = {alpha}, rho = {rho}, tE = {tE}, t0 = {t0}", fontsize=12)

    axes[0].plot(times, results[0], lw=1.0, colour='black') #  notice results[0] is the magnification
    axes[0].set_title(f"Lightcurve")
    axes[0].set_xlabel("Time")
    axes[0].set_ylabel("Magnification")

    y1 = results[1]  # source trajectory x
    y2 = results[2]  # source trajectory y

    # Plot the caustic
    caustics = vbm.Caustics(s, q)
    for cau in caustics:
        axes[1].plot(cau[0], cau[1], colour = 'green')

    # Plot the trajectory of the source
    axes[1].plot(y1, y2, colour='red')
    p0 = np.array([y1[int(len(y1)*0.5)], y2[int(len(y1)*0.5)]])
    # Plot the source size in position p0 (t0)
    axes[1].add_patch(Circle(p0, rho, colour='red', alpha=0.5))

    axes[1].set_title(f'Caustic')
    axes[1].set_xlabel('X')
    axes[1].set_ylabel('Y')
    axes[1].set_xlim(-0.3, 0.3)
    axes[1].set_ylim(-0.3, 0.3)
    axes[1].set_aspect('equal')

    # Adjust layout
    plt.tight_layout()

    # Save the figure as an image file
    plt.savefig(f"microlensing_event_q{q}_s{s}_u0{u0}_alpha{alpha}_rho{rho}_tE{tE}.png", dpi=300)

# Define your parameters
s = 1.0
q = 2.375e-3
u0 =  0.1
alpha = 0.8
rho = 0.01
tE = 20  # days
t0 = 9500 # days

plot_lightcurve_and_caustics(s, q, u0, alpha, rho, tE, t0)
\end{lstlisting}

\subsection{Triple Lens Event}
\label{sec:triple lens}

As previously mentioned, multiple objects can collectively act as a lens during a microlensing event. We have already discussed the binary lens case, where either a stellar binary or a star $+$ planet act as the lens. Now we will discuss the more complicated case of a triple lens event, in this case, a \gls{3L1S}. This can involve one planet orbiting two stars, two planets orbiting one star, or even a planet and an exomoon orbiting a star.

In the triple-lens case, a third body with mass $M_3$ is added to the binary-lens components with masses $M_1$ and $M_2$. The total deflection of light must then include the contribution from this additional lens component. Therefore, Equation~\ref{eq:binary_lens_alpha} becomes:

\begin{equation}
\label{eq:triple_lens_alpha}
\vectorize\alpha = \frac{4 G M_1}{c^2}\frac{\vectorize r - \vectorize r_{M_1}}{|\vectorize r - \vectorize r_{M_1}|^2} + \frac{4 G M_2}{c^2}\frac{\vectorize r - \vectorize r_{M_2}}{|\vectorize r - \vectorize r_{M_2}|^2} + \frac{4 G M_3}{c^2}\frac{\vectorize r - \vectorize r_{M_3}}{|\vectorize r - \vectorize r_{M_3}|^2},
\end{equation}
where we now also consider the position of a mass $M_3$ on the lens plane. As before in Equation \ref{eq:binary_lens}, for the binary lens case, we can express this as the normalized lens equation: 

\begin{equation}
\label{eq:triple_lens}
w = z - \sum_{i=3} \frac{\epsilon_i}{\overline{z} - \overline{r}_{M_i}} = z - \frac{\epsilon_1}{\overline{z} - \overline{r}_{M_1}}
        - \frac{\epsilon_2}{\overline{z} - \overline{r}_{M_2}}
        - \frac{\epsilon_3}{\overline{z} - \overline{r}_{M_3}},
\end{equation}

where, $\epsilon_3$ is the fractional mass of the third lens ($\epsilon_3 = M_3 / M_{total}$). Now, the mass parameters satisfy the constraint $\epsilon_1 + \epsilon_2 + \epsilon_3 = 1$, where 1 is the total mass. Here, $z$ and $w$ represent the complex image and source positions, respectively, while $\overline{r}_{M_i}$ denotes the complex position of each lens component on the lens plane. The complex conjugates appear naturally from the complex formulation of the lens equation. In many practical microlensing configurations, the lenses are placed along the real axis, such that $\overline{r}_{M_i} = r_{M_i}$, an assumption used in Equation \ref{eq:binary_lens}.

Similar to the binary lens case, the three-lens equation \ref{eq:triple_lens} is a two-dimensional vector equation that can be embedded into a polynomial equation. In the triple lens case, this results in a tenth-order polynomial \citep{rhie2002cumbersome}, indicating that up to ten images of the source can be generated. 
Although there is no analytical solution for this polynomial equation, much like in the binary lens case, the roots can be found numerically.}

The light curve models of triple lens events also include some additional parameters. These are the separation ($s_{13}$), the mass ratio ($q_3 = M_3/M_1$) from the primary lens, and the orientation angle of $M_3$, $\psi$. In triple lens cases, $s_{12}$ and $q_2$, are used to denote the $M_1 - M_2$ separation and the $M_2/M_1$ mass ratio, respectively. In Figure \ref{fig:triple-lens vs binary-lens} we show an example of a triple lens light curve compared to a binary lens and an FSPL. Triple lens light curves can be recognised by their complexity, with multiple peaks and troughs. 

Triple lens events still represent a relatively small fraction of detected microlensing events, and modelling techniques for these more complex systems remain computationally challenging and are still under active development. Nevertheless, detections of these systems are increasing  \citep{Gaudi2008, Gould2014, Han_2013, Han2017, Song2014, Luhn2016, Danek_2015, Danek_2019}. We discuss some examples in Sections \ref{sec:circumbinary system} and \ref{sec:multi-planet system}. 

\begin{figure}[h!]
    \centering
    \includegraphics[width=0.7\linewidth]{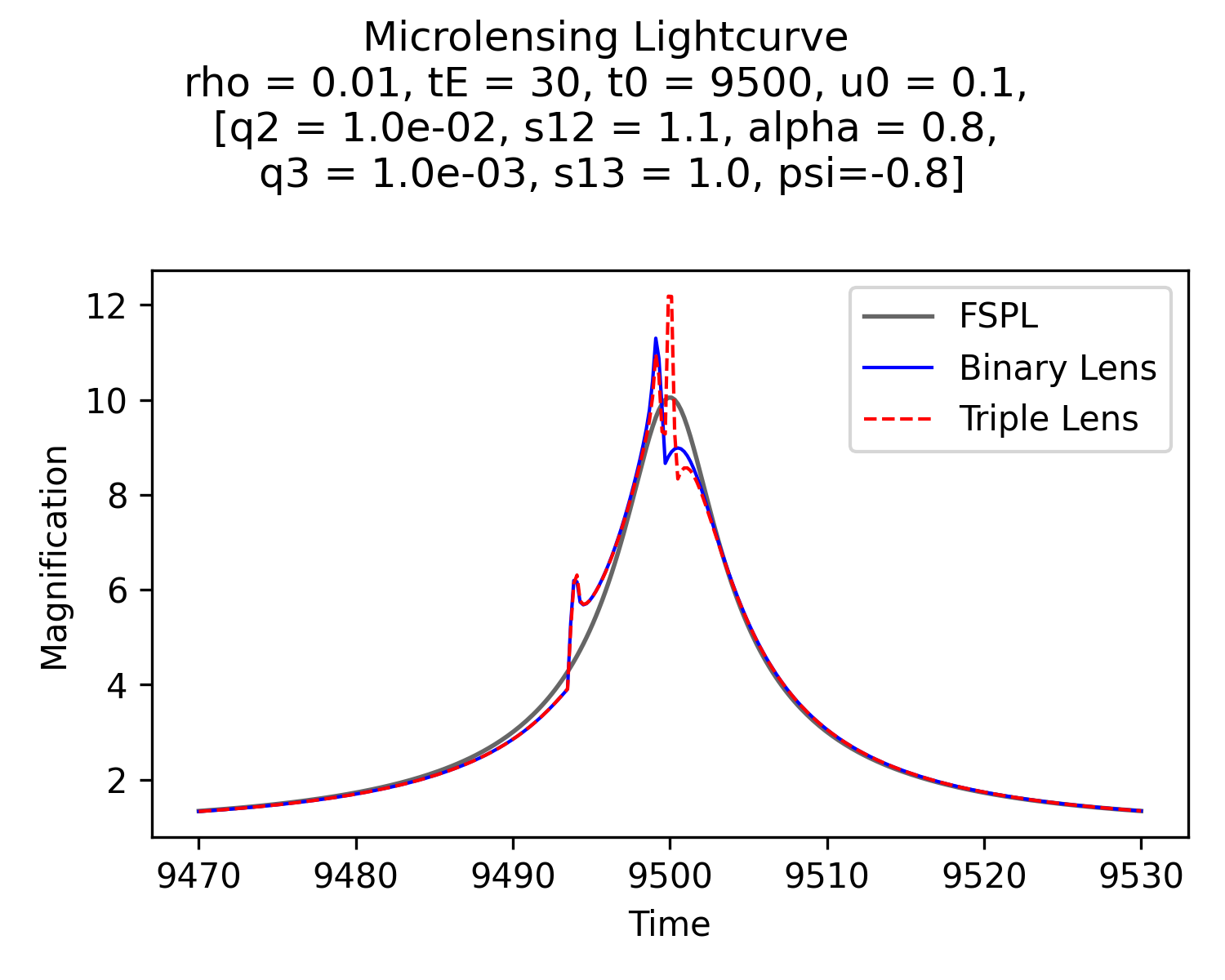}
    \caption{Comparison of a \gls{FSPL} microlensing light curve (gray) with those from \gls{2L1S} (blue) and \gls{3L1S} (red). All three models share the same parameters: $\rho = 0.01$, $t_E = 30$ days, $t_0 = 9500$ days, and $u_0 = 0.1$. The \gls{2L1S} model introduces a second lens mass with mass ratio $q_2 = 0.01$ and separation $s_{12} = 1.1$ from the primary lens, oriented at an angle $\alpha = 0.8$. The \gls{3L1S} model builds on this by adding a third lens mass with mass ratio $q_3 = 0.001$, separation $s_{13} = 1.0$, and position angle $\psi = -0.8$. Plot made with \texttt{VBMicrolensing}.}
    \label{fig:triple-lens vs binary-lens}
\end{figure}

\subsection{Characterising Lens System Parameters}
\label{sec:lens system params}

One distinguishing factor of microlensing compared to other exoplanet detection techniques is that host or lens stars are not observed directly. Additionally, these stars are often faint, or their light is blended with other stars, especially when observing towards the Galactic bulge, one of the most densely populated regions in our galaxy. This means that routine observables from the microlensing event do not directly yield physical properties, such as mass and distance, of the planetary system beyond the planet/star mass ratio $q$, and projected separation $s$. This lack of information hinders our ability to make statistical comparisons between the frequency of planets found in microlensing surveys and those detected by other planet search methods. Overcoming this limitation has been a major focus in the field of microlensing. It is now clear that more information on the host and planet can be gathered from a combination of studying second-order effects present in the microlensing light curve, and from high-angular resolution follow-up imaging. Although these methods have shown considerable success, improvements are still being made. 

Before an explanation of the methods by which we can characterise a microlensing system is given, we should revisit the routine observables for a generic microlensing event. As discussed in the previous section \ref{subsec:single_lens} in a single lens event, five parameters can be measured from the light curve photometry. These are the maximum time of magnification $t_0$, the impact parameter $u_0$, the Einstein crossing time $t_{\rm E}$, the flux from the lensed source star $F_{\rm s}$ and the blend flux $F_{\rm b}$. For binary events, two more parameters can be obtained, the mass ratio $q$, and projected separation $s$ (in units of $\theta_{\rm E}$). See Table \ref{tab:lensing_parameters} for more information.

In the following section, we describe additional parameters that can be measured when their effects are present and detected in the light curve model, and how they can lead to mass and distance measurements of the host and planet. 

\subsubsection{Second Order Effects}\label{subsec:2nd_order_effect}
Several second-order effects can be present and measured on a microlensing light curve that deviates from the classic Paczyński curve. 

\paragraph{Finite Source Effects:} One of the most common effects on the microlensing light curve is the {\it finite source effect}, which means that the angular size of the background source star is not negligible compared to the Einstein radius of the lens. Figure~\ref{fig:pspl_vs_fspl} illustrates how the shape of a microlensing light curve can change when finite-source effects are taken into account for a \gls{1L1S} case.  During a planetary microlensing event, a planet can be detected when the source star passes reasonably close to at least one of the caustics produced by the planet. When this happens, the finite angular size of the source can be measured as it deviates from the point-source approximation and is measured in units of Einstein ring radii, $\rho_* \equiv \theta_* /\theta_{\rm E}$. The angular source radius $\theta_*$ can be estimated from its magnitude and colour using empirical brightness relations \citep{Kervella2004, Boyajian2014, Adams2018}. 

The source flux can be determined by fitting its magnification model to the observed light curve, as shown in Equation \ref{eq:flux}, and by doing so it is possible to separate the varying source flux from the constant blended flux, thus deblending the source flux and allowing it to be measured. The apparent source colour can be obtained through measurements of the microlensing event in two different passbands. When both finite source effects and multicolour measurements are taken, it is possible to calculate $\theta_{\rm E}$ \citep{Gould1994a}. From there, a mass-distance relation for the lens system can be obtained: 

\begin{equation}
\label{mu_rel}
    M_{\rm L} =\frac{c^2 \theta_{\rm E}^{2}}{4G \pi_{\rm rel}}
\end{equation}
where 

\begin{equation}
    \pi_{\rm rel} = (\frac{1}{D_{\rm L}} - \frac{1}{D_{\rm S}})\ \rm AU.
\end{equation}

$\pi_{\rm rel}$ is the relative parallax between the source and the lens. The Einstein ring radius can also be related to the Einstein ring crossing time as $t_{\rm E}=\theta_{\rm E}/\mu_{\rm rel,G}$, where $\mu_{\rm rel,G}$ is the relative source-lens proper motion in the geocentric reference frame. Likewise, the source crossing time can be calculated by $t_{*}=\theta_{*}/\mu_{\rm rel,G}$.

\begin{figure}
    \centering
    \includegraphics[width=\linewidth]{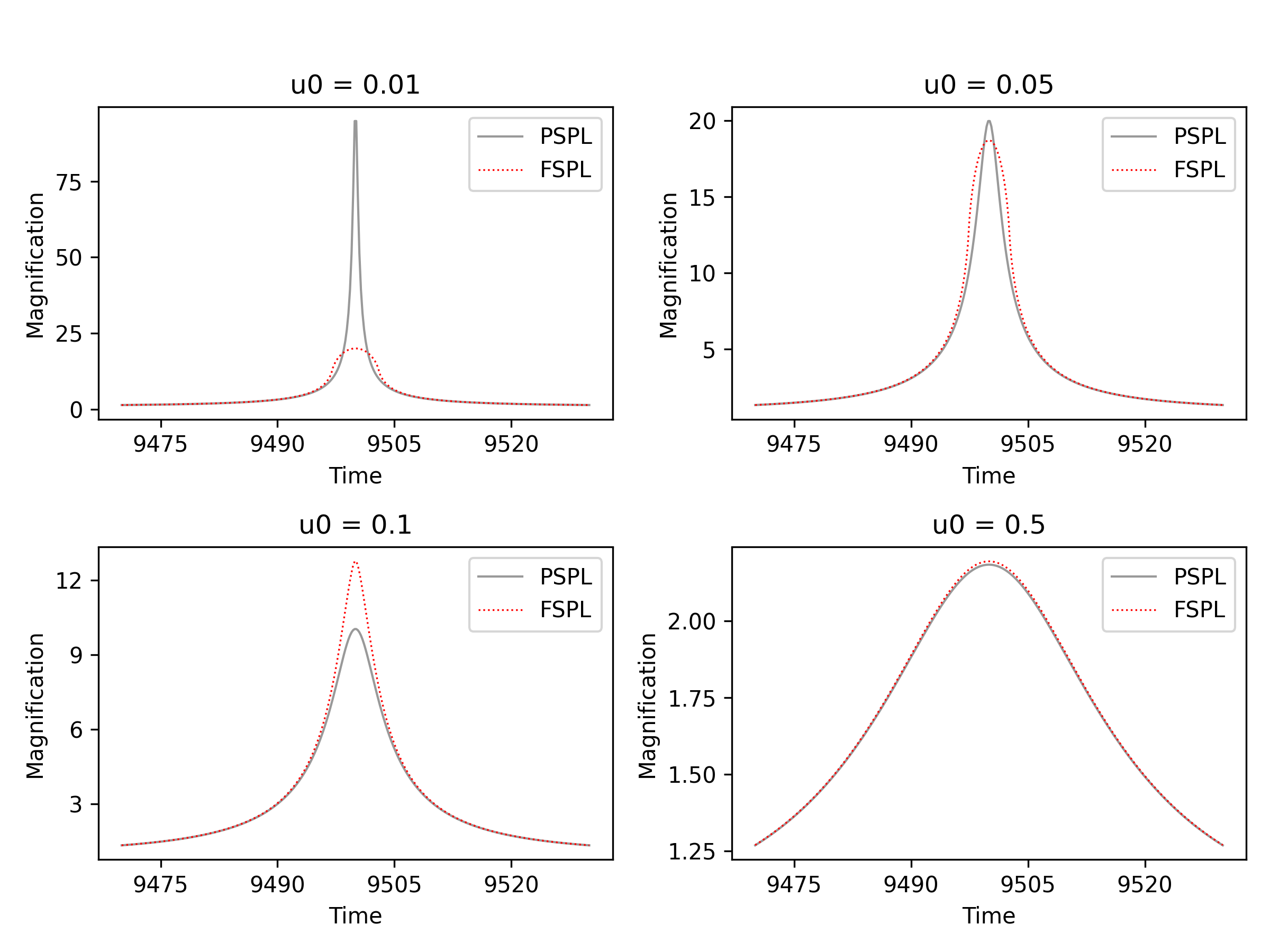}
    \caption{Comparison of a \gls{PSPL} and a \gls{FSPL} microlensing light curve for different impact parameters $u_0$. All models assume an Einstein time $t_E$ = 30 days, and a time of maximum magnification $t_0$ = 9500 days. For the \gls{FSPL} light curves, the same source radius $\rho$ = 0.1 is assumed. As the source passes closer to the lens, finite-source effects smooth the light curve peak and reduce the peak magnification. Plot made with \texttt{VBMicrolensing}.}
    \label{fig:pspl_vs_fspl}
\end{figure}

\paragraph{Microlens Parallax:}An additional effect that can change the shape of the light curve is when the assumption that the observer, the lens, and the source are all in the same place (co-spatial) and moving at constant speed, breaks down. The usual calculation for the angular separation between source and lens ($u$, measured in units of $\theta_{\rm E}$) assumes the constant motion of everything involved, and also assumes that the time of closest approach ($t_0$) and closest separation ($u_0$) are the same for all observers. When either of these assumptions fail, it is generally termed microlens parallax ($\pi_{\rm E}$) and it is a two-dimensional vector with north and east components ($\pi_{{\rm E},N}, \pi_{{\rm E},E}$). 

This can manifest in three ways: {\it orbital parallax} occurs when the timescale of a microlensing event lasts a significant fraction of a year ($>80$ days). Coupled with the acceleration of the Earth, this can lead to the a measurable change in the source-lens relative velocity. Deviations from the Paczyński light curve can be seen due to the non-linear trajectory of the lens relative to the source \citep{Gould1992}. 

{\it Terrestrial parallax} is the result of observers at different locations on Earth seeing slight shifts in the Einstein ring due to their different perspectives \citep{Gould2009}. These different perspectives will cause changes in the observed magnifications. This effect is exacerbated for sources that are close to a caustic and are therefore highly magnified \citep{HolzWald1996}. 

Finally, {\it satellite parallax} refers to the effect seen when observers are separated by a significant fraction of the Einstein ring radius projected in the observer plane. This can happen when a microlensing event is simultaneously observed from Earth and a satellite in a solar orbit \citep{Udalski2015}. The two observers will see significant differences in magnification \citep{Gould1994b}. 

A measurement of the microlens parallax can also lead to a mass-distance relation:

\begin{equation}
\label{eq:md-parallax}
    M_{\rm L} =\frac{c^2 \pi_{\rm rel}}{4G \pi_{\rm E}^2 }.
\end{equation}

In each case of microlens parallax, ambiguities can arise, leading to multiple possible interpretations of the solution. These ambiguities can stem from physical degenaracies, but also systematics in the photometry. These degeneracies limit the ability to extract useful parallax parameters. However, if the microlens parallax is robustly measured, together with finite source effects, then the mass and distance of the lens system can be calculated using the two mass-distance relations (equations \ref{mu_rel} and \ref{eq:md-parallax}).

A key characteristic of introducing parallax into the model is the asymmetry it produces in the wings (or tails) of the light curve. Often, this asymmetry is subtle and might be difficult to detect by eye. In Figure~\ref{fig:parallax_and_om_diff}, we illustrate this effect by comparing a binary lens microlensing light curve without parallax with one that includes parallax.

\begin{figure}[h]
\centering
\includegraphics[width=\linewidth]{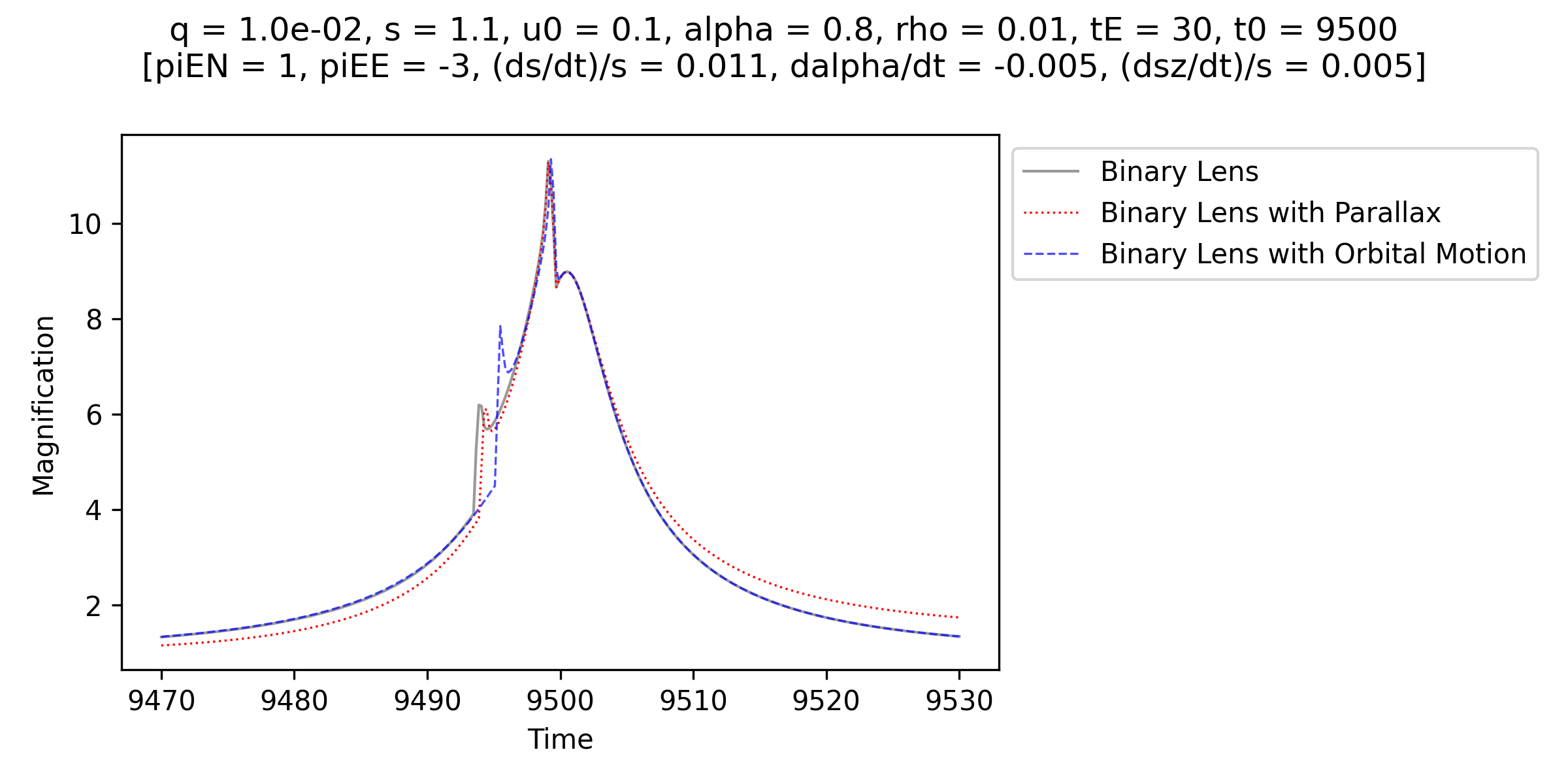}
\caption{Comparison of a binary-lens microlensing light curve without higher-order effects, and with the addition of parallax and orbital motion. Plot made using \texttt{VBMicrolensing}.}
\label{fig:parallax_and_om_diff}
\end{figure}

\paragraph{Lens Orbital Motion:} Microlensing detects companions that are located around the Einstein radius, typically around 0.5-10 AU. Therefore, the timescale for a planetary or binary stellar companion to orbit within the lens system is on the order of a few months, at least. In comparison, a microlensing event generally lasts only a few weeks, making it a small fraction of the orbital period. Therefore, the relative positions of the lensing bodies can be assumed to be fixed during the event. However, there are cases where the relative orbital motion of the various components can result in deviations in the light curve, as shown in Figure~\ref{fig:parallax_and_om_diff}, at the time of the microlensing event. The most common approach to introducing orbital motion to the light curve model is by adding two parameters. These parameters describe the instantaneous velocities of the secondary binary component relative to the primary (e.g. the planet's instantaneous velocity compared to its host's). Since the position of the binary lens changes so little during the brief microlensing event, we can assume that the relative motion is linear \citep{Dominik1998}. This assumption keeps calculations simplified while retaining the orbital motion effect. The projected instantaneous orbital velocity components of the lens can be described at the projected separation, $s(t)$, and the position angle, $\alpha(t)$, where: 

\begin{gather}
     s(t) = s_0 + \frac{{\rm d}s}{{\rm d}t}t,\\
     \alpha(t) = \alpha_0 + \frac{{\rm d}\alpha}{{\rm d}t}t.
\end{gather}
$s_0$ and $\alpha_0$ are the initial projected separation and the initial position angle at the beginning of the microlensing event, respectively. $\frac{{\rm d}s}{{\rm d}t}$ is the rate of change between the lens components projected onto the lens plane, and $\frac{{\rm d}\alpha}{{\rm d}t}$ is the angular rotation rate of the lens. A mass-distance relation can be obtained through Kepler's third law.


The linear approximation for orbital motion, however, can break down if the orbital motion is very rapid (e.g. close binaries or very short-period planets) or when the microlensing event duration is long. Then, a more complex orbital model could be required, called a ``full Keplerian" orbital model, where the elliptical motion of the bodies is taken into account. This model entails two additional parameters that describe the motion perpendicular to the plane of the sky (see \citealt{Skowron2011, Shin2011}). 

When the microlensing parallax effect is present along with orbital motion, these two effects are often degenerate due to rotational symmetry.

\paragraph{Binary Sources:} When a source has a companion and the lens passes sufficiently close to the caustics associated with each source component, then a deviation in the light curve can be seen. In the case of static binary sources, where their orbit is relatively stable during the microlensing event, the deviation depends on several factors. First, is the flux ratio between the two sources $F_{\rm R}$. When one source is brighter than the other, then its contribution to the light curve will dominate. Second, is the impact parameter ($u_{\rm 01}$ and $u_{\rm 02}$ for the primary and secondary source, respectively) which describes the closest approach of the lens to each source. Finally, there is the time of closest approach of the primary source ($t_{\rm 01}$) and the secondary source ($t_{\rm 02}$). This type of effect is known as a {\it binary source} effect, where a deviation is seen due to the magnification of the second source by the lens. The Einstein ring radius can be estimated by $\theta_{\rm E} = \theta_{*,1}/\rho_{*,1}$ or $\theta_{\rm E} = \theta_{*,2}/\rho_{*,2}$.

When the primary source's motion is affected due to the companion, additional deviations can be seen on the light curve over time. This effect is called {\it xallarap}, which is parallax spelled backward. Light curve modelling with xallarap involves the addition of five parameters; xallarap vector $\xi_{\rm E}$, and its north and east components, $\xi_{{\rm E},N}, \xi_{{\rm E},E}$; the orbital period of the sources, $P$; and the phase and inclination angles of the orbit ($\phi, i$), respectively. The xallarap parameter can be related to the Einstein ring radius by: 

\begin{equation}
    \xi_{\rm E} = \frac{a_{\rm S}}{D_{\rm S}\theta_{\rm E}},
\end{equation}
where $a_{\rm S}$ is the semi-major axis of the source companion around the centre of mass of the binary source system \citep{Han1997, Miyake2012}. From Kepler's third law: 

\begin{equation}
    a_{S}^3 = \frac{P_\xi ^2 M_{S,2}^3}{(M_{S,1} +M_{S,2})^2},
\end{equation}
where $M_{\rm S,1}$ is the mass of the primary source and $M_{\rm S,2}$ is the mass of the secondary source. $P_\xi$ is the orbital period of the binary source system, and together with the $\xi_E$ vector, it can be inferred from the light curve model. However, the individual masses of the binary sources are not often well-constrained since these are two unknowns. To get an estimate of these masses, one would have to employ a Galactic model to provide priors on the expected stellar population. 

Depending on the configuration of the source and lens system, the light curve could exhibit only binary source effects, only xallarap effects, or both. The xallarap effect is stronger for binary sources with a short orbital period, and becomes undetectable for events where $P_\xi \gg t_{\rm E}$.

\subsubsection{High-Angular Resolution Follow-up Observations}
\label{sec: high res}

Obtaining the physical properties of the lens system from second-order effects on the light curve can sometimes prove to be tricky, since they are often prone to degeneracies. Furthermore, to obtain robust measurements of the lens properties, one would need at least two mass-distance relations. high-angular resolution follow-up observations allow us to directly observe the lens star - which is not possible during the microlensing event when the source, lens and observer are aligned. These observations are typically taken with large ground-based telescopes with adaptive optics capabilities (such as Keck I and II, \citealt{Wizinowich_2000}, or Subaru, \citealt{Luo_subaru_2010}, on Mauna Kea, Hawai'i), or with space-based telescopes such as \gls{Hubble} or \gls{Webb}. These powerful telescopes can resolve or partially resolve the source and lens, enabling us to directly measure the lens flux and relative proper motion (magnitude and direction) to the source. 

\paragraph{Lens Flux: } High resolution follow-up observations are typically taken in the near-infrared bands, such as $J$ ($\lambda_{\rm central} = 1.25\mu$m), $K$ ($\lambda_{\rm central} = 1.65\mu$m), and $H$ ($\lambda_{\rm central} = 2.2\mu$m) from ground-based telescopes. These passbands suffer less extinction compared to optical wavelengths as they can more effectively penetrate the interstellar dust between us and the Galactic centre. A measured magnitude of the lens star can yield the following mass-distance relation:

\begin{equation}
\label{eq: mass vs luminosity}
    K_{\rm L} = 10 + 5{\rm log_{10}}(D_{\rm L}/1{\rm kpc}) + M_{\mathcal{K}}( M_{\rm L}) + A_{\rm K,L},
\end{equation}
where $M_{\mathcal{K}}( M_{\rm L})$ is a mass-luminosity relation in the $K$ passband (but can be substituted for any passband), and $A_{\rm K,L}$ is the dust extinction in the foreground of the lens star. Empirical mass-luminosity relations for a main sequence star are appropriate in the majority of cases. Examples of empirical mass-luminosity relations are from \cite{Henry1993} ($M_{\rm L} \geq 0.66\ M_{\odot}$), \cite{Delfosse2000} ($0.54\ M_{\odot} \geq M_{\rm L} \geq 0.12\ M_{\odot}$) and \cite{Henry1999} ($0.10\ M_{\odot} \geq M_{\rm L} \geq 0.07\ M_{\odot}$). However, for host stars $\sim 1M_{\odot}$ whose luminosity may change substantially over the age of the Galaxy, isochrones are likely to be more accurate \cite[e.g.][]{Girardi2002}. Measurements of the lens flux in multiple passbands can give independent mass and distance estimates.

\paragraph{Relative Proper Motion: }When the peak of the microlensing event has passed, the source and lens are no longer aligned. Depending on their relative proper motion (how many arcseconds per year they move apart relative to each other), follow-up observations can be taken $\sim$3-10 years after the peak of the event to resolve or partially resolve the lens. This can also happen in reverse if high-resolution images of the lens and source were taken before the microlensing event took place. In high-resolution imaging, therefore, the relative proper motion can be calculated by measuring the source-lens separation as a function of time, multiple epochs of data will give more robust measurements. When the relative proper-motion is measured in this way, it is in the {\it heliocentric reference frame} expressed as $\mu_{\rm rel, H}$. 

 The relative proper motion can also be related to light curve properties such as the angular source size and the Einstein ring radius, due to finite source effects. In this case, the measurement would be in the {\it geocentric reference frame} written as $\mu_{\rm rel, G} = \theta_{*}/t_{*} = \theta_{\rm E}/t_{\rm E}$. This relationship between proper motion and Einstein ring radius enables us to obtain the mass-distance relation found in Equation \ref{mu_rel}.

Therefore, a measurement of the relative proper motion from directly imaging the lens in high resolution data can yield mass and distance measurements. However, before this can be done, we must convert between the two reference frames, i.e. convert  $\mu_{\rm rel, H}$ to $\mu_{\rm rel, G}$. The difference between the two $\mu_{\rm rel}$ is usually insignificant, although this can change for lenses at small distances or for small proper motions. Converting between the two reference frames can be achieved by: 

\begin{equation}
\label{mu_G-mu_H}
    \mu_{\rm rel, G} = \mu_{\rm rel, H} - \frac{v_\oplus \pi_{\rm rel}}{\rm AU},
\end{equation}
where $v_\oplus$ is the projected velocity of Earth relative to the Sun at the time of peak magnification during the microlensing event. 

Direct measurement of $\mu_{\rm rel, H}$ can also help determine more precise values for the microlensing parallax. Typically, one component of the two-dimensional parallax is measured more precisely than the other. This is the component in the east-west direction $\pi_{\rm E,E}$, while the North-South component $\pi_{\rm E,N}$ is weakly constrained. Since the microlensing parallax vector is parallel to the relative proper motion vector (which is measured more precisely in high resolution follow-up observations), we can calculate a more precise $\pi_{\rm E,N}$ using: 

\begin{equation}
    \pi_{\rm E} = \frac{\pi_{\rm rel}}{t_{\rm E}} \frac{\mu_{\rm rel,G}}{|\mu_{\rm rel,G}|^2}
\end{equation}
and Equation \ref{mu_G-mu_H} \citep{Gould1994, Ghosh2004, Bennett2007}.

\begin{sidewaystable}
\centering
\begin{tabular}{|l|p{1.8cm}|p{7cm}|p{8cm}|}
\hline
\textbf{Type} & \textbf{Commonly Used Symbol} & \textbf{Parameter Name} & \textbf{Explanation} \\
\hline

\multirow{10}{*}{} 
& $A(t)$ & Total magnification or amplification of the event at any given time & The magnification of the source star's flux due to the gravitational microlensing phenomenon. \\
& $J$ & Jacobian matrix & Conversion matrix between a point in the source plane and the corresponding point in the lens plane. \\
& $F_i(t)$ & Total flux $F_i(t)$ for time $t$ for instrument $i$ & $F_i(t) = A(t)\:f_{s,i} + f_{b,i}$ \\
& $f_{s,i}$ & Source Flux & Flux from the source for instrument $i$. \\
& $f_{b,i}$ & Blend Flux & Flux from other stars that were blended with the microlensed source star. \\
& $\theta_\mathrm{E}$ & Angular Einstein radius & $\theta_\mathrm{E} = \sqrt{\frac{4GM_\mathrm{L}}{c^2D_\mathrm{S}}\left(\frac{D_\mathrm{S}}{D_\mathrm{L}}-1\right)}$ \\ 
& $G$ & Gravitational constant & \\
& $M_\mathrm{L}$ & Mass of the lens system & \\
& $c$ & Speed of light & \\ 
& $D_\mathrm{S}$ & The observer-source distance & \\
&  $D_\mathrm{L}$ & The observer-lens distance & \\
\hline
\multirow{6}{*}{\Gls{1L1S}} 
& $t_\mathrm{E}$ & Einstein crossing time & Time taken for the background source to cross the lens' Einstein radius, as seen by the observer. The time scale of the gravitational lensing event. \\
& $u_0$ & Minimum angular separation & Minimum angular separation between the source and lens as seen by the observer, normalized by the Einstein radius $\theta_\mathrm{E}$. \\
& $t_0$ & Time of the peak or minimum separation & Time when the lens-source separation is smallest (when $u = u_0$, $t = t_0$). \\
& $\theta_*$& Source angular radius & \\
& $\rho_*$ & Source angular radius in Einstein units & $\rho_* \equiv \theta_* / \theta_\mathrm{E} $ \\
& $t_*$ & Source radius crossing time & Time it takes for the source radius to cross the lens. ($t_* \equiv \rho_*  t_E$) \\
\hline
\multirow{3}{*}{\Gls{2L1S}} 
& $s$ & Separation of binary lens masses & Separation of the two lenses during the event, normalized by the Einstein radius $\theta_\mathrm{E}$. \\
& $\alpha$ & Angle between lens-source motion and binary axis & Counter-clockwise angle between the lens-source relative motion and the binary lens axis. Not to be confused with the angle of light bending. \\
& $q$ & Mass ratio & Ratio of the secondary lens mass to the primary lens mass. \\
\hline
\multirow{6}{*}{Binary source} 
& $F_\mathrm{R}$ & Flux ratio & Flux ratio of the secondary source to the primary source. \\
& $u_{01}$ & Impact parameter of primary source & Impact parameter of the primary source. \\
& $u_{02}$ & Impact parameter of secondary source & Impact parameter of the secondary source. \\
& $t_{01}$ & Closest approach time of primary source & Time of closest approach for the primary source. \\
& $t_{02}$ & Closest approach time of secondary source & Time of closest approach for the secondary source. \\
& $\rho_{*,1}$ & Source radius for primary  & Radius of the primary source. \\
\hline
\end{tabular}
\caption{Parameters for different lensing scenarios}
\label{tab:lensing_parameters}
\end{sidewaystable}

\section{Demographics of the Detected Planets}
\label{demographics}

Thousands of planets have been discovered to date using all the different detection techniques, and tens of thousands more are expected to be discovered as space-based technology improves. Only a couple hundred of these planets, however, have been discovered by gravitational microlensing since it relies on the chance alignment between the source, lens, and observer. Another limitation is that of the detected planetary systems, obtaining precise and reliable measurements of mass and distance can be challenging, as outlined in Section \ref{sec:lens system params}. A major reason for this is that the detection does not rely on the host or companion's light, and although this poses some challenges, it also comes with several advantages. 

 Microlensing observations provide important tests of planet formation models because they are sensitive to planets in wide orbits across a broad range of masses. For example, \cite{Suzuki_2018} found substantially more intermediate-mass giant planets than predicted by standard core accretion models, suggesting that current theories of planet formation may be incomplete.

Each exoplanet detection method has its strengths, with microlensing uniquely suited to detecting low-mass planets in wide orbits around a diverse range of host stars. Additionally, it is currently the only method capable of probing planetary populations far beyond the solar neighbourhood toward the Galactic centre. In the following subsections, we outline the regions of parameter space to which microlensing is sensitive.

\subsection{Wide Orbit `Cold' Planets}

The probability of detecting a planet with microlensing is maximised when the planet projected separation from its host star is approximately equal to the Einstein radius of the lensing system, $r_{\rm E}=D_{\rm L}\theta_{\rm E}$. This is the radius of the Einstein ring image, which is seen when the lens and source are in perfect alignment and where lensing effects are the strongest. When a planet is near the Einstein radius, its gravitational influence will call a deviation, or planetary anomaly, on the light curve. The Einstein radius for typical microlensing events will correspond to a physical distance of 1-5 AU for solar mass lenses towards the Galactic centre. Planets with a projected separation equal to these distances are more likely to be detected, as the sensitivity peaks at the Einstein radius of the host star \citep{Mao1991, GouldLoeb1992, Bennett1996}. Planets with smaller orbital separations lie well within the Einstein radius, and therefore their lensing signal is weaker or often indistinguishable from the host star's lensing effect. Therefore, these planets are harder to detect with gravitational microlensing. Planets that lie beyond the Einstein radius (so even greater separation from the host star) can be detected, although the probability decreases the further away the planet gets.

For many planetary systems, this Einstein radius distance (1-5 AU) is expected to coincide with the region beyond the protoplanetary disk \citep{GouldLoeb1992}. This is where temperatures are below the ideal temperature of water in its liquid form, therefore earning its name as the {\it snow line}. Planets at or beyond the snow line are typically referred to as `cold' planets. In the right panel of Figure \ref{fig:planet_list} we see that many planets found with microlensing are estimated to be orbiting beyond their system's snow line. In particular, we can use microlensing's unique capability to detect low-mass planets beyond the snow line. This parameter space complements the other detection techniques, which are sensitive to planets with shorter orbital radii, for instance transit photometry and radial velocity. 

A study by \cite{Zang2025} found that super-Earths are common in wide Jupiter-like orbits, with approximately 0.35 super-Earth planets per star occuring in these systems. While many exoplanets spanning a wide range of masses have been detected in short-period orbits, their abundance at wider separations remains poorly constrained. The results of \cite{Zang2025} suggest that the observed population of microlensing planets is inconsistent with many standard planet-formation models, particularly those that include planet migration. This further highlights the importance of microlensing surveys to the study of planetary formation and demonstrates how synergy between multiple exoplanet detection methods is necessary to obtain a complete picture of planetary systems.

\begin{figure}[h!]
    \centering
    \includegraphics[width=1\linewidth]{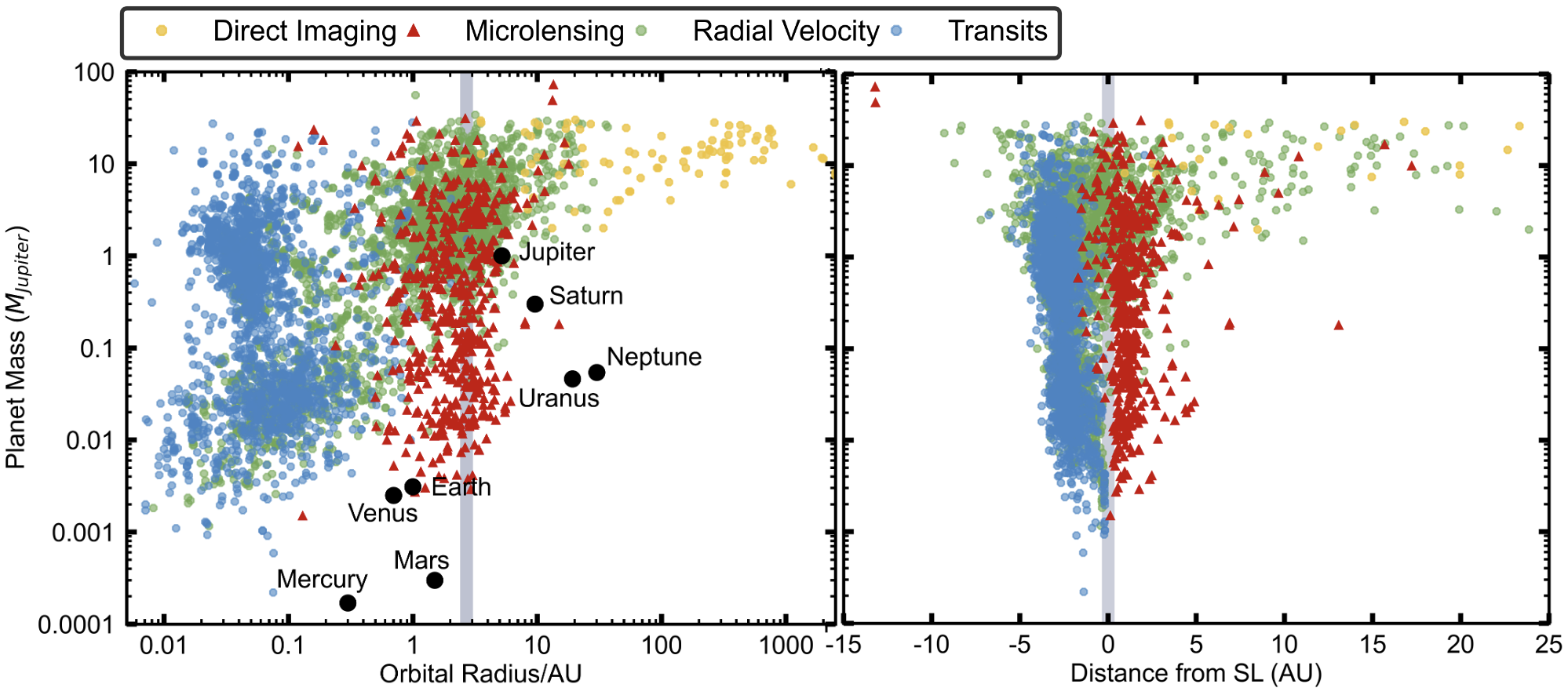}
    \caption{\textit{Left Panel:} Exoplanets from the imaging, microlensing, radial velocity and transit technique. The grey line indicates the snow line for the Solar System. \textit{Right panel:} Exoplanet mass as a function of the estimated snow line distance from their host star. Data taken from the \href{https://exoplanetarchive.ipac.caltech.edu/}{NASA Exoplanet Archive}.}
    \label{fig:planet_list}
\end{figure}

\subsection{Free Floating Planets}

Free-floating (or `rogue') planets are objects that do not orbit any star, and these elusive objects can be detected with gravitational microlensing \citep{DiStefano1999}. Rogue planets have two possible ways of forming. They could either form in isolation, where they would assemble through a similar process to star formation \citep{Whitworth2006}, or they could form around a star, initially bound, and only later escape its host's gravitational field \citep{Vera2012, sumi2011unbound}. Planets can be ejected from their hosts gravitational field in a number of ways. This can be the result of chaotic processes during planet formation \citep[e.g.][]{Rasio1996}, stellar flybys \citep[e.g.][]{Malmberg2011},  or when the host enters its post main sequence stage \citep[e.g.][]{Adams2013}.

The survey collaborations \gls{MOA}, \gls{OGLE} and \gls{KMTNet} have a sufficiently high enough cadence to detect isolated microlensing events, in addition to planetary events. Microlensing events with short time scales ($\sim$1 day long) were identified and studied, with the results indicating that there could be a substantial population of Jupiter-mass rogue planets, with an abundance of $\sim$2 rogue Jupiter-mass planets per star in the Galaxy. These short time scale events can be found in \citet{sumi2011unbound}, with the search being expanded in \cite{sumi2023free} with MOA-II survey data collected between 2006 and 2014. They found six short-duration microlensing events likely to be caused by rogue planets. \cite{Koshimoto_2023} using the same data also discovered low-mass rogue planet candidates. They analysed thousands of single lens microlensing event candidates, singling out thirteen very short-duration events which could be caused by low-mass rogue planets. Since these short duration events caused by low-mass objects are very difficult to detect, their results suggested that there may actually be a very large population of low-mass rogue planets in our galaxy.

A study of rogue planets was also conducted from \gls{KMTNet} survey data \citep{Gould2022}, further supporting the idea that rogue planets are extremely common. A comparison between the rogue planet mass-functions found in \cite{Gould2022} and in \cite{sumi2023free} are compared in \cite{Yee_2025}.

Other recent studies on the occurrence rates of rogue planets have suggested that there are fewer than 0.25 rogue Jupiter-mass planets per star in the Galaxy \citep{Mroz2023}. Moreover, a study by \cite{DeRocco2024} finds microlensing-based upper limits on the abundance of sub-terrestrial free-floating planets, constraining their number density to below $1.4 \times 10^7$ per cubic parsec for masses $10^{-4}\ M_{\oplus}$.

Short-duration microlensing events can be indicators of rogue planets because the duration of the microlensing event (the Einstein crossing time, $t_{\rm E}$) depends on the mass of the lens. Jupiter-mass objects are expected to produce microlensing events with $\sim$1 day timescales, whereas Earth-mass objects $\sim$0.1 days. However, if a bound planet has a large enough orbit, the light from the source star may only be magnified by the planet and not the host star \citep{DiStefano1999, HanKang2003, Han_2005}. This confusion can seriously impact the occurrence rates of rogue planets. In Figure \ref{fig:ffp_lc} we show three examples of candidate rogue planets. We use the word candidate because we cannot rule out the possibility of these planets orbiting a very distant companion star.

The study of rogue planets is still in its prime, but with the golden age of space-based microlensing surveys on the horizon, particularly with \gls{Roman} (see also \citealt{Johnson2020}), this niche, yet important, exoplanet demographic will be better understood. 

Some research has also been carried out on the detection of rogue planets in other galaxies. \cite{Dai2018} found that quasar microlensing could be a means of probing planets in galaxies that act as lenses. They show that a population of approximately 2000 objects ranging from Moon to Jupiter mass per main sequence star could explain the energy shifts of X-ray emission lines in the gravitationally lensed quasar RXJ 1131-1231.
 
\begin{figure}[h!]
    \centering
    \includegraphics[width=1\linewidth]{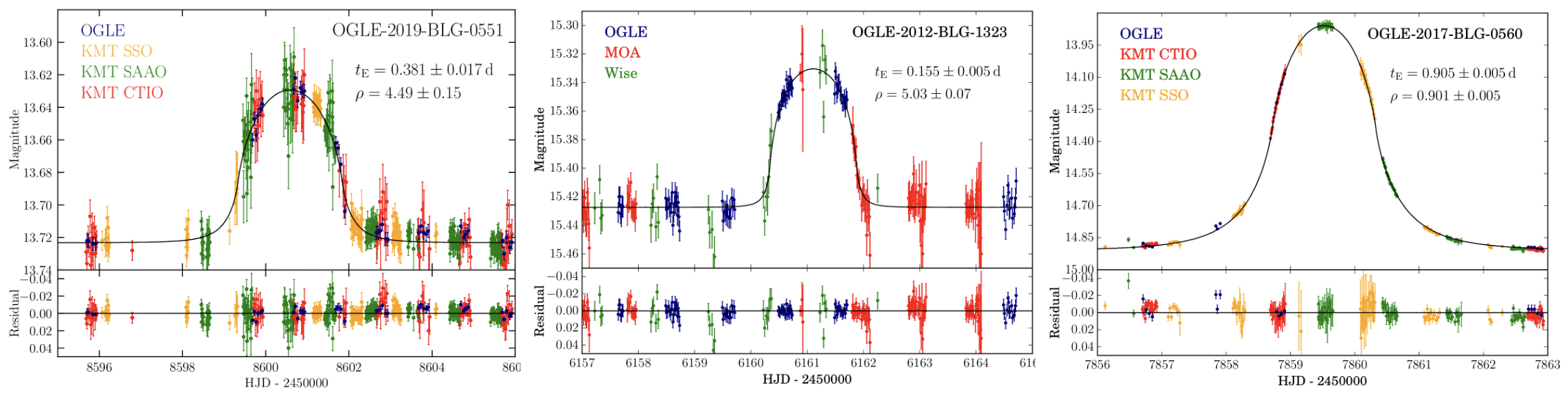}
    \caption{The light curves of candidate rogue planets OGLE-2019-BLG-0551, OGLE-2012-BLG-1323, and OGLE-2017-BLG-0560, taken from  left \cite{mroz_2020_ff} Middle and right\cite{Mroz2019}.} 
    \label{fig:ffp_lc}
\end{figure}

\subsection{Low-Mass Planets}

When a background source's light is magnified by a planetary lens system, the magnitude of the perturbation on the light curve due to the planet does not depend on the planet's mass. Instead, it depends on how close the source's light passes the planetary caustics. If the source passes very close or crosses the planetary caustics, it can cause a very large perturbation. As the mass of the planet decreases, planetary deviations also become rarer and briefer, but the magnitude of the perturbation can still be large for the background source. Therefore, microlensing can be sensitive to low-mass planets \citep{Bennett1996}. 

Although the full picture of low-mass planet demographics is yet to be seen, a number of studies have been conducted in the last decade, probing the detection efficiency and mass function of planets identified through microlensing. A study by \cite{Cassan2012} found that planets with lower masses, particularly in the super-Earth to sub-Neptune range (5-15 $M_{\oplus}$), were relatively common at wider orbits (1-10 AU). Furthermore, they estimated that planets are, in general, abundant in our Galaxy, with at least one planet per star. 

Another prominent study in microlensing is by \cite{Suzuki2016}. They studied the planet-host mass-ratio for a number of microlensing events, finding that the mass-ratio function breaks and peaks at $q\sim10^{-4}$ for planets in wide orbits. This implies that cold Neptune mass planets are the most frequently detected by microlensing, which contradicts the prediction made by \cite{Laughlin2004} of a sub-Saturn desert for wide-orbit planets. However, further investigation with more precise and independent measurements of both planet and host star masses (rather than just the mass-ratio) is essential for a clearer understanding.

 Low-mass planets, down to Earth or even Mars still remain hard to detect with ground-based microlensing. This is because ground-based observations have a limited resolution, making short or faint deviations in the light curve difficult to capture. The upcoming \gls{Roman} mission is expected to uncover a substantial population of terrestrial-mass planets through its space-based microlensing survey. 


\subsection{Host Star demographics}

Due to gravitational lensing's unique nature, microlensing can detect planets orbiting a variety of host stars, or host objects, from brown dwarfs to black holes.

This is a consequence of the technique not relying on the light emitted by the lensing object. Microlensing relies only on the gravitational field of the lens, not its luminosity. This means that even very faint or non-luminous objects, such as brown dwarfs, white dwarfs, neutron stars, and black holes, can act as lenses \citep[e.g.][]{Sahu2022, Lam_2022} and potentially host a planet. \cite{Blackman2021Nature} report the possible discovery of the first wide-orbit planet around a white dwarf, providing evidence that planets in such systems can survive both the giant and asymptotic giant phases of their host star's evolution. Such detection proved that microlensing is a method that can also detect older planetary systems that have evolved off the main sequence.  Low-mass stars are far more numerous in our Galaxy than high-mass stars, which is why microlensing surveys have detected large numbers of them, making microlensing an ideal method for their discovery. Moreover, the Einstein crossing time scales with the square root of the lens mass, $t_{\rm E} \propto \sqrt{M}$, so events caused by low-mass stars have shorter durations. Surveys such as OGLE, MOA, KMTNet, and the forthcoming Roman Space Telescope are designed with sufficiently high cadences to remain sensitive to these short-timescale events.

One characteristic that needs to be explored further is the occurrence rate of planets that orbit beyond the snow line as a function of host star mass. For example, at these wider orbits, are more massive stars more likely to host more massive planets? Studies with Kepler data, for planets in short orbits, have indicated there is a significant difference in planet occurrence for M dwarfs compared to solar type stars with spectral types F, G, and K \citep{Mulders2015}. However, a different trend is expected for planets beyond their host star's snow line. \cite{Laughlin2004} suggests that massive planets, such as gas giants, will form more easily around more massive stars. Moreover, they suggest that the protoplanetary disks of smaller stars, such as M-dwarfs, will dissipate their Hydrogen and Helium before they can be accreted onto protoplanets to form gas giants. However, microlensing has detected several systems with gas giants orbiting M dwarfs \citep[e.g.][]{Poleski_2014, Tsapras2014, Dong2009b, Bennett_2020, Bennett2024}, indicating that the formation of gas giants around M dwarfs could be more plausible than previously predicted. 

Most of the planets detected so far have been at distances $<1$ kpc from Earth. Microlensing is currently the only technique that can detect planetary systems towards the Galactic centre. This gives us the opportunity to study lenses in both the disk and Galactic bulge, which can be seen in Figure \ref{fig:host_dist}. A study by \cite{Koshimoto_2021} using microlensing data to investigate the distribution of exoplanets at varying distances to the Galactic centre found that there is no significant dependence of planet occurrence on the Galactocentric distance. This implies that cold planets—those most readily detectable via microlensing—are fairly uniformly distributed across the Galaxy, including within the Galactic bulge. Therefore, planet formation processes might be similarly effective across different galactic environments, from the dense stellar populations of the bulge to the sparser outer regions of the disk. 

Due to the limitations that microlensing faces, it is difficult to determine the stellar ages, surface temperatures, and spectral types of the host stars. Follow-up observations of the host star could address this once the source and lens are resolved. Observations would also have to be taken in multiple band-widths in both the near-infrared and visible, so that accurate colour information can be achieved. In the future, this can be achieved with \gls{Roman} since it will have the capability to conduct follow-up observations over a wide range of bands. 

\begin{figure}[h!]
    \centering
    \includegraphics[width=0.6\linewidth]{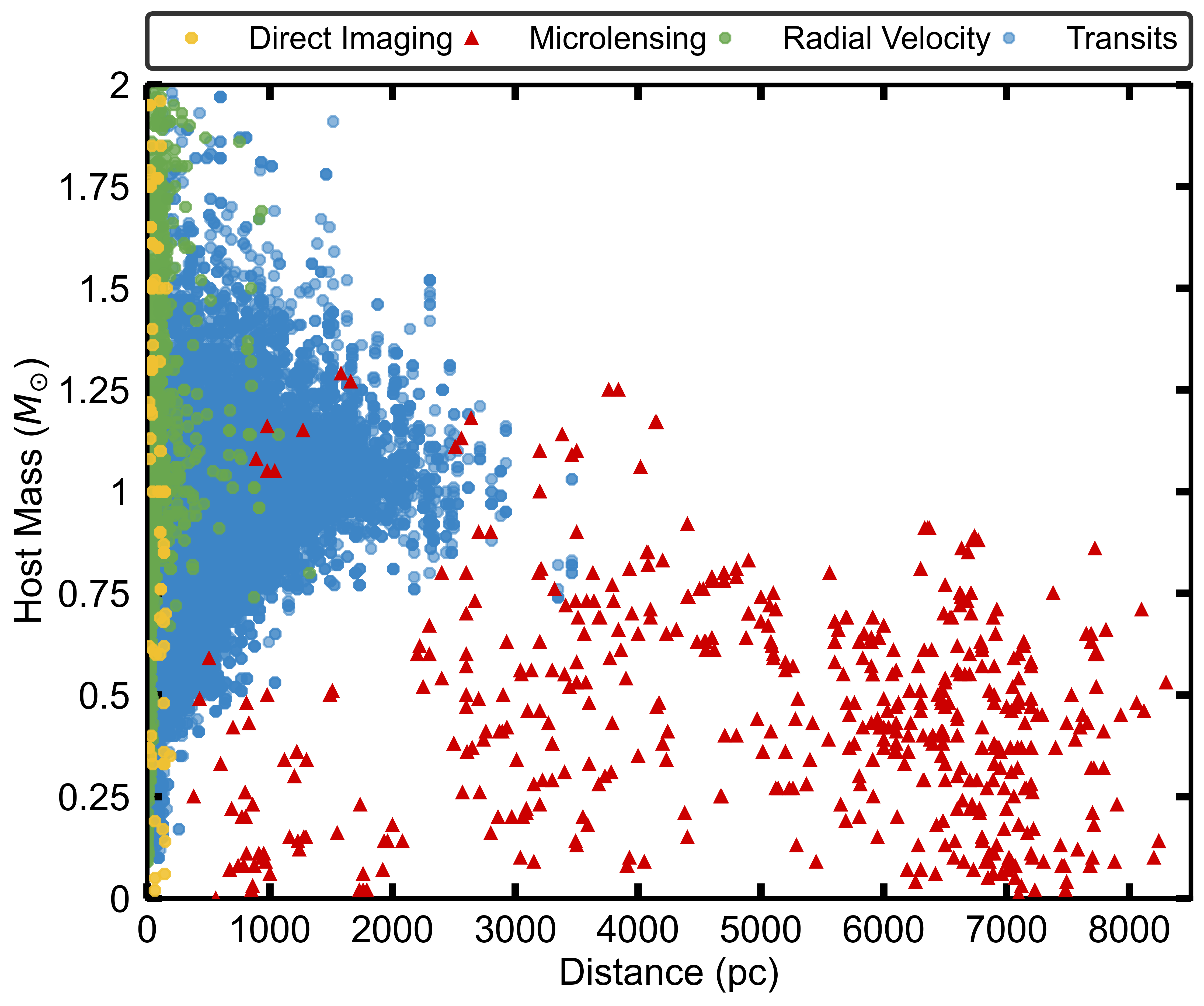}
    \caption{Host stars of detected planetary systems as a function of their distance from Earth. Data taken from the \href{https://exoplanetarchive.ipac.caltech.edu/}{NASA Exoplanet Archive}.}
    \label{fig:host_dist}
\end{figure}

\subsection{Circumbinary Systems}
\label{sec:circumbinary system}

A system that involves a planet orbiting a pair of stars in close orbit is called a circumbinary system. Although microlensing is sensitive to such systems, the detection of them can be challenging due to the complex nature of their light curves, and the precise alignment that needs to occur. The first confirmed circumbinary system detected via microlensing is OGLE-2007-BLG-349 \citep{Bennett2016}, which consists of a 80 $M_{\oplus}$ planet orbiting a pair of M-dwarfs. More recently, \cite{Han2020} analysed OGLE-2018-BLG-1700, finding that the system consists of a planet that belongs to a stellar binary system. The system's triple lens nature (two host stars and a planet) was identified from the complex anomaly pattern of the light curve (see Figure \ref{fig:circum}).

Modelling such systems requires the use of triple-lens modelling \citep{rhie1997infimum}, described in Section \ref{sec:triple lens}. The light curve models of triple-lens systems appear more complex, as all objects in the lensing system perturb the light from the background source. In Figures \ref{fig:circum} and \ref{fig:multi-planet} we show examples of the complex nature of triple-lens light curves. 

The discovery of these types of systems has large implications for understanding planet formation. 

\begin{figure}[h!]
    \centering
    \includegraphics[width=0.8\linewidth]{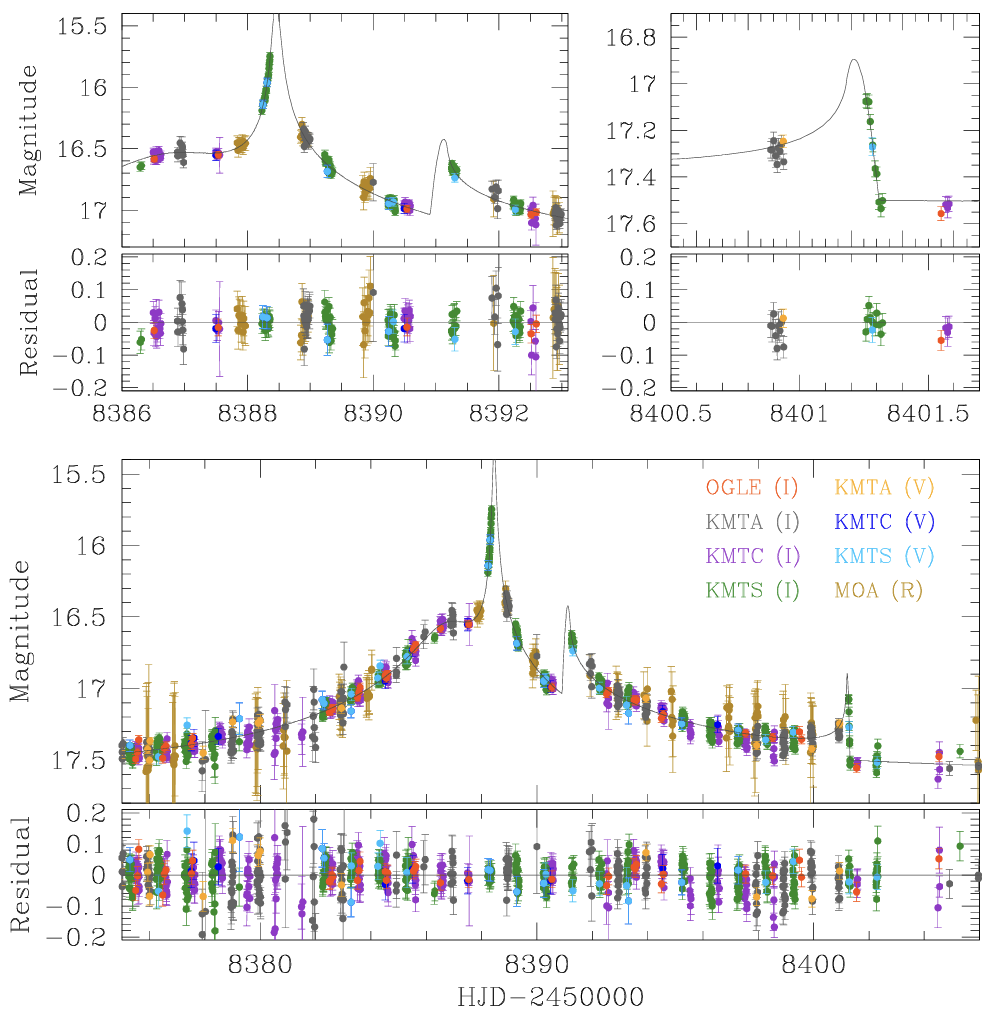}
    \caption{The light curve model for circumbinary event OGLE-2018-BLG-1700, from \cite{Han2020}. The upper panels show the zooms of the peak regions.}
    \label{fig:circum}
\end{figure}

\subsection{Multi-planet Systems}
\label{sec:multi-planet system}

Although rare, microlensing has also detected systems with multiple planets (i.e. multiple planets orbiting one star). It is rare to detect these systems with microlensing because all planets in the system would need to significantly perturb the light curve, and the chance alignment required to detect a single planet is already small. Nevertheless, such detections are possible if the planets are within the lensing zone and affect the magnification pattern of the light curve \citep{Griest1998}. 

The first published multi-planet event in microlensing is OGLE-2006-BLG-109, which has a Jupiter and Saturn mass planet orbiting a 0.5 $M_\odot$ star \citep{Gaudi2008, Bennett_2010}. Modelling these multi-planet events requires triple lens modelling, similar to what is used in circumbinary systems. However, instead of one planet orbiting two host stars, we have two planets orbiting one host star (see Section \ref{sec:triple lens}). Triple lens events are often signified by complex light curves that have many features (spikes and troughs) exhibited. 

For the multi-planet case we show the light curve model of the event OGLE-2006-BLG-109 in Figure \ref{fig:multi-planet}. This system consists of two planets, a Jovian and a Saturn mass planet orbiting a single star. The identification of these two planets by \cite{Gaudi2008} was done by studying the anomalous regions of the light curve. A single Jovian planet orbiting a star is only accounted for by features 1, 2, 3 and 5 in Figure \ref{fig:multi-planet}. The existence of feature 4 can only be explained by a second planet. For more details on the modelling of this event, see \cite{Gaudi2008} and \cite{Bennett_2010}.

The second multi-planet event, OGLE-2012-BLG-0026, was published by \cite{Han_2013}, which revealed two cold gas giants orbiting a solar mass star. Discoveries of such systems are important to theoretical work on planet formation and planet demographics. 

In the case of microlensing, the multi-planet systems discovered so far only consist of two planets. Of course, systems with more than two planets also exist; however, they have not been discovered yet with microlensing.

\begin{figure}[h!]
    \centering
    \includegraphics[width=1\linewidth]{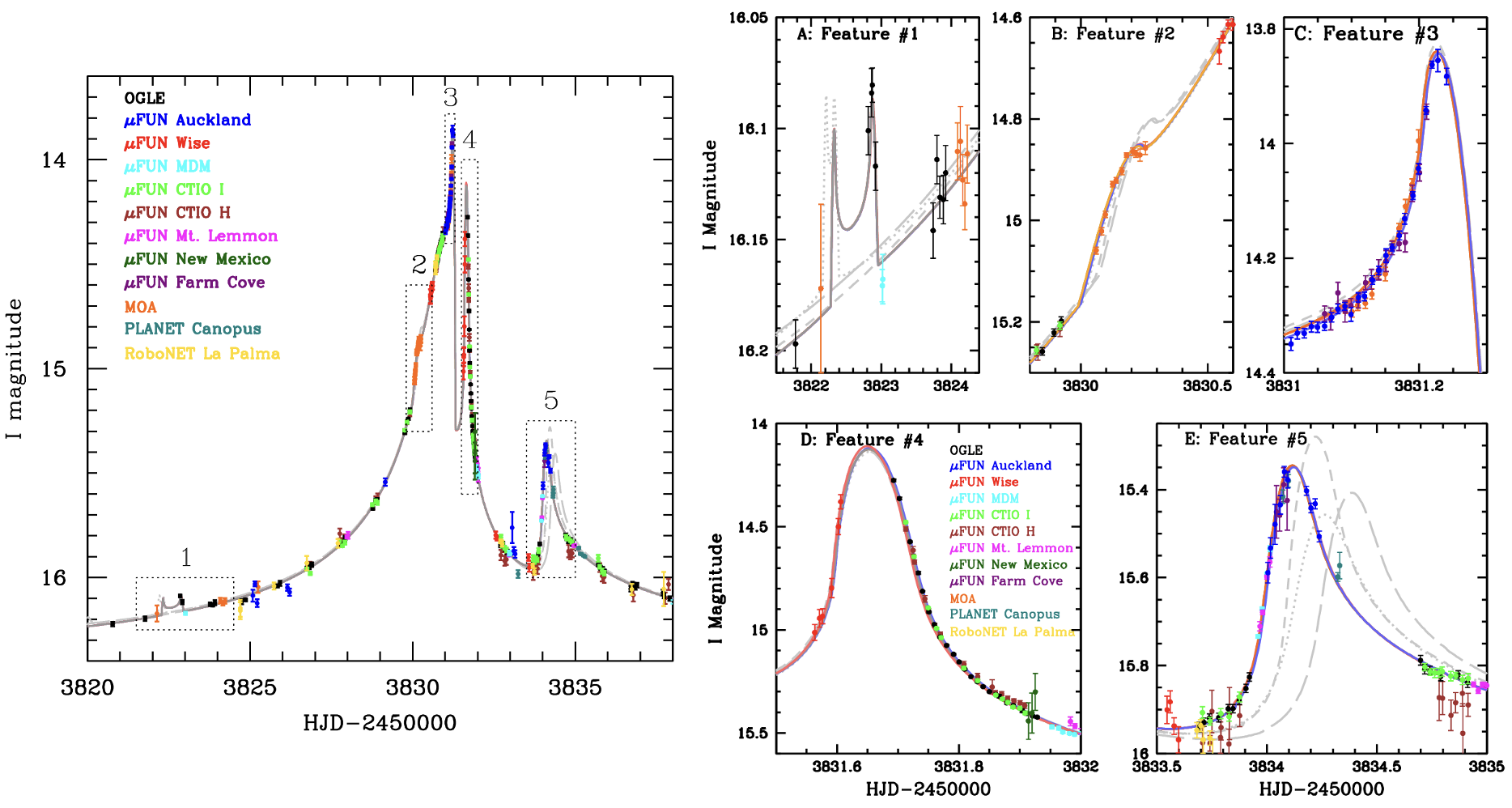}
    \caption{The light curve model for multi-planet event OGLE-2006-BLG-109, and the zoom of each feature. Image from \cite{Gaudi2008} and \cite{Bennett_2010}.}
    \label{fig:multi-planet}
\end{figure}

\section{Example of Data Analysis}
When talking about gravitational microlensing data analysis, we can summarize it in three main procedures: (1) identifying potential events through variations in stellar brightness; (2) modelling the event’s light curve to extract details about the lens; and (3) measuring the lens’ mass through follow-up or precursor observations. The third step is optional and only required when finite source effects are not enough to yield accurate estimates of the physical parameters by themselves. In this section, we provide a summary of these processes.

These three main procedures  are typically performed in a particular order (see Figure \ref{fig:work flow}). First, you must identify the event using data from the various microlensing surveys. The event selection process is described in more detail in the following subsection. Subsequently, the light curve modelling can ensue. In this step, you can determine whether the event is a single-lens or binary-lens, and find the model parameters. The physical parameters can be determined if there are significant higher order effects on the light curve, or estimated using Galactic models with priors. Alternatively, after waiting several years for the source and lens to no longer be aligned, the lens can be imaged directly high angular resolution telescopes. This can also be achieved prior to the microlensing event itself. For example, a Galactic field may be imaged with a high angular resolution telescope in anticipation of ongoing or future microlensing surveys in that region. Once a microlensing event is identified, these high-resolution precursor observations can be used to directly detect the lens star before it becomes closely aligned with the source.

Direct measurements of the lens flux, and relative proper motion to the source, can be used as constraints in the light curve model. Remodelling the light curve, with these constraints can help reveal if the light curve is the best fit model, break degeneracies in the light curve model (if they exist), and get more accurate model parameters (like, $t_E$ or $q$, for example). 

\begin{figure}[h!]
    \centering
    \includegraphics[width=1\linewidth]{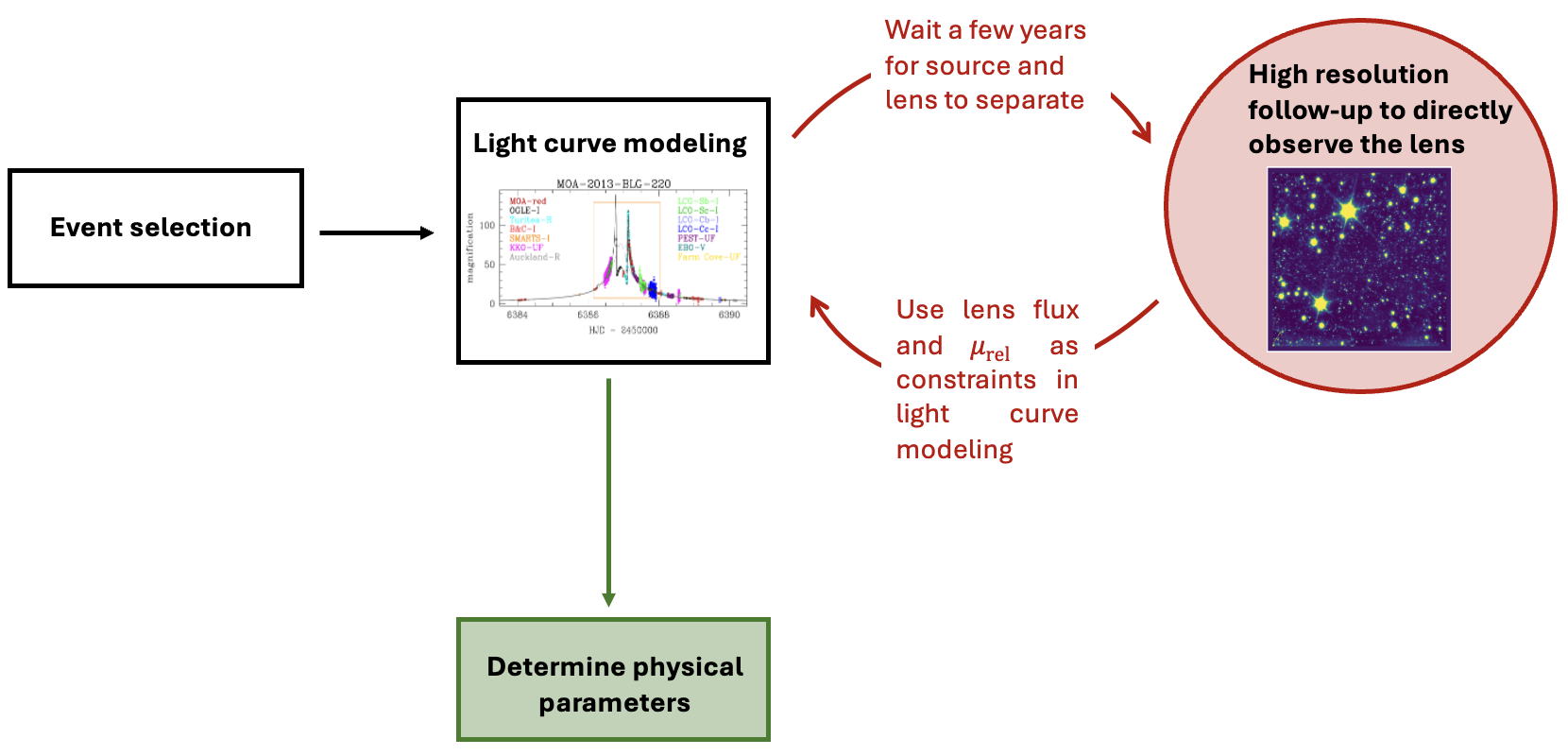}
    \caption{Work-flow order, starting with event selection, light curve modelling/detecting a planetary event, and finally determining the physical parameters of the system.}
    \label{fig:work flow}
\end{figure}

In the following sections, we present data analysis examples for microlensing. We present the light curve modelling of event MOA-2020-BLG-135, and the physical parameters which were determined using the light curve model parameters and a Galactic model with priors \citep{ishitanisilva2022moa}. Furthermore, we describe the process of analysing Keck high resolution data, using MOA-2013-BLG-220 \citep{Vandorou2020}, as an example.

\subsection{Event Selection}

Gravitational microlensing events are quite rare. 
Therefore, to detect a microlensing exoplanet, millions of stars in regions of high stellar density must be monitored simultaneously and continuously over extended periods at a high cadence. Thanks to \gls{MOA}, \gls{OGLE}, \gls{KMTNet}, and now \gls{PRIME}, hundreds of microlensing event candidates and dozens of new microlensing exoplanets are identified each year. Given the large number of stars observed, a selection method is needed to identify potential microlensing occurrences. Although different collaborations may use slightly varied approaches, the most common method involves detecting events with magnification and then fitting a \gls{PSPL} microlensing model to verify the likelihood that the light curve contains a microlensing event. From there, further analyses are conducted to identify possible planetary light curves.

In the case of the \gls{MOA} collaboration, the detection process begins by identifying significant brightness amplifications in stars, which may indicate the gravitational lensing of a background object. Since the Galactic bulge is highly crowded, \gls{MOA} uses real-time difference imaging analysis, where a high-quality reference image is subtracted from subsequent observations to isolate only variable flux \citep{bond2001real}. This technique allows small brightness variations to be detected even when individual stars are not fully resolved. It scans for positive or negative \gls{PSF} profiles, depending on brightness changes, and retains objects with a signal-to-noise ratio greater than five. This step filters out cosmic rays and other artificial effects. Cosmic rays do not exhibit \gls{PSF} shapes and do not recur as consecutive events. Next, light curves are built and checked within a window to confirm a steady baseline flux. Events that display a consistent brightening pattern across at least three consecutive measurements are retained, eliminating other non-microlensing variations. This is necessary to ensure that selected light curves show clear and steady brightening events. Then, a \gls{PSPL} model is fitted to the candidates to verify that each event's behavior is consistent with that of a real microlensing event. The \gls{PSPL} fitting is necessary to eliminate events that can resemble microlensing, such as moving objects, cataclysmic variables, and supernovae. When fitting a \gls{PSPL}, such events usually produce a high $\chi^2$ and / or unphysical values of parameters \citep{sumi2011unbound, sumi2013microlensing}. It is possible that with additional data points or further analysis, the alerted transient candidates may be identified as non-microlensing.

Collaborations like \gls{MOA}, \gls{OGLE}, and \gls{KMTNet} alert to possible microlensing events on their websites. \gls{MOA} alerts are available at: \href{https://moaprime.massey.ac.nz/moaarchive}{https://moaprime.massey.ac.nz/moaarchive}; while the \gls{OGLE} warning system is accessible at: \href{https://ogle.astrouw.edu.pl/ogle4/ews/ews.html}{https://ogle.astrouw.edu.pl/ogle4/ews/ews.html}; and \gls{KMTNet} light curves are available at: \href{https://kmtnet.kasi.re.kr/~ulens/}{https://kmtnet.kasi.re.kr/~ulens/}. These websites provide quick access to light curves with initial \gls{PSPL} fits and results. Table~\ref{tab:telescopes} provides a list of main telescopes used by the gravitational microlensing community.

Due to the large number of light curves to analyze, the quick \gls{PSPL} fit serves as a preliminary step before performing a more detailed binary or triple lens analysis. With the help of the \gls{PSPL} fit, microlensing experts often identify interesting light curves with potential planetary signatures, either by visual inspection or through computational tools, such as the publicly available \texttt{RTModel} \citep{bozza2024rtmodel}. Binary or triple lens models are then applied to these selected candidates. See Section \ref{subsection:modelling} for more modelling information.

Over the years, alternative approaches have been developed to select interesting microlensing, many of them implementing machine learning techniques \citep{ishitanisilva2023searching,wyrzykowski2016black,chu2019automatic,godines2019machine,mroz2020identifying,gezer2022early}. Nevertheless, \gls{PSPL} fitting remains a major component of most event selection methods. For more information on machine learning applications for exoplanets.

\subsection{Light Curve Modelling}\label{subsection:modelling}

The goal of light curve modelling is to determine the physical parameters that best explain the observed data. There are many open-source tools available for light curve modelling; see Table~\ref{Tools} for a list. 

In scientific modelling, it is easy to overcomplicate the model by adding unnecessary parameters. To avoid this, we always begin with the simplest model, which in the case of microlensing, is a single source and a single lens. From there, we gradually add complexity, introducing additional objects such as another lens or a secondary source. 

To illustrate the modelling process, we use the example of MOA-2020-BLG-135Lb, a planetary event whose discovery and modelling analysis were reported in \cite{ishitanisilva2022moa}. The paper is open access, and the data used for the analysis are also available within the paper, along with a link in the first figure of the paper. For this published event, the modelling was performed using the \texttt{eesunhong} software \citep{Bennett1996}, and the outputs were analyzed with the \texttt{MOAna} software \citep{moana}. Although \texttt{eesunhong} is now publicly available on GitHub (\url{https://github.com/golmschenk/eesunhong}), its documentation is still a work in progress and may be more suitable for experienced microlensing researchers. As an alternative, we have prepared a publicly available package that combines \texttt{RTModel} \citep{bozza2024rtmodel} for identifying initial conditions and \texttt{pyLIMA} \citep{Bachelet2017} for the \gls{MCMC} analysis. This package, designed for easier use by newcomers, can be found at \url{https://github.com/stelais/mb20135}. For other open-source tools, please also refer to Table~\ref{tab:open_source_tools}.

For microlensing exoplanets, the event name can provide significant information. The event was named  to indicate that it was first discovered by the \gls{MOA} collaboration using data from the \textbf{MOA} telescope, a purpose-built 1.8 metre wide-field telescope located at Mount John Observatory, New Zealand. Its first alert was issued in July \textbf{2020}. Located at Galactic coordinates $(l,\, b) = (0.15598^\circ, -1.95678^\circ)$, the event was observed as the telescope pointed toward the Galactic bulge (\textbf{BLG}). This was the \textbf{135th} alert issued by the collaboration in that year. Since microlensing events involve at least two objects, the lens and the source, it is important to specify which one is or contains the planet. In this case, the lens system is denoted by \textbf{L}, and the planet, as a secondary lens, is indicated by an additional \textbf{b}. Thus, the microlensing event is named MOA-2020-BLG-135, and the exoplanet discovered is MOA-2020-BLG-135Lb.

The observed data are shown in Figure~\ref{fig:mb20135}. In addition to the MOA-Red filter, which corresponds to a customized wide bandwidth from $600\;\mathrm{nm}$ to $900\;\mathrm{nm}$, the event was also observed in the visual band using the MOA-\textit{V} filter. Furthermore, \gls{CFHT} on Mauna Kea, Hawaii, observed the event in the Sloan Digital Sky Survey $i$-band filter. It is noteworthy that the light curve closely resembles a single-lens model with an additional bump in the interval $\mathrm{HJD'} = [9040.7, 9041.5]$\footnote{$\mathrm{HJD'} = \mathrm{HJD} - 2,450,000$}, making the event likely to trigger alert systems that rely on single-lens fitting.

\begin{figure}
    \centering
    \includegraphics[width=1.0\linewidth]{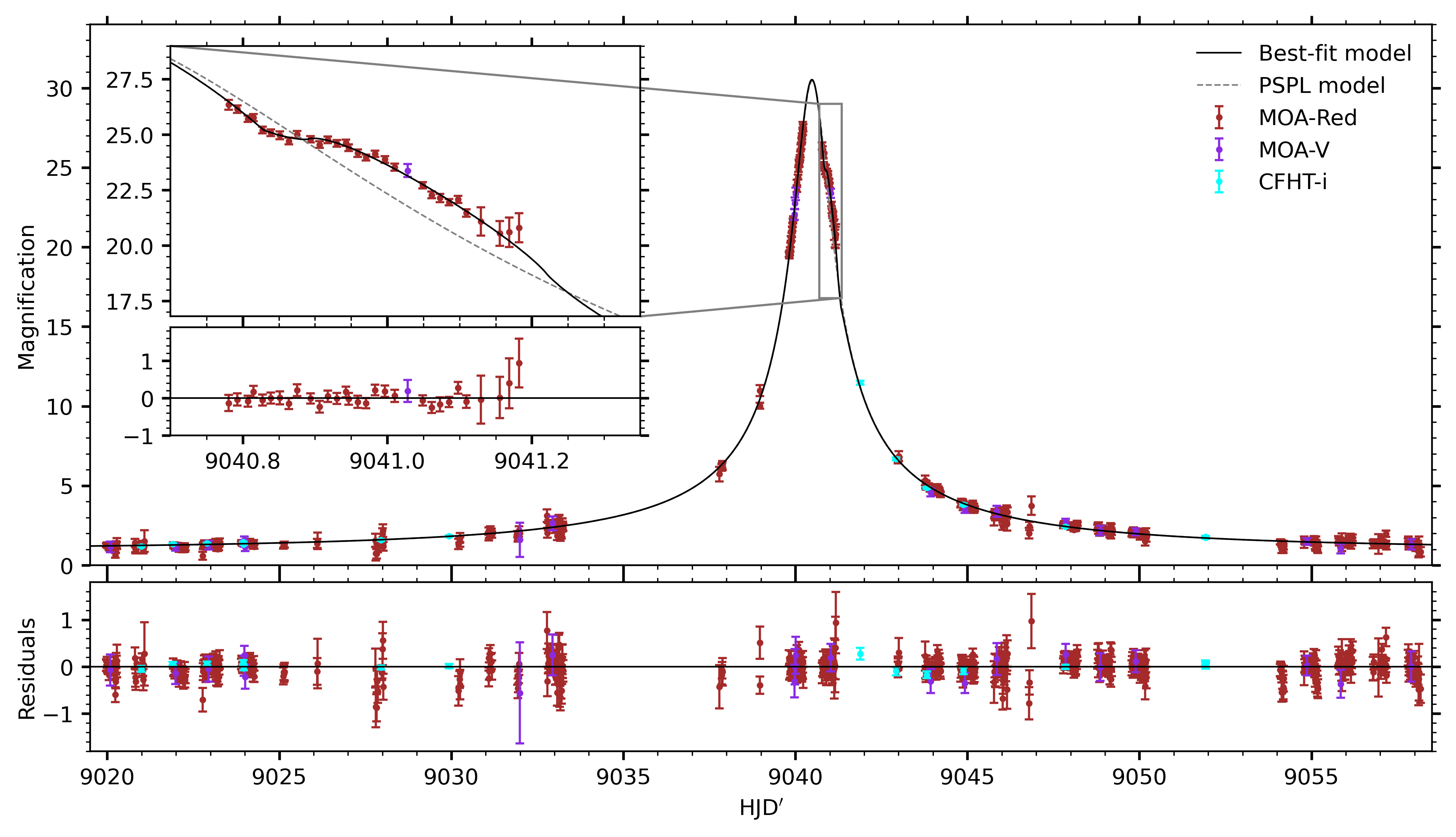}
    \caption{The best-fit model for the MOA-2020-BLG-135 light curve is shown in a magnification versus time ($\mathrm{HJD}' = \mathrm{HJD} - 2,450,000$) plot from \cite{ishitanisilva2022moa}. The data from three instruments (MOA-Red, MOA-\textit{V}, and CFHT-\textit{i}) are displayed, along with a \gls{PSPL} model and the best-fit model, which is a \gls{2L1S}.
}
    \label{fig:mb20135}
\end{figure}

\subsubsection{Single Lens, Single Source}
A \gls{PSPL} fitting is a type of \gls{1L1S}  model defined by three primary parameters: the Einstein crossing time, $t_\mathrm{E}$, the time of minimum separation, $t_0$, and the minimum angular separation, $u_0$. Although collaborations provide preliminary \gls{PSPL} models, it is important to start by rerunning them, to ensure that all potential local minima are found.

The most common approach to fitting the observed light curve is to perform a grid search over the parameters $t_\mathrm{E}$, $t_0$, and $u_0$ to identify the optimal model parameters. Because \gls{PSPL} models are computationally fast and have relatively few parameters, grid search remains a practical and effective method to optimize this model. For events similar to MOA-2020-BLG-135, \gls{PSPL} fitting results offer great initial conditions for more complex models, such as binary lens or binary source fittings.

Due to the finite size of the source and the limb-darkening effect, an additional parameter, the normalized angular radius $\rho_*$ (alternatively $t_*$, as $t_* = \rho_* t_\mathrm{E}$) — can be introduced. This model is referred to as a \gls{FSPL} model.  Both \texttt{eesunhong}, the code used in the published analysis, and \texttt{RTModel}, the software included in the online material, perform a grid search to fit \gls{FSPL} models to the light curve.

In addition to the fit parameters already cited, we include the blend flux, $f_{b,i}$, and the un-lensed source flux, $f_{s,i}$, for each instrument's specific passband $i$. We have to take into account the blend flux in our model because, in crowded stellar fields, the source star is often blended with light from nearby unlensed stars. For each instrument band $i$, we determine the total flux that minimizes the $\chi^2$. The total flux $F_i(t)$ at time $t$ is given by Equation~\ref{eq:flux} ($F_i(t) = A(t,\textbf{x}) \, f_{s,i} + f_{b,i}$), where $A(t,\textbf{x})$ is the event magnification at time $t$ for a given set of non-linear parameters $\textbf{x} = (t_\mathrm{E}, t_0, u_0, [\rho_* \; \text{or} \; t_*])$.

In Figure~\ref{fig:mb20135}, the \gls{PSPL} model is shown with dashed lines. Although it fits most of the light curve well, the zoomed-in inset highlights a region where it fails to capture an anomaly.
It is important to note that outside the anomaly interval ($\mathrm{HJD'} = [9040.7, 9041.5]$), the \gls{PSPL} and binary-lens light curves are almost the same and therefore this deviation indicates that the lens might be composed of two masses, with the less massive component potentially being a planet-mass object. The next step is to investigate the cause of this anomaly, initially by attempting to add a second lens to the model.

\subsubsection{Binary Lens, Single Source}
A binary-lens single-source model, also referred to as \gls{2L1S}, represents a lens system composed of two masses. The simplest form of the \gls{2L1S} model introduces three additional parameters: the ratio of the secondary lens mass to the primary lens mass, $q$ (or alternatively, the mass fraction $\epsilon$), the separation between the two lens components, $s$, and the counter-clockwise angle between the lens-source relative motion and the binary lens axis, $\alpha$. 

When fitting a \gls{2L1S} model to the observed light curve, different researchers use varying approaches as the first step.  A common first step is to perform a grid search again, but now with $t_\mathrm{E}$, $t_0$, $u_0$, and $\rho_*$ (or $t_*$) fixed based on the \gls{1L1S} best-fit model, while scanning over $s$, $\alpha$, and $q$. This process helps to find initial conditions for the local minima. To ensure that no optimal model is overlooked, different strategies are employed. For instance, \cite{ishitanisilva2022moa} ensures that the mass ratios within the range $q \in [10^{-5}, 10^{-1}]$ are included in the set of initial conditions. In contrast, software such as \texttt{RTModel} \citep{bozza2024rtmodel} takes a different approach to the grid search for this step. Instead, first, it automatically identifies a secondary peak in the light curve. If the anomaly is small, \texttt{RTModel} calculates initial values for the separation $s$ and mass ratio $q$ based on the angular distance of the anomaly from the primary lens. If the anomaly is large, it likely indicates a binary star lens system, where the mass ratio $q$ is large. In such cases, \texttt{RTModel} matches pre-defined templates with the characteristic double-peaked structure. Once an initial guess is selected, \texttt{RTModel} performs a downhill minimization using the Levenberg–Marquardt algorithm \citep{marquardt1963algorithm} to locate a local minimum of the $\chi^2$ function \citep{bozza2024rtmodel}.

Once the initial conditions are identified, a \gls{MCMC} algorithm is used to explore the full parameter space and perform a complete fit, with all parameters allowed to vary freely. Both \texttt{eesunhong}, which was used for the published event analysis, and \texttt{pyLIMA}, available in the online material, implement \gls{MCMC} methods with a variety of samplers. Then, the best-fit model is selected by looking for the lowest $\chi^2$, which in that case indicated a model with $q \sim 10^{-4}$. Figure~\ref{fig:mcmc_mb20135} shows the full parameter space exploration for the binary-lens single-source model and Figure~\ref{fig:mb20135} displays a \gls{2L1S} model as the best-fit solution found. For more information on the published parameters, see \cite{ishitanisilva2022moa}.

It is worth noticing the degeneracy in $s$, which appears in Figure~\ref{fig:mcmc_mb20135}. Degeneracies that follow the $s_A\leftrightarrow 1/s_B$ relationship, where A and B represent different solutions, predicted by \citet{dominik1999binary} are relatively common and are often referred to as the close-wide degeneracy. Recently, \citet{zhang2022ubiquitous} demonstrated that the close-wide degeneracy is strictly held only in the specific case of $u_0=0$. For the more general case of $u_0>0$, an alternative name was proposed calling it the ``offset degeneracy'', which more exactly follows:
\begin{equation}
s_A = \dfrac{1}{2}\left(2 (u_0/\sin(\alpha)) - (s_B - 1/s_B)+\sqrt{{\left[2 (u_0/\sin(\alpha)) - (s_B - 1/s_B)\right]}^2+4}\right) .    
\end{equation}
Since degenerate solutions are common in light curve modelling, it is always good to explore those scenarios, identifying all possible solutions. For the analysis of MOA-2020-BLG-135Lb, the \gls{MCMC} was able to jump between the degenerate solutions. However, in some cases it may be necessary to explore degenerate solutions manually by setting the equivalent degenerate case as an alternative initial condition. Some modelling platforms, such as \texttt{RTModel}, account for the offset degeneracy within the code \citep{bozza2024rtmodel}.

High-order effects, such as microlensing, annual parallax, and orbital motion discussed in Section~\ref{subsec:2nd_order_effect}, could be included in the model if necessary, but they were not clearly detected for this event. \cite{ishitanisilva2022moa} claimed the event was short ($t_\mathrm{E} \approx 17 \; \mathrm{days}$) and faint, and the timing of its peak magnification coincided with Earth's acceleration toward the Sun's projected position into the lens plane being near its minimum. Furthermore, the \gls{2L1S} model with 7 parameters fitted the data well.

\begin{figure}
    \centering
    \includegraphics[width=1.0\linewidth]{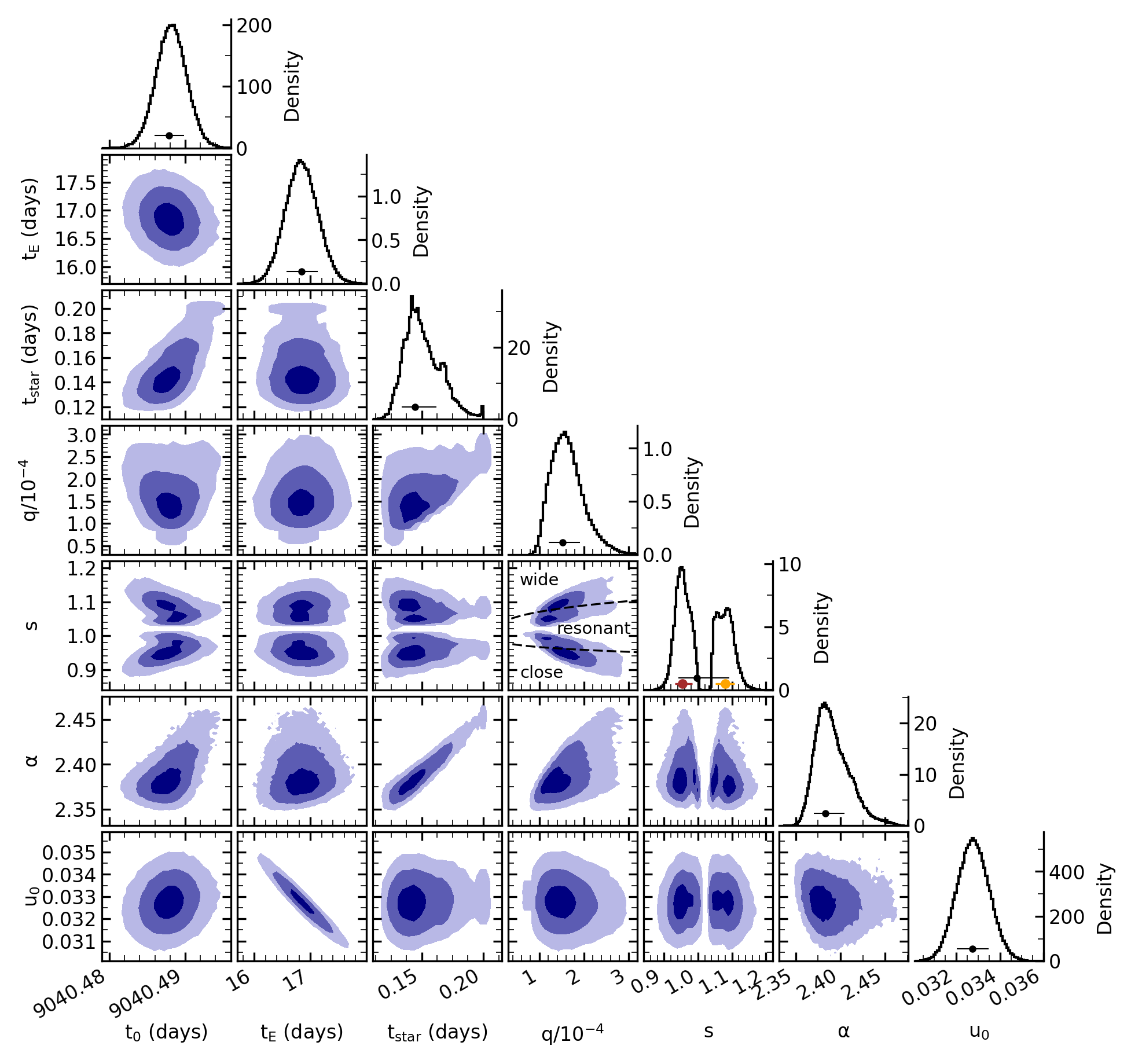}
    \caption{The marginalized posterior distributions from \cite{ishitanisilva2022moa}'s \gls{MCMC} runs, correlating the seven parameters for the binary-lens single-source model (\gls{2L1S}), $t_0$, $t_\mathrm{E}$, $t_*$, $q$, $s$, $\alpha$, $u_0$, and, in the diagonal, the one-dimensional probability density function of each parameter. The 68.3\%, 95.5\%, and 99.7\% confidence intervals are shown in dark blue, medium blue, and light blue, respectively, in the posterior distribution plots. In the \gls{PDF} plots, the black dot points out the median, and the thin line marks the 68.3\% confidence interval. The degeneracy in $s$ is recognizable by the presence of two 99.7\% confidence intervals regions in the plot.}
    \label{fig:mcmc_mb20135}
\end{figure}

\subsubsection{Binary Source Possibility}

The possibility of a binary source is also commonly explored, as it can produce features resembling planetary signals. This scenario can be investigated using different lens systems, such as a single lens and binary lens referred to as \gls{1L2S}  and \gls{2L2S}, respectively.

Because there are now two sources, we must take into consideration the impact parameter of the primary source $ u_{01} $ and the impact parameter of the secondary source $ u_{02} $, as well as the time of the minimum separation for the primary source $ t_{01} $ and for the secondary source $ t_{02} $. The new model still includes $t_\mathrm{E}$ and now adds the flux ratio $ F_\mathrm{R} $ of the secondary source to the primary source for each of the observed passbands. The fitting procedure is similar to the other models, usually starting with a grid search and followed by an \gls{MCMC}. 

It is important to note that adding more parameters to a model can often reduce the $\chi^2$ value simply due to increased model flexibility. As a result, a lower $\chi^2$ does not necessarily indicate a better or more appropriate model when comparing models with different numbers of parameters. To properly evaluate model quality, additional model selection criteria are necessary. In addition to the $\chi^2$ comparison, \citep{ishitanisilva2022moa} used the Bayesian Information Criterion \citep{schwarz1978estimating}, which penalizes models with more parameters. Alternatively, the software \texttt{RTModel} applies Wilks' theorem \citep{wilks1938large} to evaluate the statistical significance of the $\chi^2$ improvement, and then determine the minimum threshold differences for model selection.

\subsubsection{Lens System Parameters Post-light Curve modelling}
Light curve models of planetary events provide the mass ratio $q$, but they do not directly provide the planet's mass or its distance from Earth. To solve that, one approach is to determine the host star's mass by observing the system when the separation between the source and lens is large enough to make such observations possible. This method is discussed in Section~\ref{subsection:mass_measurement}. However, during the detection of the event, the lens and source are too close to allow for such measurements. Consequently, alternative methods must be employed to estimate the properties of the lens system. Galactic models, such as those presented by \citet{bennett2014moa} and \citet{Koshimoto_2021Galactic}, offer a valuable tool for estimating key lens system parameters, including its distance and mass.

The work for the presented example reported the results using \citet{bennett2014moa}'s Galactic model with two different priors. Assuming that all stars have equal probability of hosting a planet, \cite{ishitanisilva2022moa} found a planet mass of $m_\mathrm{planet} = 11.3_{-6.9}^{+19.2} \;M_\oplus$ and a host-star mass of $M_\mathrm{host} = 0.23_{-0.14}^{+0.39} \;M_\odot$, located at a distance $D_\mathrm{L} = 7.9_{-1.0}^{+1.0} \;\mathrm{kpc}$. Under the assumption that the planet-hosting probability scales proportionally with stellar mass, they instead estimated $m_\mathrm{planet} = 25_{-15}^{+22} \;M_\oplus$, $M_\mathrm{host} = 0.53_{-0.32}^{+0.42} \;M_\odot$, and $D_\mathrm{L} = 8.3_{-1.0}^{+0.9} \;\mathrm{kpc}$. Although the equal-probability prior is often adopted as a simple reference assumption, recent microlensing studies suggest that higher-mass stars may be more likely to host the types of planets to which microlensing is most sensitive \citep{bhattacharya2021moa}, somewhat consistent with earlier findings from radial velocity \citep{johnson2010giant} and direct imaging \citep{nielsen2019gemini} surveys. For more details, refer to \cite{ishitanisilva2022moa}. The difference between these estimates illustrates how the adopted Galactic prior can influence the inferred physical properties of the lens system when the lens mass and distance are not directly measured. 

It's worth noting that in most microlensing studies, the Galactic prior is applied after the light curve modelling stage rather than incorporated directly into the parameter-space exploration. While including the Galactic model self-consistently during the fit would be more rigorous mathematically, this is computationally challenging because the prior depends on parameters such as $t_\mathrm{E}$, which vary throughout the exploration of parameter space. As a result, Galactic Bayesian analyses are typically performed as a post-processing step following the determination of the best-fit microlensing model. More sophisticated analyses incorporating additional constraints, such as high-resolution imaging, can instead perform a more self-consistent joint modelling of the light curve and lens properties.

\subsection{Lens System Mass Measurement}\label{subsection:mass_measurement}

Once a microlensing event has been modelled, the next step is to try and determine the lens system's physical parameters. These include host star mass, distance, planet mass, and projected separation. In Section \ref{sec:lens system params} we go over all the possible ways in which a mass-distance relation can be obtained. Many of these include the presence of higher order effects on the light curve - such as, parallax, orbital motion, xallarap or binary source effects. Additionally, one can obtain mass-distance relations from directly observing the lens flux and measuring its relative proper motion to the source, once they are no longer aligned. In this section, we outline the steps required to make direct lens flux and relative proper motion measurements with high-angular resolution data.

We take the example of MOA-2013-BLG-220, a planetary event initially modelled and published in \cite{Yee2014}. High-angular resolution observations were taken with Keck 6 years after the peak of the microlensing event, resolving the source and lens. The data and analysis are presented in \cite{Vandorou2020}. The following example uses publicly available Keck data from \href{https://koa.ipac.caltech.edu/cgi-bin/KOA/nph-KOAlogin}{KOA} - the Keck Observatory Archive. Example python notebooks using the MOA-2013-BLG-220 data can be found at \url{https://github.com/KatieVandorou/MB13220-Example-Analysis}.

\subsubsection{Keck Data}

Follow-up observations for microlensing events are typically taken with Keck I's \gls{OSIRIS} instrument, or with Keck II's \gls{NIRC2} instrument, in the $K$, $H$, or $J$ passbands. MOA-2013-BLG-220 was first observed in 2015 with the \gls{NIRC2} laser guide adaptive optics system, using the wide camera in the $K$ passband ($\lambda_{\rm c} \sim 2.1 \mu \rm m$). Adaptive optics capabilities are crucial for ground-based observations, as it accounts for the distortion effects due to the Earth's atmosphere. In 2019, additional data was taken with the narrow \gls{NIRC2}'s narrow camera in the $K$ passband. 

\paragraph{Wide camera data:} Data taken with \gls{NIRC2}'s wide camera have a plate scale of 39.69 milliarcsecond (mas) per pixel (one frame is $4000\times 4000$ pixels). This plate scale is typically too large to resolve the small separations between source and lens, however, it is almost necessary to obtain calibrated apparent magnitudes of the stars in the frame. The wide frame can capture more stars, which can then be matched to known catalogues such as the \gls{VVV}. Keck data taken with the \gls{OSIRIS} instrument can also be used for stellar calibrations; however, the frame is not as wide relative to \gls{NIRC2}'s wide camera. \gls{OSIRIS} has a plate scale of 9.94 mas per pixel (one frame is $2000\times2000$ pixels).

\paragraph{Narrow camera data:} Data taken with \gls{NIRC2}'s narrow camera (plate scale = 9.94 mas per pixel; one frame is $1000\times1000$ pixels) or with \gls{OSIRIS} to resolve the lens and source. The relative source lens proper motion and lens flux measurements are conducted using this data. In Figure \ref{fig:wide_vs_narrow} we show a comparison of a \gls{NIRC2} wide versus narrow frame for the MOA-2013-BLG-220 field.

\begin{figure}[h!]
    \centering
    \includegraphics[width=0.8\linewidth]{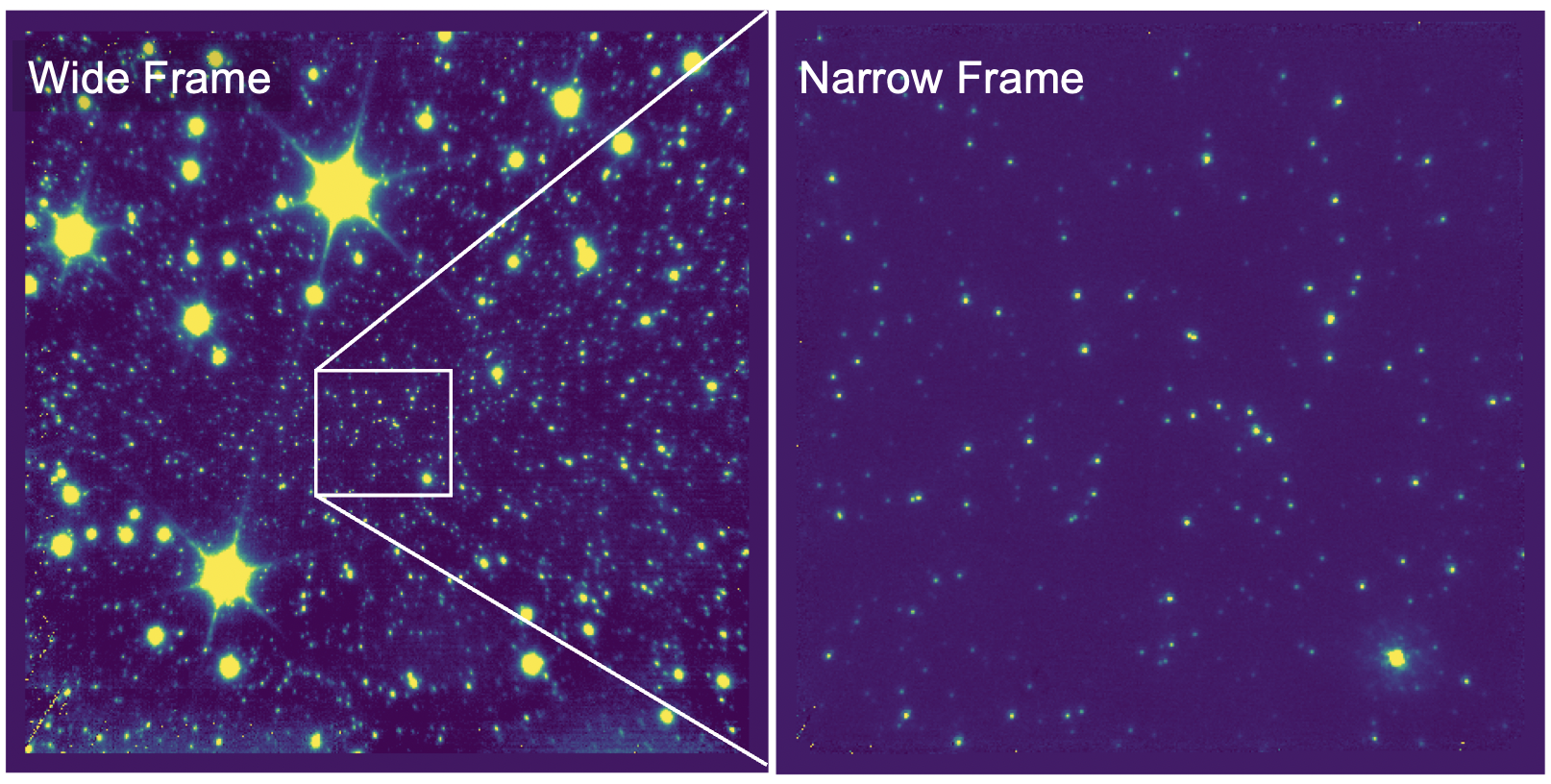}
    \caption{The Keck \gls{NIRC2} wide frame compared to the \gls{NIRC2} narrow frame for the MOA-2013-BLG-220 field.}
    \label{fig:wide_vs_narrow}
\end{figure}

\paragraph{}Once observations have taken place, the raw images need to be corrected for dark current and flat-field. They also need to be sky corrected before co-added into a final stack. At the time of observation, dark frames, flat-field images, and sky images are taken. The standard procedure for this is to use the \gls{KAI} reduction pipeline \citep{Lu2022} which is an open source software which can be found on GitHub, (see Table \ref{tab:open_source_tools}). Correcting for instrumental aberrations and geometric distortions improves the image quality, enabling for a much more accurate analysis. Each raw image is corrected in this way and then co-added, or stacked, into one final master frame. An example of a raw frame compared to a master frame is shown in Figure \ref{fig:before_vs_after}. Alternative tools that can be used are the SWarp software (which stacks images) \citep{Bertin2010} and the SExtractor software \citep{Bertin1996} which can be used in the astrometry and photometry process. \citep[e.g.][]{Batista_2014, Beaulieu2016}

\begin{figure}[h!]
    \centering
    \includegraphics[width=0.8\linewidth]{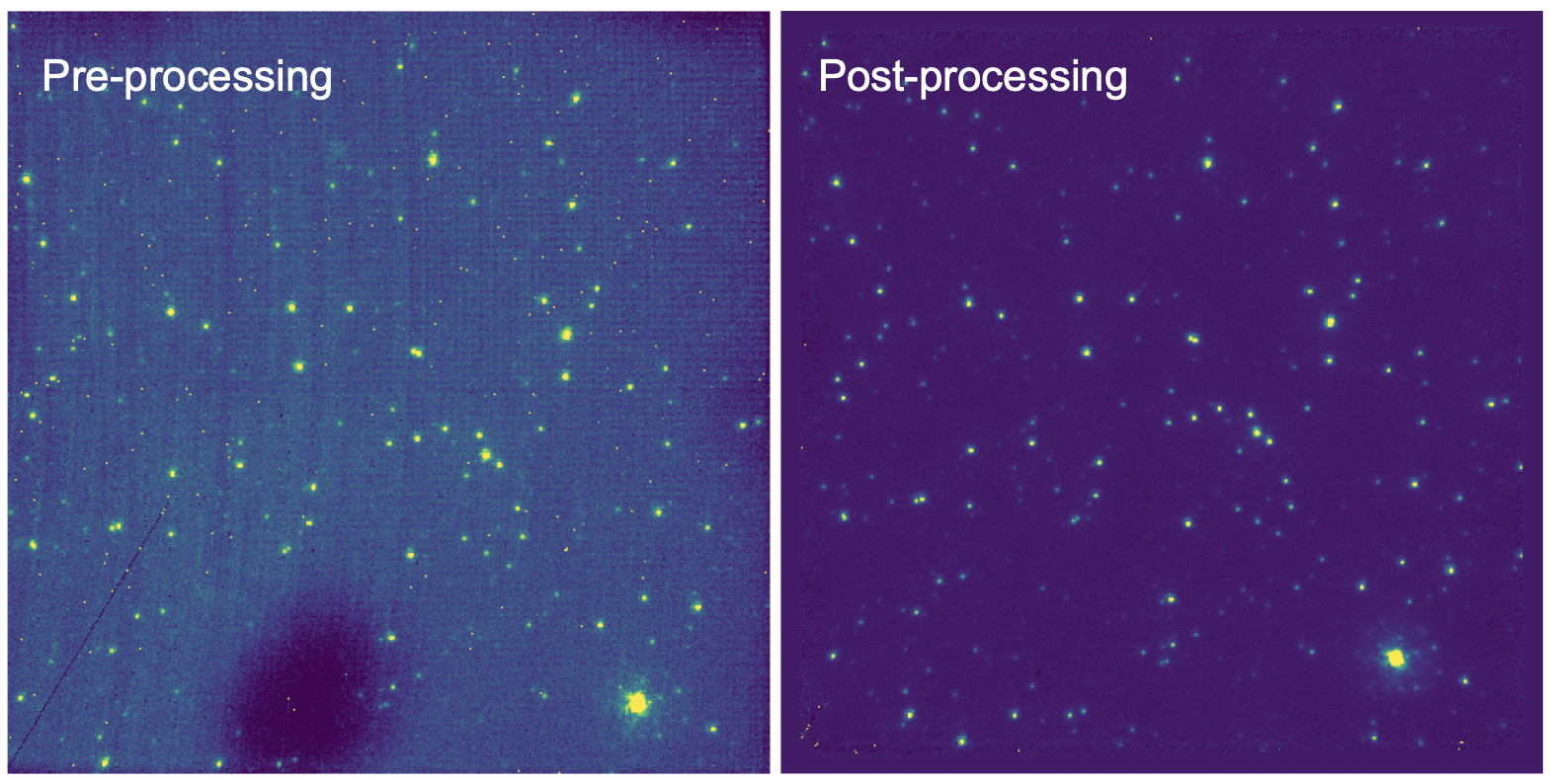}
    \caption{Comparison of a Keck Narrow frame of MOA-2013-BLG-220 before and after processing. The left frame is a `raw' Keck image, while right one has been dark current corrected, flat-fielded, sky corrected, and co-added.}
    \label{fig:before_vs_after}
\end{figure}

\subsubsection{Lens Detection}

Once the images have been reduced using the procedure described above, we are ready to detect the lens. This involved several steps and can either be a `simple' or more complicated process depending on the magnitude of separation between the source and lens, quality of the high resolution images, and the level of flux contribution from unrelated nearby stars.

\paragraph{Target Identification:}  The first step is to identify the target (source and lens blend) in the Keck images. This is done by using the R.A. and Dec. coordinates, as well as matching the Keck images to the original \gls{MOA}, \gls{OGLE}or \gls{KMTNet} data using a process called `difference imaging' \citep[e.g.][]{Bond2001}. When the target has been identified, it can first be inspected by eye to see whether the source and lens are resolved. In the case of MOA-2013-BLG-220, there is a clear and visible separation between the two stars, as shown in Figure \ref{fig:lens-detection}. 

\begin{figure}[h!]
    \centering
    \includegraphics[width=0.8\linewidth]{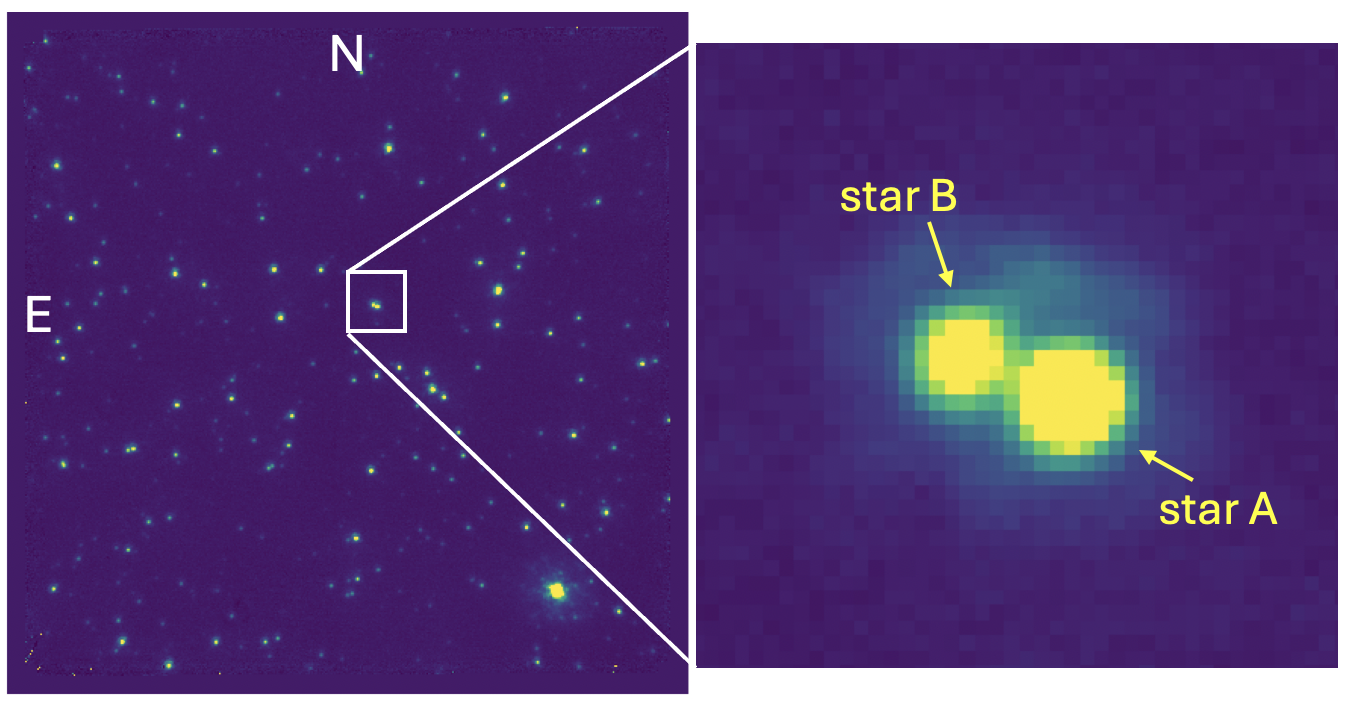}
    \caption{A zoom of the source and lens candidates for MOA-2013-BLG-220. We label them `star A' and `star B' for now.}
    \label{fig:lens-detection}
\end{figure}

We label the two stars of interest as `star A' and `star B'. Discovery papers report a source magnitude and colour in the $I$ and $V$ passbands - and in some cases a predicted $K$ band magnitude. However, if they do not, a $K$ band magnitude for the source can be estimated from the $(V-I), I$ data. Using MOA-2013-BLG-220 as an example, \cite{Yee2014} report: 

\begin{gather*}
    I_{\rm S,0} = 17.975 \pm 0.100\ \rm mag. \\
    (V-I)_{\rm S,0} = 0.585 \pm 0.050\ \rm mag.
\end{gather*}
Here, the zero notation represents a dereddened magnitude and colour. Using the colour-colour relations from \cite{Bessell1988}, we can find the corresponding $V-K = 1.53$ colour. Using this, we calculate a dereddened source magnitude of $K_{\rm S,0} = 17.03 \pm 0.11$ mag. However, to compare it to the Keck data we need a magnitude not corrected for extinction. Therefore, we have to add extinction. This can be done in several ways; the method we present here is using the \gls{OGLE} extinction calculator (\href{https://ogle.astrouw.edu.pl/cgi-ogle/getext.py}{https://ogle.astrouw.edu.pl/cgi-ogle/getext.py}). With the MOA-2013-BLG-220 R.A. and Dec. coordinates the extinction calculator returns: 

\begin{gather*}
    A_{\rm I} = 1.178 \\
    E(V-I) = 0.988 \pm 0.131 \\
    R_{\rm JKVI} = 0.3248
\end{gather*}
Where $A_{\rm I}$ is the extinction in $I$ band, $E(V-I)$ is the colour excess, and $ R_{\rm JKVI}$ is the reddening-free colour index. 
To calculate the $K$ band extinction, $A_{\rm K}$, we proceed with the following steps: 

\begin{gather*}
    E(J-K) = R_{\rm JKVI} E(V-I) \\
    A_{\rm K} = 0.53 E(J-K)
\end{gather*}
Therefore, we find $A_{\rm K} = 0.17 \pm 0.02$ and $K_{\rm S} = 17.20 \pm 0.11$ mag. It is important to note that this is just an estimate from light curve data and colour-colour relations. The relation presented for extinction $A_{\rm K}$ follows from measurements of near-infrared extinction ratios. For Galactic bulge studies, the coefficient of 0.53 comes from \citet{Nishiyama2008, Nishiyama2009}, derived from red clump stars towards the Galaxy. The next step is to identify which star, `A' or `B', is the source and lens, and to accomplish this, we need to measure their respected magnitude in the Keck frames.

\paragraph{Photometry and Astrometry:} This is process of obtaining accurate star positions (in X and Y pixel coordinates) and stellar magnitudes. The standard procedure is to use a software called \texttt{DAOPHOT-II} \citep{Stetson1987} and DAOPHOT-MCMC (a modified version of \texttt{DAOPHOT-II} by \cite{Terry_2021}). For code access and installation instructions, see \href{https://github.com/skterry/daophot_mcmc}{https://github.com/skterry/daophot\_mcmc}. In addition, one should read the \texttt{DAOPHOT-II} manual (available online) before continuing. An alternative method is presented in \cite{Vandorou2020}, however, this code is not available to the public.

The first steps involve running the commands \texttt{ATTACH, FIND, PHOT, PICK, PSF} and \texttt{GROUP}. Once the main frame image is attached, a list of stars is created. The \texttt{PICK} command allows you to select which stars you want to use to build an empirical \gls{PSF} model. This model acts like a blueprint for what a single star should `look like'. To build the model, you should pick bright (but not saturated) isolated stars close to the target star (see Figure \ref{fig:PSF and residuals}). Running \texttt{PSF} will go through these stars, checking for bad pixels, and output a \gls{PSF} model. The step of selecting stars and running the \texttt{PSF} module can be done several times to get the best possible \gls{PSF} model. Once you are satisfied with your \gls{PSF} model, run the \texttt{GROUP} command.  A good PSF model will appear featureless in a residual image (see Figures \ref{fig:PSF and residuals} and \ref{fig:good vs bad psf star}). The output file is edited so that it only has star A and star B. This output file contains the X and Y pixel positions, and instrumental magnitudes of each star. This is essentially all the information we need.

\paragraph{What if I can't see the lens?} In many cases \citep[e.g.][]{Terry_2021, Terry2022, Bhattacharya2021}, the source and lens are not as visibly separated as in MOA-2013-BLG-220. This requires some additional steps after running the six commands mentioned above. Once a single-star \gls{PSF} model has been built, the \texttt{DAOPHOT-II} command \texttt{ALLSTAR} can be run. This program refines the stellar photometry by improving the fit of point spread functions to the stars in the field, and outputs this information into catalogue. In addition, it creates a residual image by subtracting the \gls{PSF} model from the original image. In the residual images, all single stars should have a flat featureless residual. If this is not the case and a star is a blend of two (or even three), a signal will be seen in the residual image. In Figure \ref{fig:PSF and residuals}, we show the residual image of MOA-2013-BLG-220. In Figure \ref{fig:good vs bad psf star} we show a close-up of two \gls{PSF} star examples. \gls{PSF} star 1 shows a featureless residual, which means that it is a single star and that we should keep it when building our empirical \gls{PSF} model. On the other hand, \gls{PSF} star 2 is not featureless, and we can see that it is likely composed of two stars. We do not want to keep this star to build our \gls{PSF} model.

When working with a microlensing event where the source and lens are not completely resolved, and they may still look like `one-star' or one elongated star, studying the residual image will reveal the fainter star hidden behind the brighter one. The pixel positions of this residual signal can be added to the output file created by the \texttt{FIND} module. \texttt{PHOT, PSF} and \texttt{GROUP} are run again, in addition to \texttt{ALLSTAR}. The reason for rerunning \texttt{ALLSTAR} is to check whether this residual signal is actually a star. If it matches the \gls{PSF} model, then the second residual image produced will be featureless. 

\begin{figure}[h!]
    \centering
    \includegraphics[width=0.8\linewidth]{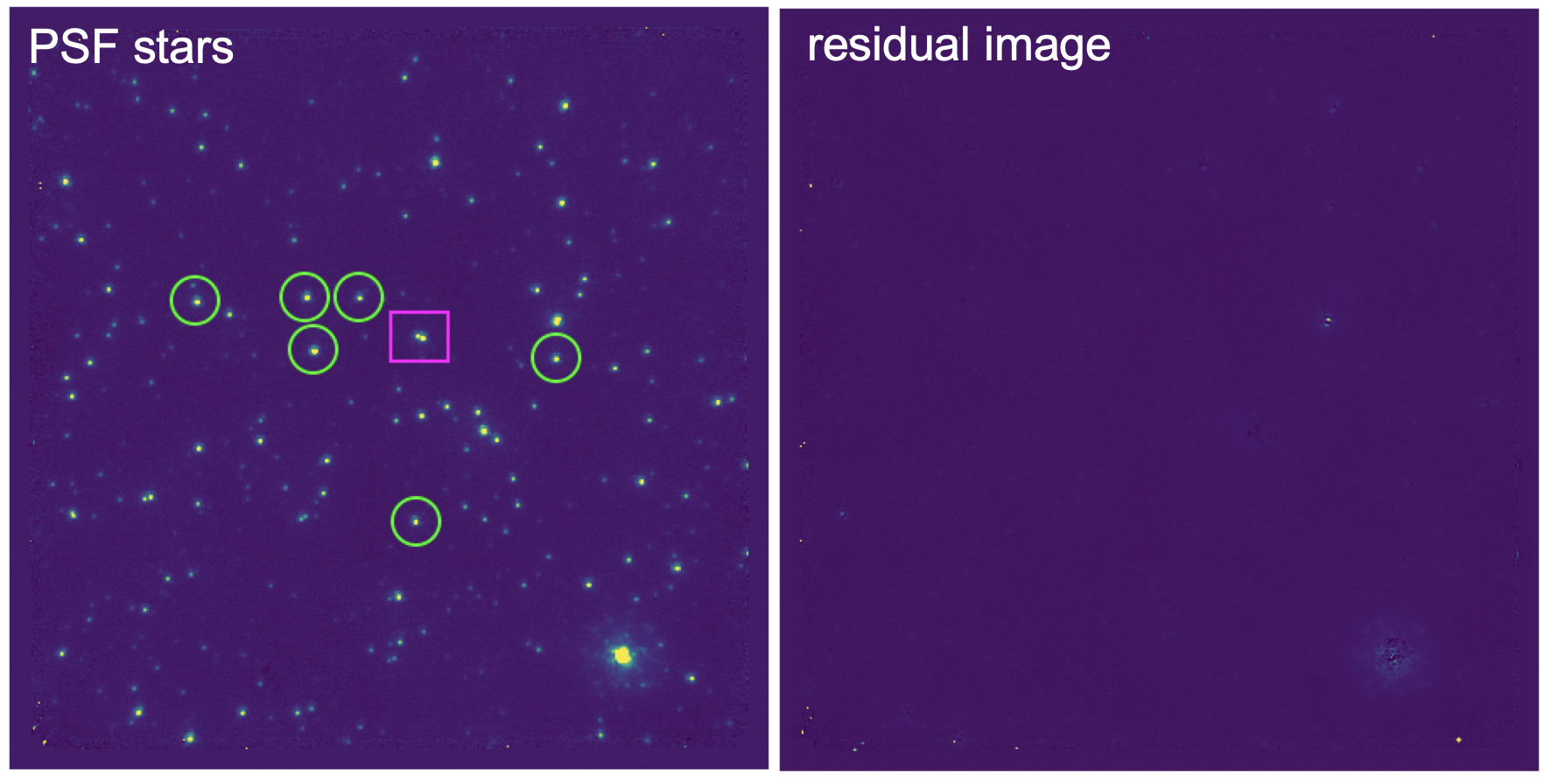}
    \caption{\textit{Left:} An example of \gls{PSF} star selection (green) and the target (magenta). \textit{Right:} The residual image produced for MOA-2013-BLG-220.}
    \label{fig:PSF and residuals}
\end{figure}

\begin{figure}[h!]
    \centering
    \includegraphics[width=0.6\linewidth]{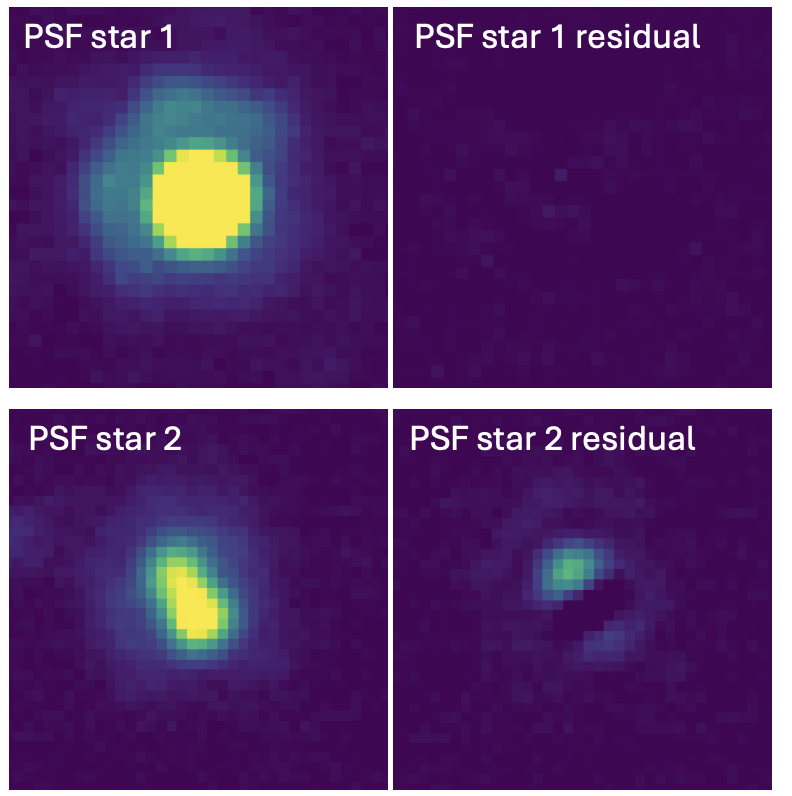}
    \caption{Example of a `good' and `bad' \gls{PSF} star. Good stars, like \gls{PSF} star 1, produce featureless residuals. Bad stars, like \gls{PSF} star 2, will have features in the residual image which could indicate that it is two stars.}
    \label{fig:good vs bad psf star}
\end{figure}

\paragraph{Errors with \gls{MCMC} and Jackknife:} The stellar positions and instrumental magnitudes alone are not enough information for a strong and robust conclusion to be drawn. We need to know the uncertainties attached to these parameters. By using \texttt{DAOPHOT-MCMC} a \gls{MCMC} routine can be run on the targets (outlined in the \texttt{GROUP} output file) to find the best fit location for each star, and their flux ratio and the associated standard deviation error (see \cite{Terry_2021} for more details). 

Although the \gls{MCMC} routine is a powerful way to determine the best-fit parameters and uncertainties, another source of uncertainty must be considered, particularly when analysing high resolution data from ground-based observatories. These uncertainties involve how the stellar \gls{PSF} across each individual frame differs due to atmospheric distortions. To account for this, the Jackknife method can be adopted \citep{Quenouille1949, Quenouille1956, Tukey1958}. For further implementation of this method, see \cite{Bhattacharya2021}.

\subsubsection{Lens Flux and $\mu_{\rm rel}$ measurements}

Now that we have precise positions from running the \texttt{DAOPHOT-MCMC} routine, we can calculate the separation between star A and star B. In pixel coordinates, we find:

\begin{gather*}
    X_{\rm A} - X_{\rm B} = 7.354 \pm 0.043\ \rm pixels\\
    Y_{\rm A} - Y_{\rm B} = 2.962 \pm 0.044\ \rm pixels,\\
\end{gather*}
where a positive X value is in the east direction and a positive Y value is in the north direction. To convert these values to physical units, we multiply them by the plate scale of the instrument. In this case, $9.942$ mas/pixel. To find the relative proper motion, we need to know the time that has elapsed between the microlensing event and the Keck observation. The event took place on $1^{\rm st}$ April 2013, while the narrow camera Keck observation took place on the $27^{\rm th}$ May 2019. Therefore, the time elapsed is approximately 6.156 years. Dividing the separation by the time elapsed, we find the relative proper motion between the two stars (see Table \ref{220table}). Therefore, we find that the magnitude of the relative proper motion is $\mu_{\rm rel,H}=12.804 \pm 0.0717$ mas/yr. We can convert this to the geocentric proper motion using equation \ref{mu_G-mu_H}. We find $\mu_{\rm rel,G}=12.62 \pm 0.11$ mas/yr, which is consistent with the value predicted in \cite{Yee2014}.

From the Keck instrumental magnitudes of the two stars, we can measure the flux ratio of star B/star A. To get the real apparent magnitudes of the two stars from the flux ratio, we can use the following equations, solving them simultaneously:

  \begin{equation}
 	K_{\rm Lens} - K_{\rm Source} = -2.5\log_{10}(f_{\rm L}/f_{\rm S}),
 	\label{sim1}
 \end{equation}
 
 \begin{equation}
 	\label{sim2}
 	K_{\rm S+L}= -2.5\log_{10}(10^{-0.4K_{\rm Lens}}+10^{-0.4K_{\rm Source}}).
 \end{equation}

$K_{S+L}$ is the total flux of the source and lens, which we obtain from the calibrated wide images, where the source and lens cannot be resolved. For MOA-2013-BLG-220 we find $K_{S+L} = 16.75 \pm 0.02$ mag. We find that star A = 17.197 $\pm$ 0.021 mag and star B = 17.930 $\pm$ 0.024 mag. Since the source has a predicted $K$ magnitude of 17.20 from the light curve, we conclude that star A is the source and star B is the lens, as shown in Figure \ref{fig:lens+source}.

\begin{figure}[h!]
    \centering
    \includegraphics[width=0.5\linewidth]{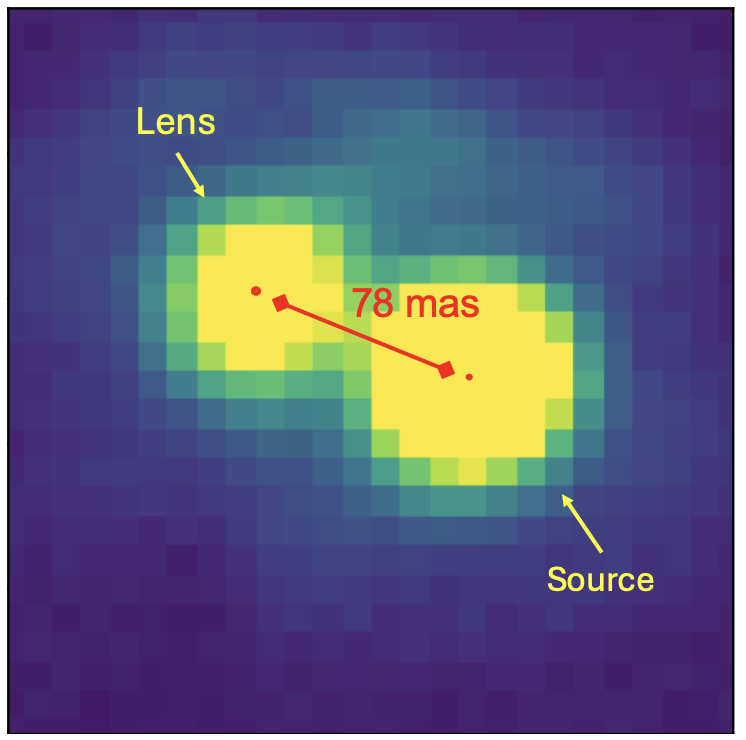}
    \caption{The lens and the source for MOA-2010-BLG-220 identified in the narrow NIRC2 Keck frame, with a separation of $\sim$ 78 mas.}
    \label{fig:lens+source}
\end{figure}

 {\setlength{\tabcolsep}{25pt}
\begin{table}[h!]
\centering
\caption{Separation, relative proper motion, flux ratio, and source and lens $K$ band magnitudes from the 2019 \gls{NIRC2} Keck data.}
\label{220table}
\begin{tabular}{l c c c}
\hline
\hline
Parameter & Units & Value & $\pm$ Error \\
\hline
$S_{\rm East}$ & mas & 73.112 &  0.431 \\   
$S_{\rm North}$ & mas & 29.449 &  0.439 \\
$S_{\rm Total}$ & mas & 78.821 &  0.423 \\
\hline
$\mu_{\rm rel, East}$ & mas/yr & 11.877 &  0.073 \\
$\mu_{\rm rel, North}$ & mas/yr & 4.784 &  0.072 \\
$\mu_{\rm rel, Total}$ & mas/yr & 12.804 & 0.072 \\
\hline
$f_{\rm L}/f_{\rm S}$ & - & 0.509 & 0.008 \\
$K_{\rm Lens}$ & mag & 17.930 & 0.024 \\
$K_{\rm Source}$ & mag & 17.197 & 0.021 \\
\hline
\end{tabular}
\end{table}}

\subsubsection{Mass and Distance measurements}

With a relative proper motion value and a lens flux measurement, we can now calculate the host mass and distance, and planet mass for MOA-2013-BLG-220. Looking back at section \ref{sec: high res} we can use equations \ref{eq: mass vs luminosity} and \ref{mu_rel} to get mass-distance relations. We plot both relations in Figure \ref{fig:220md}. Since the MOA-2013-BLG-220 host is a solar-type star, we use isochrones instead of an empirical mass luminosity relation. The isochrones used here are from the 6.4 Gyr population from \cite{Girardi2002}. We obtain the mass and distance from where they intersect. 

There is still one unknown parameter, however, the distance to the source star. Source stars are often assumed to be in the bulge and are therefore designated an approximate distance of $\sim8$ kpc. For a more precise measurement, a Galactic model can be used. A Galactic model provides a statistical framework for understanding the distribution of stars and their distances in our Galaxy. The source distance can be included in the light curve as a model parameter, using a Galactic model as a prior \cite{Koshimoto_2021Galactic}. 

To find the mass of the planet, we need to know the host-planet mass ratio, $q$, which is a property derived from the light curve model. It is standard practice to remodel the light curve using constraints from high resolution imaging. We can use constraints on the source flux and the relative proper motion. This was performed in \cite{Vandorou2020}, where they found a new value of $q = (2.99 \pm 0.03)\times 10^{-3}$. In this case, the light curve properties do not change much with the additional constraints from the Keck data. However, for some events, the difference can be large, indicating either that not all light curve scenarios were explored, or a problem with the high resolution imaging data.  

Finally, we estimate the MOA-2013-BLG-220 host to be a solar-type star with mass $M_{\rm L} = 0.88 \pm 0.05\ M_{\odot}$ at a distance of $D_{\rm L} = 6.72 \pm 0.59$ kpc, with a $M_{\rm p} = 2.74 \pm 0.17\ M_{\rm J}$ companion at a projected separation of $3.03 \pm 0.27$ AU.

\begin{figure}[h!]
    \centering
    \includegraphics[width=0.7\linewidth]{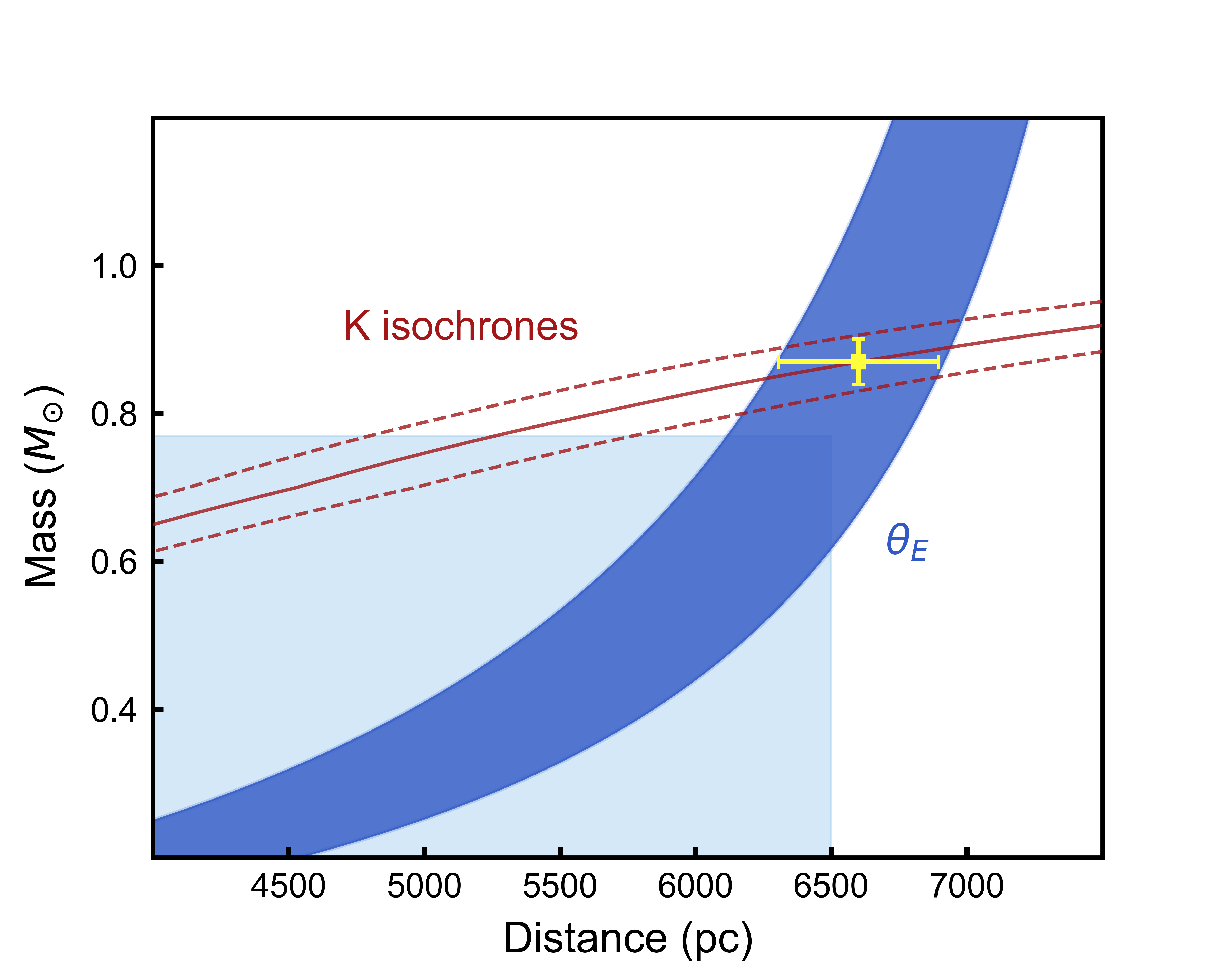}
    \caption{Mass as a function of distance for the host star of microlensing event MOA-2013-BLG-220. We show two mass distance relations, one using $\theta_{\rm E}$ (blue), and the other using isochrones and the measure lens $K$ magnitude (red). The mass and distance for this event is where these relations intersect (yellow). The faintly filled in region indicate the upper limits ($M_{\rm L} < 0.77\ M_{\odot}$, $D_{\rm L} < 6.5$ kpc) from \cite{Yee2014}.}
    \label{fig:220md}
\end{figure}

\clearpage
\section{Table of Available Instruments}
\begin{table}[h!]
\centering
\begin{tabular}{|p{3cm}|p{3cm}|p{2cm}|p{6cm}|}
\hline
\textbf{Project} & \textbf{Lead Organization} & \textbf{Start Date} & \textbf{Status} \\ \hline
\hline
Optical Gravitational Lensing Experiment (OGLE) & University of Warsaw & 1992 & 1.3-meter telescope in Chile. \\ \hline
Microlensing Observations in Astrophysics (MOA) & Nagoya University and University of Auckland & 1995 & 1.8-meter telescope in New Zealand. \\ \hline
Korea Microlensing Telescope Network (KMTNet) & Korea Astronomy and Space Science Institute & 2009 & 3 sets of 1.6-meter telescopes in the Southern Hemisphere.\\ \hline
PRime-focus Infrared Microlensing Experiment (PRIME) & Osaka University & 2024 & 1.8-meter near-infrared telescope at South Africa’s Sutherland Observatory, designed to complement  \gls{Roman}. \\ \hline
Vera C. Rubin Observatory (Rubin) & U.S. National Science Foundation and the U.S. Department of Energy & 2025 & The 8.4-meter telescope in Chile released first light images on 23 June 2025. Full survey operations begin later 2025, with data public after two years. \\ \hline
\textit{Nancy Grace Roman Space Telescope} & NASA & 2027 (scheduled) & 2.4-meter mirror and advanced infrared detectors to conduct wide-field surveys of the galactic bulge. \\ \hline
Earth 2.0 telescope & Chinese Academy of Sciences & 2028 (scheduled) & One of the seven telescopes of this space-based observatory will complement  \gls{Roman}’s observations. Under development in China. \\ \hline
\end{tabular}
\caption{Summary of various telescopes and their details. Adapted from https://spectrum.ieee.org/rogue-planet with additional information added.}
\label{tab:telescopes}
\end{table}

\section{Table of Available Open Source Tools}

\begin{table}[h!]
    \centering
    \caption{List of available Open Source Tools. LC = Light Curve. }
    \label{Tools}
    \renewcommand{\arraystretch}{1.2} 
    \begin{tabular}{|p{2cm}|p{2.0cm}|p{8.5cm}|p{1.5cm}|}
        \hline
        \textbf{Tool name} & \textbf{Function} & \textbf{GitHub Repository} & \textbf{Publ.} \\
        \hline
        \hline
        \small{PyLIMA} & LC model & \makecell[tl]{\scriptsize \url{https://github.com/ebachelet/pyLIMA}} & \scriptsize{\cite{Bachelet2017}} \\   
        \hline
        \small{RTModel} & \small{LC model} & \scriptsize{\url{https://github.com/valboz/RTModel}}& \scriptsize{\cite{Bozza2024}}\\
        \hline
        \small{VBMicrole-nsing} & \small{Microlensing event computation} & \makecell[tl]{\scriptsize \url{https://github.com/valboz/VBMicrolensing}} & \scriptsize{\cite{Bozza2010}} \\
        \hline
        \small{eesunhong} & \small{LC model} & \makecell[tl]{\scriptsize \url{https://github.com/golmschenk/eesunhong}} & \scriptsize{\cite{Bennett1996}} \\
        \hline
        \small{MOAna} & \small{\texttt{eesunhong} reader} & \makecell[tl]{\scriptsize \url{https://github.com/clementranc/moana}} &  \scriptsize{\cite{moana}} \\
        \hline
        \small{muLAN} & \small{LC model} & \makecell[tl]{\scriptsize \url{https://github.com/muLAn-project/muLAn}} &  \scriptsize{\cite{ranc2018mulan}} \\
        \hline
        \small{MuLens Model} & \small{LC model} & \makecell[tl]{\scriptsize \url{https://github.com/rpoleski/MulensModel/releases/tag/v1.0.0}} & \scriptsize{\cite{Poleski2018}} \\
        \hline
        Bagel & \small{LC model} & \makecell[tl]{\scriptsize \url{https://github.com/MovingUniverseLab/BAGLE_Microlensing}} & \\
        \hline
        \small{PopSyCLE} & \small{Simulation} & \makecell[tl]{\scriptsize \url{https://github.com/jluastro/PopSyCLE/tree/main}} & \scriptsize{\cite{Lam2020_popcycle}} \\
        \hline
        \small{KAI} & \small{Keck pipeline} & \makecell[tl]{\scriptsize \url{https://github.com/Keck-DataReductionPipelines/KAI}} & \scriptsize{\cite{Lu2022}} \\
        \hline
        \small{DAOPHOT-MCMC} & \small{Astrometry \& photometry} & \makecell[tl]{\scriptsize \url{https://github.com/skterry/daophot_mcmc}} & \scriptsize{\cite{Terry_2021}} \\
        \hline
    \end{tabular}
    \label{tab:open_source_tools}
\end{table}

\newpage

\section{Acknowledgments}
We would first like to thank the reader for making it this far in the book and for taking the time to learn about microlensing. We hope it has been an enjoyable and informative experience. We would also like to thank the reviewer for their valuable comments, as well as the senior scientists in the microlensing community whose historical perspective and scientific expertise helped bring this chapter to fruition. Finally, we would like to thank our friends, families, and partners, whose support, encouragement, and patience help sustain us throughout our scientific endeavours.


\clearpage
\bibliographystyle{yahapj}
\bibliography{sample}

\newpage
\glsnogroupskiptrue
\printglossary[type=\acronymtype,title={List of Abbreviations}]
        
\end{document}

%% file: abbreviations.tex
\makeglossaries

\newcommand{\eg}{\textit{e.g.}}

\newacronym{1L1S}{1L1S}{single-lens single-source}
\newacronym{2L1S}{2L1S}{binary-lens single-source}
\newacronym{3L1S}{3L1S}{triple-lens single-source}
\newacronym{1L2S}{1L2S}{single-lens binary-source}
\newacronym{2L2S}{2L2S}{binary-lens binary-source}
\newacronym[user1={\citep{schwarz1978estimating}}]{BIC}{BIC}{Bayesian Information Criterion}
\newacronym{CCD}{CCD}{charged-coupled device}
\newacronym{CV}{CV}{cataclysmic variables}
\newacronym{FOV}{FOV}{field-of-view}
\newacronym{FFI}{FFI}{full-frame image}
\newacronym{FFP}{FFP}{free-floating planet}
\newacronym{FSPL}{FSPL}{finite-source point-lens}
\newacronym{LC}{LC}{light curve}
\newacronym{LMC}{LMC}{Large Magellanic Cloud}
\newacronym{PDF}{PDF}{probability density function}
\newacronym{PSF}{PSF}{point spread function}
\newacronym{PSPL}{PSPL}{point-source point-lens}
\newacronym{SNR}{SNR}{signal-to-noise ratio}
\newacronym{WFI}{WFI}{Wide Field Instrument}

\newacronym[description={Canada–France–Hawaii Telescope}]{CFHT}{CFHT}{the Canada–France–Hawaii Telescope}
\newacronym[user1={\citealp{auvergne2009corot}}, sort={CoRoT}]{CoRoT}{\textit{CoRoT}}{\textit{Convection, Rotation et Transits Plan\'etaires}}
\newacronym[description={Centre National d'\'Etudes Spatiales (the French national space agency)}]{CNES}{CNES}{the French national space agency}
\newacronym[description={Cerro Tololo Inter-American Observatory}]{CTIO}{CTIO}{the Cerro Tololo Inter-American Observatory}
\newacronym[user1={\citealp{Aubourg1993}}]{EROS}{EROS}{Exp\'erience de Recherche d'Objects Sombres}
\newacronym{ESA}{ESA}{the European Space Agency}
\newacronym[user1={\citealp{exofoptess}}]{ExoFOP}{ExoFOP-\textit{TESS}}{the Exoplanet Follow-up Observing Program for \textit{TESS}}
\newacronym[user1={\citealp{brown2018gaia}}, sort={Gaia}]{Gaia}{\textit{Gaia}}{the \textit{Gaia Mission}}
\newacronym[user1={\citealp{zimmerman2010universe}}, sort={Hubble}, description={\textit{Hubble Space Telescope}}]{Hubble}{\textit{Hubble}}{the \textit{Hubble Space Telescope}}
\newacronym{KAI}{KAI}{Keck AO Imaging }
\newacronym[user1={\citealp{borucki2010kepler}}, sort={Kepler}]{Kepler}{\textit{Kepler}}{the \textit{Kepler Space Telescope}}
\newacronym[user1={\citealp{howell2014k2}}, sort={K2}]{K2}{\textit{K2}}{the \textit{Kepler's Second Light Mission}}
\newacronym[user1={\citealp{Kim2016}}, description={Korea Microlensing Telescope Network}]{KMTNet}{KMTNet}{the Korea Microlensing Telescope Network}
\newacronym{LSST}{LSST}{Legacy Survey of Space and Time}
\newacronym{MACHO}{MACHO}{MAssive Compact Halo Object}
\newacronym[user1={\citealp{Gould2010}}]{muFUN}{$\mu$FUN}{Microlensing Follow Up Network}
\newacronym[user1={\citealp{Diminik2010}}]{MiNDSTEp}{MiNDSTEp}{Microlensing Network for the Detection of Small Terrestrial Exoplanets}
\newacronym[user1={\citealp{sumi2013microlensing}}, description={Microlensing Observations in Astrophysics}]{MOA}{MOA}{Microlensing Observations in Astrophysics}
\newacronym{NASA}{NASA}{the National Aeronautics and Space Administration}
\newacronym{NIRC2}{NIRC2}{Near Infrared Camera Second Generation}
\newacronym[user1={\citealp{Udalski2003}}, description={Optical Gravitational Lensing Experiment}]{OGLE}{OGLE}{the Optical Gravitational Lensing Experiment}
\newacronym{OSIRIS}{OSIRIS}{OH-Suppressing Infrared Imaging Spectrograph}
\newacronym[user1={\citealp{Albrow1998}}]{PLANET}{PLANET}{Probing Lensing Anomalies NETwork}
\newacronym[user1={\citealp{sumi2025prime}}]{PRIME}{PRIME}{PRime-focus Infrared Microlensing Experiment}

\newacronym[user1={\citealp{spergel2015}}, sort={Roman}, description={\textit{Nancy Grace Roman Space Telescope} \small{(formerly the \textit{Wide Field InfraRed Survey Telescope})}}]{Roman}{\textit{Roman}}{the \textit{Nancy Grace Roman Space Telescope} \small{(formerly the \textit{Wide Field InfraRed Survey Telescope})}}

\newacronym[description={South African Astronomical Observatory}]{SAAO}{SAAO}{the South African Astronomical Observatory}
\newacronym[description={Siding Spring Observatory}]{SSO}{SSO}{the Siding Spring Observatory}
\newacronym[user1={\citealp{ricker2014transiting}}, sort={TESS}]{TESS}{\textit{TESS}}{the \textit{Transiting Exoplanet Survey Satellite}}
\newacronym[user1={\citealp{stassun2018tess}}]{TIC}{TIC}{\textit{TESS} input catalog}
\newacronym[user1={\citealp{gardner2006james}}, sort={Webb}, description={\textit{James Webb Space Telescope}}]{Webb}{\textit{Webb}}{the \textit{James Webb Space Telescope}}

\newacronym[user1={\citealp{kingma2014adam}}]{Adam}{Adam}{Adaptive Moment Estimation}
\newacronym{AI}{AI}{artificial intelligence}
\newacronym[user1={\citealp{lecun1998gradient,krizhevsky2012imagenet}}]{CNN}{CNN}{convolutional neural network}
\newacronym[user1={\eg, \citealp{lecun2015deep}}]{DNN}{DNN}{deep neural network}
\newacronym{GPU}{GPU}{graphics processing unit}
\newacronym[user1={\citealp{clevert2015fast}}]{ELU}{ELU}{exponential linear unit}
\newacronym[user1={\citealp{maas2013rectifier}}]{Leaky ReLU}{Leaky ReLU}{Leaky rectified linear unit}
\newacronym[user1={\citealp{hochreiter1997long}}]{LSTM}{LSTM}{long short-term memory network}
\newacronym{MCMC}{MCMC}{Markov Chain Monte Carlo}
\newacronym{ML}{ML}{machine learning}
\newacronym{NN}{NN}{neural network}
\newacronym[user1={\citealp{nair2010rectified,glorot2010understanding,sun2015deeply}}]{ReLU}{ReLU}{rectified linear unit}
\newacronym[user1={\citealp{elman1990finding}}]{RNN}{RNN}{recurrent neural network}
\newacronym{VVV}{VVV}{Vista Variables in the Via Lactea}